\documentclass[12pt]{article}%
\usepackage[nosort]{cite}
\usepackage{graphicx}
\usepackage{multicol}
\usepackage{amsfonts}
\usepackage{amssymb}
\usepackage{amsmath}
\usepackage{dsfont}
\usepackage{heck}
\usepackage{afterpage}
\usepackage{setspace}
\usepackage{verbatim}
\usepackage{color}
\usepackage{longtable}
\usepackage{float}
\usepackage{subcaption}
\usepackage{epsfig}
\usepackage{epstopdf}
\usepackage{adjustbox}
\usepackage{todonotes}
\usepackage[makeroom]{cancel}
\usepackage{tikz}
\usepackage[margin=1in]{geometry}
\usepackage{titletoc}
\usepackage{hyperref}%
\providecommand{\U}[1]{\protect\rule{.1in}{.1in}}
\pdfoutput=1
\newsavebox{\mysavebox}

\usetikzlibrary[shapes.geometric]
\usetikzlibrary{positioning}
\usetikzlibrary{calc,intersections,through,backgrounds}
\tikzset{>=stealth}
\usetikzlibrary{decorations.pathmorphing}
\usetikzlibrary{decorations.markings}
\tikzset{>=stealth}
\hypersetup{colorlinks,citecolor=black,filecolor=black,linkcolor=black,urlcolor=black,pdftex}
\usetikzlibrary{decorations.markings}

\numberwithin{equation}{section}

\newcommand{\ba}{\begin{eqnarray}}
\newcommand{\ea}{\end{eqnarray}}

\newcommand{\be}{\begin{equation}}
\newcommand{\ee}{\end{equation}}

\def\zb{{\overline{z}}}

\def\zb{{\overline{z}}}
\newcommand{\ttwoD}{\theta^+}
\newcommand{\tbtwoD}{\overline{\theta}{}^+}
\tikzstyle{startstop} = [rectangle, rounded corners, minimum width=3cm, minimum height=1cm,text centered, draw=black, fill=blue!10]
\tikzstyle{startstop} = [rectangle, rounded corners, minimum width=3cm, minimum height=1cm,text centered, draw=black, fill=blue!10]
\tikzstyle{io} = [trapezium, trapezium left angle=70, trapezium right angle=110, minimum width=3cm, minimum height=1cm, text centered, draw=black, fill=blue!30]
\tikzstyle{process} = [rectangle, minimum width=3cm, minimum height=1cm, text centered, draw=black, fill=orange!30]
\tikzstyle{decision} = [diamond, minimum width=3cm, minimum height=1cm, text centered, draw=black, fill=green!30]
\tikzstyle{arrow} = [thick,->,>=stealth]
\begin{document}

\date{October 2016}

\title{From 6D SCFTs to Dynamic GLSMs}

\institution{UNC}{\centerline{${}^{1}$Department of Physics, University of North Carolina, Chapel Hill, NC 27599, USA}}

\institution{JMU}{\centerline{${}^{2}$Department of Physics, James Madison University, Harrisonburg, VA 22807, USA}}

\authors{Fabio Apruzzi\worksat{\UNC}\footnote{e-mail: {\tt fabio.apruzzi@unc.edu}},
Falk Hassler\worksat{\UNC}\footnote{e-mail: {\tt fhassler@unc.edu}},\\[4mm]
Jonathan J. Heckman\worksat{\UNC}\footnote{e-mail: {\tt jheckman@email.unc.edu}},
and Ilarion V. Melnikov\worksat{\JMU}\footnote{e-mail: {\tt melnikix@jmu.edu}}}

\abstract{Compactifications of 6D superconformal field theories (SCFTs) on four-manifolds generate
a large class of novel 2d quantum field theories.
We consider in detail the case of the rank one simple non-Higgsable cluster 6D SCFTs.
On the tensor branch of these theories, the gauge group is simple and there are no matter fields.
For compactifications on suitably chosen K\"ahler surfaces, we present
evidence that this provides a method to realize
2d SCFTs with $\mathcal{N} = (0,2)$ supersymmetry. In particular, we find that
reduction on the tensor branch of the 6D SCFT yields a description of the same 2d fixed point
that is described in the UV by a gauged linear sigma model (GLSM)
in which the parameters are promoted to dynamical fields, that is, a ``dynamic GLSM'' (DGLSM).
Consistency of the model requires the DGLSM to be coupled to additional non-Lagrangian sectors obtained from
reduction of the anti-chiral two-form of the 6D theory. These extra sectors include both chiral
and anti-chiral currents, as well as spacetime filling non-critical strings of the 6D theory.
For each candidate 2d SCFT, we also extract the left- and right-moving central charges in terms of data of
the 6D SCFT and the compactification manifold.}

\maketitle

\tableofcontents

\enlargethispage{\baselineskip}

\setcounter{tocdepth}{2}

\newpage

\section{Introduction \label{sec:INTRO}}

One of the notable predictions of string theory is the existence of 6D superconformal field theories (SCFTs)
\cite{Witten:1995zh, Strominger:1995ac, Seiberg:1996qx}.
Though a microscopic formulation of these theories remains elusive, it is especially remarkable that upon
compactification to lower dimensions, simple geometric operations of the
compactification space lead to highly non-trivial dualities. This includes the
well-known case of compactification of the $(2,0)$ SCFTs on a $T^{2}$, and the
corresponding S-duality of $\mathcal{N}=4$ Super Yang-Mills theory
\cite{Montonen:1977sn, Vafa:1997mh}. Compactifying on Riemann
surfaces with punctures leads to $\mathcal{N}=2$ dualities \cite{Witten:1997sc, Argyres:2007cn,
Gaiotto:2009we}. Similar considerations hold for compactification to three,
two and one dimension (for a partial list of examples,
see \cite{Maldacena:1997de, Witten:2011zz, Dimofte:2011ju, Gadde:2013sca, Gukov:2016gkn}).

It is natural to ask whether similar structures persist for compactifications
of 6D\ SCFTs with minimal, i.e., $(1,0)$ supersymmetry. Recent work on
6D\ SCFTs with a tensor branch \cite{Heckman:2013pva, Gaiotto:2014lca, DelZotto:2014hpa,
Heckman:2014qba, DelZotto:2014fia, Heckman:2015bfa, Bhardwaj:2015xxa} has
produced a classification of nearly all such theories.\footnote{More
precisely, this classification involves geometric phases of F-theory
backgrounds. It is expected that the small number of non-geometric backgrounds
recently discussed in \cite{Tachikawa:2015wka} (see \cite{Hanany:1997gh}
for the type IIA realization of these theories) can be included through a
suitable generalization of these earlier results \cite{Bhardwaj:2015oru}.}
See also \cite{Apruzzi:2013yva, Apruzzi:2015wna} for a classification
of theories with a holographic dual.

Compactifications of these $(1,0)$ SCFTs to lower dimensions have the
potential to provide access to strong coupling dynamics in the resulting
lower-dimensional theories. One complication is that due to the reduced amount
of supersymmetry, any such theory will be subject to more quantum corrections
than their $(2,0)$ counterparts. Nevertheless, it is still possible to extract
some data for the 4d theories obtained from compactification, as in references
\cite{DelZotto:2015isa, Gaiotto:2015usa, Ohmori:2015pua, Franco:2015jna, DelZotto:2015rca,
Hanany:2015pfa, Aganagic:2015cta, Ohmori:2015pia, Coman:2015bqq, Morrison:2016nrt, Heckman:2016xdl}.

\begin{figure}[t!]
\begin{center}
\scalebox{1}[1]{
\includegraphics[scale=0.5]{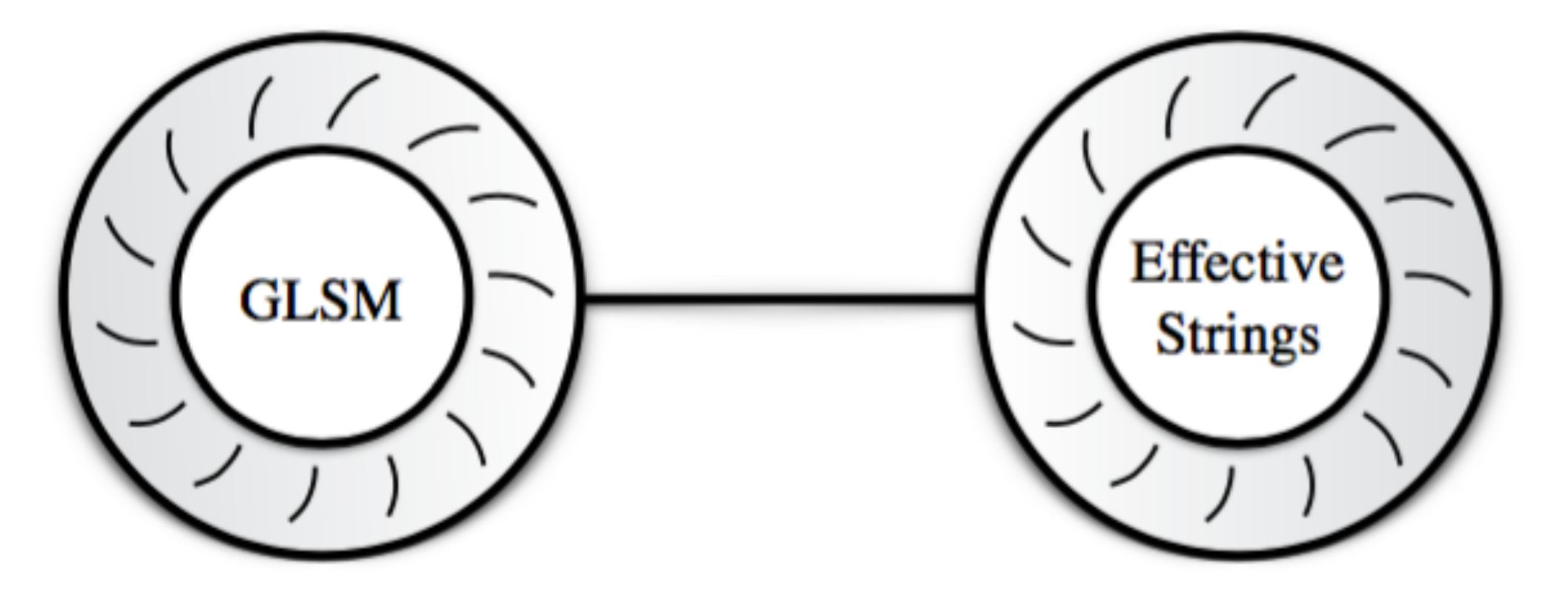}}
\end{center}
\caption{Compactification of the tensor branch deformation of a 6D SCFT on a four-manifold yields a DGLSM, i.e. a
gauged linear sigma model in which the couplings are dynamical. These theories are also coupled
to chiral / anti-chiral currents and spacetime filling strings.}
\label{fig:genquiver}
\end{figure}

In this work we begin the study of the resulting 2d effective theories
obtained by compactification of $(1,0)$ SCFTs on a four-manifold. We find that
the resulting 2d theories are often 2d\ SCFTs, and moreover, are characterized
by an appropriate UV gauged linear sigma model (GLSM) in which some of the
parameters of the model are promoted to dynamical chiral and
anti-chiral bosons. These gauge theories are often coupled to additional chiral / anti-chiral
currents as well as spacetime filling strings of the 6D theory (see figure \ref{fig:genquiver} for a depiction).
We also use this geometric characterization to produce parametric families of 2d\ SCFTs.

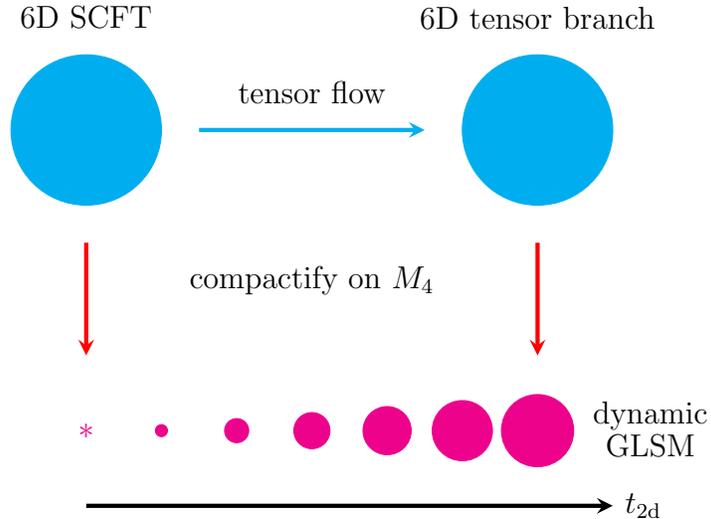
\begin{figure}[t]
\centering
\begin{tikzpicture}
\draw  (0,1.5) node {6D SCFT};
\filldraw[cyan] (0,0) circle[radius=1];
\draw (6,1.5) node {6D tensor branch};
\filldraw[cyan] (6,0) circle[radius=1];
\draw (3,0.5) node {tensor flow};
\draw[cyan, ultra thick,->] (1.5,0) -- (4.5,0) ;
\draw (3,-2) node {compactify on $M_4$};
\draw[red, ultra thick, ->] (0,-1.5) -- (0,-3);
\draw[red, ultra thick, ->] (6,-1.5) -- (6,-3);
\draw[ultra thick,->] (0,-5) -- (7,-5) node[right] {$t_{\text{2d}}$};
\draw[magenta] (0,-4) node {$\ast$};
%
\foreach \x in {6,5,4,3,2,1,0}
\filldraw[magenta](\x,-4) circle[radius = 0.08*\x];
%
\draw (7.5,-3.8) node {dynamic};
\draw(7.5,-4.2) node {GLSM};
\end{tikzpicture}
\caption{Depiction of the different theories starting from compactification of
a 6D theory. We can either begin with a 6D SCFT and compactify directly to two
dimensions, obtaining a strongly coupled candidate 2d SCFT indicated by
{\color{magenta}{$\ast$}}. Alternatively, we can instead consider
compactification of the 6D theory far out on its tensor branch, arriving at a
weakly coupled dynamic GLSM (the coupling is indicated by the radius of the
circle) fibered over the tensor non-linear sigma model. The match in anomalies
for the two 2d theories so obtained suggests that {\color{magenta}{$\ast$}} is
at finite distance in the appropriate sigma model metric.}%
\label{anomomatch}%
\end{figure}

Now, although these $(1,0)$ theories have reduced supersymmetry when compared
with their $(2,0)$ counterparts, this is compensated by the fact that the
theory on the tensor branch typically has more structure. One of the central
points of this work will be to exploit this description on the tensor branch
to achieve a better understanding of the microscopic ingredients of the
resulting 2d effective field theories. Figure \ref{anomomatch} shows the
different RG\ flow trajectories:\ we can either directly compactify our
6D\ SCFT on a four-manifold, reaching a 2d effective theory, or we can first
flow on the tensor branch and then compactify the theory on this branch,
yielding an a priori different 2d effective theory. In spite of their differences in six dimensions,
we will present evidence that the two a priori different 2d theories obtained from compactification
are in fact one and the same.

Indeed, for $(1,0)$ theories, the tensor branch is governed by a weakly
coupled 6D gauge theory coupled to tensor multiplets and possibly exotic
matter fields. The scalar of the tensor multiplet promotes the gauge coupling
to a dynamical field, and the 6D anti-chiral two-form potential is crucial for 6D
anomaly cancellation via the Green-Schwarz mechanism \cite{Green:1984sg,
Green:1984bx, Sagnotti:1992qw, Sadov:1996zm}.

Upon compactification on a four-manifold, the 6D\ vector multiplet and matter fields reduce
to fields present in a 2d non-abelian\ GLSM. More striking, however, is that
the gauge coupling of the GLSM\ will be dynamical, and the anti-chiral
two-forms will reduce to 2d chiral and anti-chiral bosons which couple to the
fields of the GLSM. Additionally, the presence of spacetime filling effective
strings (as dictated by the tadpole for the anti-chiral two-form)\ provides
additional chiral sectors which couple to the GLSM.

What then is the relation of this 2d theory to the one obtained by directly
compactifying the 6D\ SCFT on a four-manifold? We propose the following
answer. There is one theory with dynamical couplings that depends on fields of
an underlying non-linear sigma model with non-compact target space. The target
space is non-compact due to unsuppressed contributions of large values of the
tensor scalar field, and precisely in this regime we have a description in
terms of a weakly coupled theory fibered over the non-linear sigma model
target space. Of course no such weakly coupled description can be given at the
origin of the tensor branch, and experience with similar strongly coupled
systems might at first suggest these two theories are at infinite distance
from one another. On the contrary, we will present evidence that the origin is
at finite distance in the target space metric.

The primary evidence we provide comes from compactifying on a K\"ahler surface,
where we expect to get a 2d SCFT with $(0,2)$ supersymmetry. In this
case, we can calculate the anomaly polynomial for the 2d theory by dimensional
reduction of the 6D answer. We compare this with the anomaly polynomial
obtained from reduction of the 6D theory on the tensor branch, and we obtain a
perfect match.

The fact that this picture hangs together in a consistent fashion also helps
to address questions of interest in a purely 2d context. For
example, although $(0,2)$ GLSMs are widely studied, the case where the
parameters of the model are promoted to dynamical fields (let alone chiral
fields) is not well understood, even though, as we discuss below, it is
actually the generic case! While there are many such theories one might write
down, a priori it is entirely unclear which of them will yield sensible
results. The top--down approach provides us with candidates with which we can
begin to explore this wider world of 2d dynamics, and even in
these relatively simple theories, we find a rich structure including the
following features:


\begin{enumerate}
\item Gauge degrees of freedom with dynamical couplings.

\item A non-compact target space and therefore no normalizable vacuum state.

\item A rich current algebra, part of which is gauged, that originates from
reduction of the 6D anti-chiral two-form.

\item Marginal couplings that originate from the geometry of the compactification manifold.
\end{enumerate}

To provide more details of this picture, our focus in this work will be on
compactification of a class of 6D\ SCFTs we refer to as
\textquotedblleft simple non-Higgsable clusters.\textquotedblright\ The
non-Higgsable clusters (NHCs) of reference \cite{Morrison:2012np}
are the building blocks used in the construction of all 6D\ SCFTs
via F-theory \cite{Heckman:2013pva, Heckman:2015bfa}. In F-theory terms, these theories are
characterized by a collection of $\mathbb{P}^{1}$'s in the base such that the
generic elliptic fibration over these curves is singular. Non-Higgsable
clusters consist of up to three $\mathbb{P}^{1}$'s, with the specific
configuration of gauge groups and matter fully dictated by the
self-intersection of the curve. In particular, the simple NHCs (SNHCs)
consist of a single $\mathbb{P}^{1}$ of self-intersection $-n$ with
$n\in\{3,4,6,8,12\}$. For these choices, the gauge group is simple and there are no
matter fields. An additional feature of these cases is that the
F-theory model can be written (for appropriate moduli) as an orbifold $\mathbb{C}^{2}\times T^{2}/%
\mathbb{Z}
_{n}$.

Reduction of the 6D\ Green-Schwarz terms leads to an intricate anomaly
cancellation mechanism in two dimensions for these theories. In addition to
the chiral and anti-chiral bosons obtained from reduction of the
6D\ anti-chiral two form, we also find spacetime filling strings. These
theories are in turn defined by the dimensional reduction of a strongly
coupled $\mathcal{N}=2$ superconformal field theory on $\mathbb{R}^{1,1}%
\times\mathbb{P}^{1}$. The number of such spacetime filling strings is
dictated by the precise coefficients appearing in the 6D Green-Schwarz terms.
For recent work on the structure of these spacetime filling strings, see
references \cite{Gadde:2015tra, Kim:2016foj, Shimizu:2016lbw,
DelZotto:2016pvm}.

To perform precision checks on our proposal, we also specialize to the case of
a K\"{a}hler surface so that the resulting 2d effective theory enjoys $(0,2)$
supersymmetry. In this case, a putative superconformal field theory will have
a $U(1)$ R-symmetry. Assuming the absence of emergent abelian symmetries in
the infrared, this R-symmetry descends from symmetries already present in the
6D description. For the simple NHCs there are in fact no continuous flavor
symmetries -- abelian or non-abelian -- in six dimensions. This in particular
makes an analysis of the infrared R-symmetry particularly tractable. Matching
all of the data of the anomalies including the infrared R-symmetry, we obtain
a non-trivial match between the two a priori different 2d theories.

The fact that the anomalies of the two theories match is a strong indication
that we are simply describing the same branch of a single SCFT. While it is
indeed quite plausible that compactifying a 6D\ SCFT will produce a 2d\ SCFT,
it is quite non-trivial that the massive tensor branch flows back to the same
fixed point upon further compactification. We take this to mean that the
distinction between these two theories is erased once we integrate out the
massive Kaluza-Klein states.

Indeed, from the perspective of the F-theory realization of these models, in
both phases we have a seven-brane wrapped over a K\"{a}hler threefold
$M_{4}\times\mathbb{P}^{1}$. To reach the compactification of the 6D\ SCFT, we
first collapse the $\mathbb{P}^{1}$ factor to zero size. To instead reach the
DGLSM description, we first collapse $M_{4}$ to zero size. This also motivates a
physical picture in which the complexified geometric moduli of $M_{4}$
parameterize a family of 2d SCFTs.

The rest of this paper is organized as follows. First, in section
\ref{sec:NHCreview}, we introduce the class of 6D SCFTs we shall study, i.e. the simple NHCs.
We also study the twist of the 6D SCFT, and the reduction of the anomaly polynomial to two dimensions.
One of our goals will be to reproduce this answer directly from the
2d theory obtained from reduction on the tensor branch. Next, in section \ref{sec:TWOD} we
discuss some general aspects of DGLSMs, and in particular how the 6D perspective
helps in understanding various aspects of the quantum dynamics. In section \ref{sec:SEVEN} we
turn to the explicit F-theory realization of DGLSMs. Section \ref{sec:ANOMO} shows that
all gauge anomalies for the DGLSMs vanish, and moreover, calculates the anomaly polynomial for the resulting theories
wrapped on a K\"ahler surface. We present our conclusions and directions for future investigation in section \ref{sec:CONC}.
Some additional details on the reduction of free 6D supermultiplets to two dimensions are given in Appendix \ref{app:FREE}. Additional details on the spacetime filling strings of the simple NHCs are provided in Appendix \ref{app:MN}.

\section{6D SCFTs on a Four-Manifold \label{sec:NHCreview}}

In this section we review some of the salient points of 6D\ SCFTs, and their
compactification on a four-manifold. The primary evidence for the existence of
higher-dimensional conformal fixed points comes from string theory, so we
shall mainly adhere to stringy conventions in our construction and study of
such models. An important feature of all known 6D\ SCFTs is that they admit a
flow onto a tensor branch where all stringlike excitations have
finite tension. This is actually where we realize a string
construction, and then by taking an appropriate degeneration limit, we pass to
the conformal fixed point.

Our aim will be to study the effective field theory obtained from
compactifying a 6D\ SCFT on a four-manifold. Compactifications of 6D SCFTs
with $\mathcal{N}=(2,0)$ supersymmetry have been studied for example in
references \cite{Gadde:2013sca, Ganor:1996xg, Assel:2016lad}.

Experience from higher-dimensional examples strongly suggests that for
suitable compactification manifolds (for example, when the curvature of the
internal space is negative), we should expect to realize a 2d\ superconformal
field theory. This is also borne out by the fact that 6D SCFTs with a holographic dual \cite{Apruzzi:2013yva, Apruzzi:2015wna}
realize, upon compactification on a negatively curved space, an $AdS_3$
holographic dual (see e.g. \cite{Gauntlett:2000ng, Gauntlett:2006ux, Figueras:2007cn, Passias:2015gya, Karndumri:2015sia}).
Even so, due to the non-Lagrangian nature of the UV fixed point,
this provides only indirect data about the resulting 2d theory such as
various anomalies of global symmetries.

Along these lines, we will also study the tensor branch deformation of the 6D
theory. On the tensor branch, the strings of the model couple to anti-chiral
two-form potentials with anti-self-dual three-form field strengths. Reduction of this
theory to two dimensions thus leads to a 2d supersymmetric gauge theory
coupled to additional sectors such as chiral and anti-chiral currents,
and spacetime filling strings.

Even though the 6D\ SCFT and tensor branch are distinct in six
dimensions, it is natural to ask how their compactifications are related in
two dimensions. Here, the peculiariaties of 2d systems show themselves. Since
we expect the tensor branch system to also flow to a 2d\ SCFT, we will
actually get two a priori distinct 2d\ SCFTs. However, using the 6D perspective, we will also
see that the anomalies for these two theories are the same, suggesting that in spite of appearances,
we are dealing with a single 2d SCFT. This will lead us to the
notion of a \textquotedblleft dynamic GLSM,\textquotedblright\ a construct we
discuss further in section \ref{sec:TWOD}.

The rest of this section is organized as follows. First, we discuss in general terms how to construct 6D SCFTs via F-theory. We then
discuss the partial twist of 6D SCFTs on a four-manifold, and then
for the tensor branch deformation of these theories. We then compute the reduction of the 6D anomaly polynomial to two dimensions.

\subsection{6D\ SCFTs from F-theory}

To provide concrete examples of 6D\ SCFTs, we will find it convenient to use
the F-theory realization of these models. Recently there has been substantial
progress in understanding the general structure of 6D\ SCFTs via
compactifications of F-theory on elliptically fibered Calabi-Yau threefolds.

Recall that in F-theory, we work with a ten-dimensional spacetime, but one in
which the axio-dilaton of type\ IIB\ string theory can have non-trivial
position dependence. For the purposes of engineering 6D
superconformal field theories, our primary interest will be spacetimes of
the form: $\mathbb{R}^{5,1}\times\mathcal{B}$, where the base $\mathcal{B}$ is
a non-compact K\"{a}hler surface. To reach a 6D\ SCFT, we must take some
collection of $\mathbb{P}^{1}$'s in $\mathcal{B}$ and simultaneously contract
them to zero size at finite distance in moduli space. This in turn requires
that the intersection pairing for these $\mathbb{P}^{1}$'s is negative definite
\cite{Heckman:2013pva}. The condition that we have a consistent F-theory
background requires that $\mathcal{B}$ is actually the base of a (non-compact)
elliptically fibered Calabi-Yau threefold. An important result from
\cite{Morrison:2012np} is that when the self-intersection $-n$ of the
$\mathbb{P}^{1}$ satisfies:%
\begin{equation}
3\leq n\leq12~, \label{ninequ}%
\end{equation}
then the elliptic fibration is singular over the curve. The upper bound of
$12$ comes about from the condition that we can place the elliptic fiber in
standard Kodaira-Tate form. For $n=1,2$ we instead have a generic fiber type
which is smooth. Collapsing this $\mathbb{P}^{1}$ generates a 6D SCFT
\cite{Heckman:2013pva,Witten:1996qb}.

Each value of $n$ determines a minimal singularity type of the elliptic fiber
and thus a 6D effective field theory with a UV\ cutoff. The
effective field theory consists of a single tensor multiplet because there is
just one $\mathbb{P}^{1}$. The volume of this $\mathbb{P}^{1}$ is the vev of
$t$, the scalar in the corresponding tensor multiplet. Additionally, we have a gauge group,
and in some limited cases, matter fields. Let us remark that in
addition to the rank one non-Higgsable clusters, there are also NHCs with two
and three curves, consisting of related intersection patterns $3,2$ as well as
$2,3,2$ and $3,2,2$.

All of the bases which appear in an F-theory model are constructed from
\textquotedblleft gluing\textquotedblright\ these NHCs together to form larger
structures. In physical terms, this comes from taking the theory on a $-1$
curve, which has $E_{8}$ flavor symmetry, and gauging a product subalgebra.
For further details on this gluing procedure in the context of 6D\ SCFTs, see
\cite{Heckman:2013pva, Heckman:2015bfa}.

We shall primarily focus on this class
of theories, and in particular the ones with values $n=3,4,6,8,12$. These are the
cases where the seven-brane has a simple gauge group, and there are no matter fields.
An additional feature of these cases is that for appropriate moduli, the elliptically fibered Calabi-Yau threefold
is simply the orbifold $\mathbb{C}^2 \times T^2 / \mathbb{Z}_n$ \cite{Heckman:2013pva,Witten:1996qb}.
See table \ref{tab:simp} for a list of the simple NHCs.

\begin{table}[t]
\begin{center}%
\begin{tabular}
[c]{|l|l|l|l|l|}\hline
$n=3$ & $n=4$ & $n=6$ & $n=8$ & $n=12$\\\hline
$SU(3)$ & $SO(8)$ & $E_{6}$ & $E_{7}$ & $E_{8}$\\\hline
\end{tabular}
\end{center}
\caption{List of Simple NHCs}%
\label{tab:simp}
\end{table}

\subsection{Twisting a 6D\ SCFT \label{ssec:TWISTSCFT}}

Suppose then, that we have realized a 6D SCFT via a compactification of F-theory.
We now turn to the study of their compactification on a four-manifold $M_{4}$. At first, we
consider a general Riemannian four-manifold, though we shall see that to
preserve at least $\mathcal{N}=(0,2)$ supersymmetry we will need to specialize to the
case of a K\"{a}hler surface.

To begin, recall that since we have a field theory, there is a local stress
energy tensor. As such, we should expect that with sufficient supersymmetry,
it is possible to perform a topological twist of the theory.
Recall that a 6D\ SCFT has an $\mathfrak{su}(2)\simeq\mathfrak{sp}(1)_{R,6D}$
R-symmetry, so in compactifying on a four-manifold, we seek possible ways to
retain a covariantly constant spinor on the internal space.  The
supercharges of the 6D $\mathcal{N}=(1,0)$ theory transform as
a symplectic Majorana-Weyl spinor in the representation $(\mathbf{4}%
,\mathbf{2})$ of $\mathfrak{so}(5,1)\times\mathfrak{sp}(1)_{R,6D}$. Making the
further decomposition to $\mathfrak{so}(1,1)\times\mathfrak{so}(4)\times
\mathfrak{sp}(1)\simeq\mathfrak{so}(1,1)\times\mathfrak{su}(2)_{L}%
\times\mathfrak{su}(2)_{R}\times\mathfrak{sp}(1)_{R,6D}$, we have the
decomposition:%
\begin{align}
\mathfrak{so}(5,1)\times\mathfrak{sp}(1)_{R,6D}  &  \supset\mathfrak{so}%
(1,1)\times\mathfrak{su}(2)_{L}\times\mathfrak{su}(2)_{R}\times\mathfrak{sp}%
(1)_{R,6D}\\
(\mathbf{4},\mathbf{2})  &  \rightarrow(+1/2,\mathbf{2},\mathbf{1}%
,\mathbf{2})\oplus(-1/2,\mathbf{1},\mathbf{2},\mathbf{2}).
\end{align}
In accord with our 2d $\mathcal{N}=(0,2)$ conventions used in reference \cite{Apruzzi:2016iac},
we refer to fermions with spin $+1/2$ by a $-$ subscript, i.e. $\lambda_{-}$, and fermions with
spin $-1/2$ by a $+$ subscripts, i.e. $\psi_{+}$. To avoid overloading the
notation, we shall also sometimes drop these subscripts and respectively write
$\widetilde{\lambda}$ and $\psi$, respectively.

To perform a twist, we consider activating a background value for the gauge
field associated with the $\mathfrak{sp}(1)$ R-symmetry. Our aim is to
activate this field strength in such a way that the combined holonomy for the
spin connection and the R-symmetry connection retains a covariantly constant
spinor of the uncompactified 2d effective theory. On a general four-manifold,
we have available to us the standard \textquotedblleft Donaldson
twist,\textquotedblright\ introduced in reference \cite{Witten:1988ze}. In
terms of representations, we take the diagonal subalgebra of $\mathfrak{su}%
(2)_{R}\times\mathfrak{sp}(1)_{R,6D}$ which we refer to as $\mathfrak{su}%
(2)_{\text{diag}}$. The structure group retained is now $\mathfrak{su}%
(2)_{L}\times\mathfrak{su}(2)_{\text{diag}}$. The 6D supercharge now
decomposes as:%
\begin{align}
\mathfrak{so}(5,1)\times\mathfrak{sp}(1)_{R,6D}  &  \supset\mathfrak{so}%
(1,1)\times\mathfrak{su}(2)_{L}\times\mathfrak{su}(2)_{\text{diag}}\\
(\mathbf{4},\mathbf{2})  &  \rightarrow(+1/2,\mathbf{2},\mathbf{2}%
)\oplus(-1/2,\mathbf{1},\mathbf{3})\oplus(-1/2,\mathbf{1},\mathbf{1}),
\end{align}
so that now, our twisted supercharges transform as a vector, and a self-dual
two form, and a scalar. We thus conclude that on a general four-manifold, we
can twist to retain $\mathcal{N}=(0,1)$ supersymmetry in two dimensions.

To retain $\mathcal{N}=(0,2)$ supersymmetry, we can specialize further to the
case where $M_{4}$ is a K\"{a}hler surface. Since the structure algebra of the
holonomy is now reduced to $\mathfrak{u}(2)=\mathfrak{su}(2)_{L}%
\times\mathfrak{u}(1)_{R}$, this is tantamount to simply taking a further
decomposition of $\mathfrak{su}(2)_{\text{diag}}$ to its Cartan subalgebra:%
\begin{align}
\mathfrak{so}(5,1)\times\mathfrak{sp}(1)_{R,6D}  &  \supset\mathfrak{so}%
(1,1)\times\mathfrak{su}(2)_{L}\times\mathfrak{u}(1)_{\text{diag}%
}\label{KahlerTwist}\\
(\mathbf{4},\mathbf{2})  &  \rightarrow(-1/2,\mathbf{1},\mathbf{0}%
)\oplus(+1/2,\mathbf{2,+1})\oplus(+1/2,\mathbf{2,-1})\\
&  \oplus(-1/2,\mathbf{1},\mathbf{+2})\oplus(-1/2,\mathbf{1},\mathbf{0}%
)\oplus(-1/2,\mathbf{1},\mathbf{-2})
\end{align}
so as claimed, we now see two real supercharges which are internally scalars,
which are also of the same chirality, i.e. we have $(0,2)$ supersymmetry. We
now have a continuous 2d R-symmetry with algebra $\mathfrak{so}%
(2)_{R,2d}$.

Specializing further, we can also consider the case where $M_{4}$ is
Calabi-Yau, (i.e. for compact manifolds we have a K3 surface). In this case,
we need not twist the theory at all since the Calabi-Yau space already admits
covariantly constant spinors. Indeed, since the structure algebra of the
holonomy is now further reduced to $\mathfrak{su}(2)_{L}$, we can effectively
perform the same decomposition, but with an additional degeneracy due to the
$\mathfrak{sp}(1)$ R-symmetry:%
\begin{align}
\mathfrak{so}(5,1)\times\mathfrak{sp}(1)_{R,6D}  &  \supset\mathfrak{so}%
(1,1)\times\mathfrak{su}(2)_{L}\times\mathfrak{sp}(1)_{R,6D}\\
(\mathbf{4},\mathbf{2})  &  \rightarrow(+1/2,\mathbf{2},\mathbf{2}%
)\oplus(-1/2,\mathbf{1},\mathbf{2})\oplus(-1/2,\mathbf{1},\mathbf{2}),
\end{align}
that is, we have a system with $\mathcal{N}=(0,4)$ supersymmetry and the
R-symmetry algebra is now $\mathfrak{so}(4)_{R,2d}$.

For additional details on the explicit decomposition of the various spinor
representations, and in particular the relation to 6D $\mathcal{N}=(1,0)$
supermultiplets, we refer the interested reader to Appendix \ref{app:FREE}.

As a brief aside, given the fact that our starting point is a theory with
conformal symmetry, one might naturally ask whether there are more general
possibilities for retaining a conformal Killing spinor. We leave this
interesting possibility for future work, perhaps along the lines of references
\cite{Festuccia:2011ws, Klare:2012gn, Dumitrescu:2012ha}.

\subsection{Twisting on the Tensor Branch}

For suitable choices of four-manifold, we expect the compactification of the
6D\ SCFT\ to realize a 2d\ SCFT. It is also natural to consider the class of 2d theories obtained by
compactifying the tensor branch deformation of this SCFT, which will also lead to
a 2d SCFT.

An important aspect of the tensor branch is that it retains the $\mathfrak{sp}%
(1)_{6D,R}$ R-symmetry of the 6D theory, so that the
twisting procedure introduced in the previous section can also be applied
in this case as well.\footnote{Note that this also singles
out this class of possibilities for preserving some supersymmetry over the a
priori more general possibility afforded by just preserving conformal Killing
spinors.}

On the tensor branch, we have a 6D effective field theory built from
consistently combining tensor multiplets, vector multiplets and
hypermultiplets. The field content of these modes is:%
\begin{align}
\text{Tensor Multiplet}  &  \text{: }t_{6D}\oplus\psi_{L}\oplus B_{MN}^{(-)}\\
\text{Vector Multiplet}  &  \text{: }A_{M}\oplus\mu_{R}\\
\text{Hypermultiplet}  &  \text{: }Q\oplus Q^{c}\oplus\chi_{L},
\end{align}
where here, bosons are indicated by latin letters and fermions by greek
letters. For the hypermultiplet, the $c$ superscript indicates a complex scalar
transforming in the conjugate gauge group representation to $Q$. The subscript
$L$ and $R$ refers to whether the fermion has the same (L) or opposite (R)
chirality to the 6D supercharge.

In the tensor multiplet, we have $t_{6D}$ a real scalar, $B_{MN}^{(-)}$ an anti-chiral two-form and $\psi_{L}$ a
left-handed spinor with the same chirality as our $\mathcal{N}=(1,0)$
supercharge which transforms in the $(\mathbf{4},\mathbf{2})$ of
$\mathfrak{so}(5,1)\times\mathfrak{sp}(1)_{6D,R}$. Geometrically, the vacuum expectation value (vev)
of $t_{6D}$ is associated with the volume of the $\mathbb{P}^{1}$, and the origin of
the tensor branch (where the 6D\ SCFT resides) comes about from collapsing the
$\mathbb{P}^{1}$'s to zero size. The anti-chiral two-form $B_{MN}^{(-)}$ has a
three-form field strength which satisfies $H^{(3)}=-\ast_{6}H^{(3)}$.

In a string construction this mode comes about from reduction of the four-form
potential of type IIB string theory on this collapsing $\mathbb{P}^{1}$. The
relative sign convention is fixed by IIB\ conventions; there we have a
self-dual five-form, so since the $(1,1)$ form $\beta_{\mathbb{P}^{1}}%
^{(1,1)}$ which is Poincar\'{e} dual to the collapsing $\mathbb{P}^{1}$ in the
local geometry $\mathcal{O}(-n)\rightarrow\mathbb{P}^{1}$ is anti-self-dual,
we can expand:%
\begin{equation}
F^{(5)}=H^{(3)}\wedge\beta_{\mathbb{P}^{1}}^{(1,1)},
\end{equation}
i.e. $F^{(5)}=\ast_{10}F^{(5)}$ descends to our sign convention.

Let us now consider the effects of the twist on the content of the various 6D
supermultiplets. We mainly focus on the case of reduction of the tensor
multiplet, since this is the most unfamiliar case. We focus on the
case of compactification on a general four-manifold, preserving just
$\mathcal{N}=(0,1)$ supersymmetry. The remaining cases of $\mathcal{N}=(0,2)$
and $\mathcal{N}=(0,4)$ supersymmetry are obtained from further specialization
on the internal index structure of our modes. To this end, we denote by $M$ and $N$
internal vector indices, and $\mu$ a vector index on $\mathbb{R}^{1,1}$. Here,
then, is the resulting form content for a tensor multiplet after applying the
twist on a general four-manifold:
\begin{equation}
\text{Tensor Multiplet}\text{: }B_{\mu\nu}\oplus(t_{6D}\oplus\psi
)\oplus(B_{\mu M}\oplus\widetilde{\psi}_{M})\oplus\left(  B_{MN}\oplus
\psi_{MN}^{(+)}\right)  .
\end{equation}
where in the above, we denote left-handed 2d spinors with decoration by a
tilde, and right-handed ones by no tilde. Here, $\psi_{MN}^{(+)}$
transforms as a right-handed fermion and internally as a self-dual
two-form on $M_{4}$. In the above, we have grouped the terms according to
$\mathcal{N}=(0,1)$ supermultiplets, as indicated by the various parentheses.
Note that $B_{\mu\nu}$ does not really pair with a fermion. This is acceptable
because it is non-dynamical in 2d. Rather, it functions as a chemical
potential for spacetime filling strings of the 6D theory, and each of these
sectors comes with its own set of complete $\mathcal{N}=(0,1)$ supermultiplets.

The case of $\mathcal{N} = (0,2)$  supersymmetry amounts to making further restrictions on the index
structure of the model. We defer the analysis of the corresponding supermultiplet structure to section \ref{sec:SEVEN}
where we place it in the context of couplings to modes of a DGLSM.

\subsection{Anomaly Reduction}

One of the calculable quantities of all known 6D SCFTs is its anomaly
polynomial. In six dimensions, the anomaly polynomial is specified by a
formal eight-form constructed from the characteristic classes of the
associated curvatures for these background fields. It has the general form:%
\begin{equation}
I_{6D}^{(8)}=\alpha c_{2}(R_{6D})^{2}+\beta c_{2}(R_{6D})p_{1}(T)+\gamma
p_{1}(T)^{2}+\delta p_{2}(T)+\ldots~.
\end{equation}
where $c_{2}(R_{6D})$ denotes the second Chern class of an $SU(2)$ R-symmetry
bundle, $p_{1}(T)$ is the integrated first Pontryagin class of the tangent bundle, and
$p_{2}(T)$ is the second Pontryagin class. The ellipsis \textquotedblleft%
...\textquotedblright\ refers to the possibility of having additional global
flavor symmetries.

The reason this is calculable is that it can also be determined on the tensor
branch of the theory. In particular, for theories with an equal number of simple gauge group factors and tensor multiplets, anomaly cancellation via the Green-Schwarz mechanism yields a unique answer for the result. Now, from the classification results of references \cite{Heckman:2013pva, DelZotto:2014hpa, Heckman:2015bfa}, all known 6D SCFTs admit a partial tensor branch description in terms of just such a generalized quiver gauge theory (with conformal matter), so it follows that the anomaly polynomial can be calculated in all these theories
\cite{Ohmori:2014pca, Ohmori:2014kda}. This has been used for example to analyze various RG flows, see for example \cite{Heckman:2015ola, Heckman:2015axa, Heckman:2016ssk}.

Now, the very fact that we get the same answer at the conformal fixed point
and the tensor branch means that we are guaranteed to also get the same answer
for the two theories obtained from dimensional reduction. Indeed, the 2d
theory will also have an anomaly polynomial, and by activating an appropriate
background field strength for the $Sp(1)$ R-symmetry bundle as dictated by the
twist, we reduce to a formal four-form:%
\begin{equation}
I_{2d}^{(4)}=\underset{M_{4}}{\int}I_{6D}^{(8)}(\text{twist)}.
\end{equation}
Here then is the point. A priori we get two distinct 2d theories, with anomaly
polynomials $I_{2d}^{\text{SCFT}}$ and $I_{2d}^{\text{Tensor}}$. What anomaly
matching guarantees is that we actually have:%
\begin{equation}
I_{2d}^{\text{SCFT}}=I_{2d}^{\text{Tensor}}.
\end{equation}
The crucial difference from the 6D case, however, is that in many cases, we
expect the reduction of the tensor branch theory to also produce a 2d SCFT.
What this means is that our two candidate SCFTs actually have identical anomaly
polynomials. This is non-trivial, and has to do with the fact that we are in two dimensions. Indeed,
in higher-dimensional vacua obtained from a 6D parent,
there is no guarantee that compactification of the 6D tensor branch and conformal fixed
point will lead to identical anomaly polynomials.

Additionally, we can also see how these two theories may in fact be one and the same.
 In 6D, we move between the SCFT to the tensor branch by activating a
vev for $t_{6D}$, the tensor multiplet scalar. This scalar persists as a
non-compact real scalar $t_{2d}$ in two dimensions.\footnote{The crucial point
here is that our boson is non-compact. This is very analogous to the situation
in Liouville theory, a point we return to in section \ref{sec:TWOD}.}  Its
vev parametrizes the target space of a (strongly coupled) 2d NLSM,
and different choices of $t_{6D}$ simply correspond to different points in this target space.
The matching of the anomaly polynomials provides strong evidence that these candidate theories are actually part
of the same connected branch of a single SCFT!  This is particularly persuasive since
the match works for all choices of the background K\"ahler manifold as well as starting 6D theory that
we consider.

This fact is perhaps the single most important distinction between the heavily
studied case of compactifications of the $(2,0)$ theories and the
comparatively less studied $(1,0)$ theories. In the latter, we have less
supersymmetry, but as compensation, we have a potentially tractable
description of the SCFT for large non-zero values of $t_{2d}$.

Turning the discussion around, we can use this relation to extract various
details about the resulting 2d SCFTs. Consider, for example the anomaly
polynomial of a 2d SCFT with $\mathcal{N}=(0,2)$ supersymmetry. This depends
on the background values of the $U(1)$ R-symmetry and the background tangent
bundle. Written as a formal four-form on a four-manifold $Z$, we have:%
\begin{equation}
I_{2d}=\frac{k_{RR}}{2}\times c_{1}(R_{2d,R})^{2}-\frac{k_{\text{grav}}}%
{24}p_{1}(TZ)+...,
\end{equation}
where the additional terms will depend on various background flavor field
strengths. Here, the quantity :%
\begin{equation}
k_{\text{grav}}=-(c_{L}-c_{R})
\end{equation}
is the gravitational anomaly. The infrared R-symmetry will typically be a
linear combination of the $U(1)$ associated with the UV quantity $R_{2d,R}$
inherited from the Cartan of the $Sp(1)_{6D,R}$ R-symmetry, and other abelian
flavor symmetries. In the cases where there is no mixing, i.e. where there are
no additional abelian flavor symmetries, we have $c_{R}=3k_{RR}$. When there
is mixing, one must use other methods such as $c$-extremization
\cite{Benini:2012cz, Benini:2013cda} to extract this information.

\subsubsection{Twisting and Reduction}

Let us now turn to the explicit form of the background field strengths needed
to reduce the anomaly polynomial of a 6D SCFT. To this end, we note that the
characteristic classes of the 6D theory are formally defined on an
eight-manifold $W$, and the characteristic classes of the 2d theory are
formally defined on a four-manifold $Z$. The reduction to two dimensions
amounts to making the further restriction $W=Z\times M_{4}$. We need to
specify for each choice of twist how the different field strengths break up.
For simplicity we switch off all flavor symmetry field strengths. The
generalization to this case is also straightforward, but will not concern us
in the context of reduction of the SNHCs.

As a warmup, let us review the analogous calculation for the reduction of
Pontryagin classes. Recall that for a general curvature two-form $\Omega$ of a
real vector bundle $E$, the Pontryagin classes are defined via the formal
polynomial:%
\begin{equation}
p(E)=\left[  1-\frac{\text{Tr}(\Omega^{2})}{2}+\frac{\text{Tr}(\Omega^{2}%
)^{2}-2\text{Tr}\left(  \Omega^{4}\right)  }{8}+...\right]
\end{equation}
where in the above, we have absorbed a factor of $(2\pi)$ into the definition
of the curvature to adhere with 6D\ SCFT conventions for the anomaly
polynomial. Letting $\Omega_{W}$ denote the curvature form on $W$ for the
tangent bundle, we have the decomposition:%
\begin{equation}
\Omega_{W}=\Omega_{Z}+\Omega_{M},
\end{equation}
so, since Tr$(\Omega)=0$, we get:%
\begin{equation}
p(TW)=1+p_{1}(TZ)+p_{1}(TM)+p_{1}(TZ)p_{1}(TM).
\end{equation}

Let us now turn to the reduction of the 6D R-symmetry field strength. This is
fully determined by the choice of twist we enact. We shall first consider the
case of compactification on a general four-manifold, i.e., we retain just
$\mathcal{N}=(0,1)$ supersymmetry. Then, we turn to the case of
compactification on a K\"{a}hler surface, i.e., we retain $\mathcal{N}=(0,2)$ supersymmetry.

Consider first the twist which preserves $\mathcal{N}=(0,1)$ supersymmetry.
Here, the 2d anomaly polynomial consists of a single term associated with the
gravitational anomaly:%
\begin{equation}
I_{2d}=\frac{c_{L}-c_{R}}{24}p_{1}(TZ),
\end{equation}
where $c_{L}-c_{R}$ are the left- and right-moving central charges. We now
show how to compute this for reduction on a four-manifold. In what follows, we
switch off all flavor symmetry contributions. To proceed, we use the
isomorphism $SO(4)\simeq SU(2)_{L}\times SU(2)_{R}/\mathbb{Z}_{2}$ to
decompose the field strength on $M_{4}$ as:\footnote{This amounts to working with the complexified bundle $TM \otimes\mathbb{C}$; we will suppress the complexification in what follows.}%
\begin{equation}
\Omega_{M}=\Omega_{M}^{(L)}+\Omega_{M}^{(R)}.
\end{equation}
We can express the relevant field strengths in terms of the first Pontryagin
class and the Euler density:%
\begin{align}
p_{1}(TM)  &  =-2c_{2}(TM^{(L)})-2c_{2}(TM^{(R)})\\
\chi(TM)  &  =c_{2}(TM^{(L)})-c_{2}(TM^{(R)}),
\end{align}
in the obvious notation. In the twist to preserve $\mathcal{N}=(0,1)$
supersymmetry, we make the specific choice:%
\begin{equation}
c_{2}(R_{6D})=c_{2}(TM^{(R)})=-\frac{1}{2}\chi(TM)-\frac{1}{4}p_{1}(TM).
\end{equation}
Reduction of the 6D anomaly polynomial to 2d is now straightforward. Since
there is no continuous R-symmetry for these models, it suffices to consider
the term proportional to the first Pontryagin class on $Z$. We find:%
\begin{equation}
\underset{M_{4}}{\int}I_{6D}\text{(twist}_{(0,1)}\text{)}=\left(  -\frac{1}%
{4}\left(  P+2\chi\right)  \beta+2P\gamma+P\delta\right)  p_{1}(TZ).
\end{equation}
Here, $\chi$ is the topological Euler character of and $p_{1}(TZ)$
is the first Pontryagin class of $M_{4}$.

Let us now turn to the case of models with $\mathcal{N}=(0,2)$ supersymmetry,
i.e., compactification on a K\"{a}hler surface. Here, the structure group of
the manifold is reduced to $U(2)=SU(2)_{L}\times U(1)_{R}/\mathbb{Z}_{2}$. The
twist now involves activating an abelian field strength valued in the Cartan
subalgebra of $Sp(1)_{6D,R}$. In terms of bundles, the rank two bundle with
$Sp(1)_{6D,R}$ structure group decomposes as the sum of two line bundles:%
\begin{equation}
R_{6D}=\left(  R_{2d,R}\otimes K_{M_{4}}^{1/2}\right)  \oplus\left(
R_{2d,R}\otimes K_{M_{4}}^{1/2}\right)  ^{\vee}.
\end{equation}
Although the above expressions involve $K_{M_4}^{1/2}$, the formulae given below
for counting spectra and anomaly computations continue to apply even when $M_4$
fails to be $Spin$ and is just $Spin_C$.  For the spectrum this is the case because the fields continue to be
sections of well--defined bundles (we will comment on this again below); for anomalies this is the case because only characteristic
classes of $K_{M_4}$ enter into the anomaly polynomials, as can be seen below.

Now, decompose the second Chern class of $R_{6D}$ by expanding in the
associated field strengths. This yields:%
\begin{equation}
c_{2}(R_{6D})=-\left(  c_{1}(R_{2d,R})+\frac{1}{2}c_{1}(K_{M_{4}})\right)
^{2}.
\end{equation}
We now turn to the reduction of the anomaly polynomial to two dimensions.
Integrating the Pontryagin classes yields the same result as for the $(0,1)$
twist. Integrating over the internal space, we have:%
\begin{equation}
\underset{M_{4}}{\int}c_{2}(R_{6D})^{2}=\frac{3}{2}c_{1}(R_{2d,R}%
)^{2}\underset{M_{4}}{\int}c_{1}(K_{M_{4}})^{2}%
\end{equation}
and:%
\begin{equation}
\underset{M_{4}}{\int}c_{2}(R_{6D})p_{1}(TW)=-c_{1}(R_{2d,R})^{2}%
\underset{M_{4}}{\int}p_{1}(TM)-\frac{1}{4}p_{1}(TZ)\underset{M_{4}}{\int%
}c_{1}(K_{M_{4}})^{2}.
\end{equation}
Now, since we also have:%
\begin{equation}
\underset{M_{4}}{\int}c_{1}(K_{M_{4}})^{2}=P+2\chi,
\end{equation}
we learn that the anomaly polynomial integrates to:%
\begin{equation}
\underset{M_{4}}{\int}I_{6D}\text{(twist}_{(0,2)}\text{)}=\left(  \frac{3}%
{2}(P+2\chi)\alpha-P\beta\right)  c_{1}(R_{2d,R})^{2}+\left(  -\frac{1}%
{4}\left(  P+2\chi\right)  \beta+P(\gamma+\delta)\right)  p_{1}(TZ).
\end{equation}
So in other words:%
\begin{align}
k_{RR}  &  =(3P+6\chi)\alpha-2P\beta\\
k_{\text{grav}}  &  =\left(  6P+12\chi\right)  \beta-24P(2\gamma+\delta).
\end{align}

\subsubsection{Simple NHCs}

To illustrate the above calculation, we now consider the anomaly polynomial
for the simple NHCs and their reduction to two dimensions.

Consider first the 6D anomaly polynomial. The key feature which makes this
quantity calculable is that, as noted in reference \cite{Ohmori:2014kda} (see also \cite{Sadov:1996zm}),
when the number of tensor multiplets and irreducible gauge group factors is the
same, there is a unique answer compatible with anomaly cancellation via the
Green-Schwarz mechanism. As we shall need it later, let us summarize the
structure of the anomaly polynomial for all of the simple NHCs. The total
anomaly polynomial is a sum of two terms:%
\begin{equation}
I_{6D}(\text{SNHC})=I_{\text{1-loop}}(\text{SNHC})+I_{\text{GS}}(\text{SNHC}),
\end{equation}
where:%
\begin{align}
I_{\text{1-loop}}(\text{SNHC})  &  =\frac{1}{5760}\left(
\begin{array}
[c]{l}%
240(1-d_{G})c_{2}(R_{6D})^{2}-120(d_{G}-1)c_{2}(R_{6D})p_{1}(T)\\
+(23-7d_{G})p_{1}(T)^{2}+4(d_{G}-29)p_{2}(T)\\
-120h_{G}^{\vee}(12c_{2}(R_{6D})+p_{1}(T))\text{Tr}F^{2})-180n\text{Tr}F^{2}%
\end{array}
\right) \\
I_{\text{GS}}(\text{SNHC})  &  =\frac{n}{2}\left(  \frac{1}{4}\text{Tr}%
F^{2}+\frac{h_{G}^{\vee}}{n}\left(  c_{2}(R_{6D})+\frac{p_{1}(T)}{12}\right)
\right)  ^{2}, \label{IGS}%
\end{align}
where $d_{G}$ is the dimension of the gauge group, $h_{G}^{\vee}$ is the dual
Coxeter number of the group, and $-n$ is the self-intersection of the $\mathbb{P}^1$ in the base of the F-theory
model. The list of values encountered for the rank one simple NHCs is:%
\begin{equation}%
\begin{tabular}
[c]{|c|c|c|c|c|c|}\hline
$G$ & $SU(3)$ & $SO(8)$ & $E_{6}$ & $E_{7}$ & $E_{8}$\\\hline
$n$ & $3$ & $4$ & $6$ & $8$ & $12$\\\hline
$d_{G}$ & $8$ & $28$ & $78$ & $133$ & $248$\\\hline
$h_{G}^{\vee}$ & $3$ & $6$ & $12$ & $18$ & $30$\\\hline
$h_{G}^{\vee}/n$ & $1$ & $3/2$ & $2$ & $9/4$ & $5/2$\\\hline
\end{tabular}
\ \ \ \ \ \ \ . \label{tabadabadooooo}%
\end{equation}
Summing up the contributions from $I_{\text{1-loop}}$ and $I_{\text{GS}}$, we
get the 6D anomaly polynomial:
\begin{equation}
I_{6D}(\text{SNHC})=\alpha c_{2}(R_{6D})^{2}+\beta c_{2}(R)p_{1}(T)+\gamma
p_{1}(T)^{2}+\delta p_{2}(T)
\end{equation}
with:%
\begin{equation}
\alpha = -\frac{d_{G}-1}{24}+\frac{\left(  h_{G}^{\vee}\right)  ^{2}}{2n},\text{\,\,\,}
\beta  = -\frac{d_{G}-1}{48}+\frac{\left(  h_{G}^{\vee}\right)  ^{2}}{12n},\text{\,\,\,}
\gamma = -\frac{7d_{G}-23}{5760}+\frac{\left(  h_{G}^{\vee}\right)  ^{2}}{288n},\text{\,\,\,}
\delta = \frac{d_{G}-29}{1440}.
\end{equation}
Reduction on a four-manifold in the case of the $(0,1)$ twist yields:%
\begin{equation}
k_{\text{grav}}=-d_{G}\left(  \frac{P+3\chi}{12}\right)  +\frac{5P+3\chi}%
{12}+\frac{\left(  h_{G}^{\vee}\right)  ^{2}}{3n}\left(  P+3\chi\right)  ,
\label{kgrav}%
\end{equation}
and in the case of the $(0,2)$ twist (when compactified on a K\"{a}hler
surface), we also get:%
\begin{equation}
k_{RR}=-(d_{G}-1)\left(  \frac{P+3\chi}{12}\right)  +\frac{\left(  h_{G}%
^{\vee}\right)  ^{2}}{3n}\left(  4P+9\chi\right)  . \label{krr}%
\end{equation}
In the case of the simple NHCs, where there are no additional abelian $U(1)$'s
present, and assuming no emergent IR\ symmetries, we can therefore also
extract the central charges $c_{L}$ and $c_{R}$:
\begin{align}
c_{L}  &  =-\frac{\left(  d_{G}+1\right)  P+3(d_{G}-1)\chi}{6}+\frac{\left(
h_{G}^{\vee}\right)  ^{2}}{6n}\left(  22P+48\chi\right) \\
c_{R}  &  =-\frac{(d_{G}-1)(P+3\chi)}{4}+\frac{\left(  h_{G}^{\vee}\right)
^{2}}{n}\left(  4P+9\chi\right)  .
\end{align}
One of our goals will be to match to these cases using the formulation of the DGLSM.

\section{Dynamic GLSMs \label{sec:TWOD}}

In the previous section we discussed some general aspects of compactifications
of 6D\ SCFTs on four-manifolds. We also saw that the dimensional reduction on
the tensor branch provided a potentially more direct way to access data about
the resulting 2d effective theories. These theories are also intrinsically
interesting from a purely 2d perspective. Indeed, as we shortly explain, we
find a natural generalization of the standard gauged linear sigma model
construction to one in which the parameters of the theory are promoted to
dynamical, possibly chiral fields. We refer to such a theory as a
\textquotedblleft dynamic GLSM\textquotedblright\ (DGLSM).

The main reason we encounter DGLSMs in compactifying our 6D theories has to do
with the novelties of the 6D tensor multiplet.
First of all, the real scalar in 6D functions as a dynamical gauge coupling. Additionally,
the anti-chiral two-form descends, upon reduction, to chiral and anti-chiral
2d bosons, as well as a chemical potential for spacetime filling strings.
These in turn also couple to the reduction of the rest of the modes of the GLSM.

One might therefore be tempted to treat this DGLSM\ as a theory in its own
right. This is basically correct, but much as in other contexts, gauge
anomalies will only cancel with appropriate matter content. Here, we uncover a
rather rich generalization of the standard weakly coupled analysis of anomalies.

From a 2d perspective, the interplay of these ingredients would at first
appear to be rather ad hoc and arbitrary. From a 6D perspective, however, we
see that this rich structure is completely automatic!

The rest of this section is organized as follows. First, we discuss some
general aspects of dynamic GLSMs. Next, we turn to the chiral / anti-chiral
bosons of the DGLSM. After this, we outline the general structure of anomaly
cancellation in DGLSMs, which we follow with a general discussion of the
parameter space of the models. In subsequent sections we will present quite
explicit examples of how all of these ingredients intricately fit together.

\subsection{General Gauge Theories}

General 2d gauge theories are constructed in the same fashion as
their higher-dimensional counterparts. The starting point is some non-linear
sigma model with target space $\mathcal{M}$, local coordinate fields $\phi
^{a}$, and a Lagrangian with leading order terms in the derivative expansion
of the form%
\begin{equation}
\mathcal{L}_{0}\supset G_{ab}(\phi)\partial_{\mu}\phi^{a}\partial^{\mu}%
\phi^{b}+iB_{ab}(\phi)\epsilon^{\mu\nu}\partial_{\mu}\phi^{a}\partial_{\nu
}\phi^{b}~.
\end{equation}
We work in Euclidean signature with world sheet coordinates $y^{1},y^{2}$, and
$G$ and $B$ denote, respectively, the metric and B-field pulled back from
$\mathcal{M}$.\footnote{For simplicity we are just considering the bosonic
terms.}

If the geometric data $(\mathcal{M},G,B)$ has an isometry group $G$, then we
can gauge a subgroup of $G$ by introducing a set of 2d gauge
fields $A_{\mu}$ with field-strengths $F_{\mu\nu}$ and appropriate covariant
derivatives for the fields:%
\begin{equation}
\mathcal{L}_{\text{gauge}}\supset G_{ab}(\phi)D_{\mu}\phi^{a}D^{\mu}\phi
^{b}+iB_{ab}(\phi)\epsilon^{\mu\nu}D_{\mu}\phi^{a}D_{\nu}\phi^{b}+\frac
{1}{4M^{2} g(\phi)^{2}}\tr F_{\mu\nu}F^{\mu\nu} + M^{2}V(\phi)~.
\end{equation}
In this action we introduced two additional terms: a field--dependent kinetic
term for the gauge fields and a potential field for the bosons. Both of these
involve a scale $M$, and while the former, if omitted classically, will
typically be induced by quantum corrections, the latter is often required by
classical supersymmetry --- e.g. the classical D-terms in the scalar
potential. If $R$ denotes the radius of curvature of the NLSM metric, then we
might naively expect that for energies $E\gg M,1/R$ the theory will be
well-described by the gauged non-linear sigma model Lagrangian, while for
$E\approx M,1/R$ the theory will be strongly coupled.

There are a number of important differences that such 2d theories
have from their higher-dimensional counterparts. The most critical originate
from the rather different role played by scalar expectation values. Suppose
$V(\phi)$ has some flat direction with a coordinate $t$ such that $\partial
g / \partial t \neq0$. If the dimensionality of spacetime $d>2$, then we
have a family of theories parameterized by the expectation value $\langle t
\rangle$. Of course for $d=2$ the situation is rather different: we must
integrate over the zero mode of $t$! So, for instance, if the kinetic term
for the gauge fields vanishes at say $t=0$, the theory defined by $\mathcal{L}%
_{\text{gauge}}$ is not weakly coupled for any $E$! Unlike in the $d>2$ case,
we cannot choose $\langle t \rangle$ to simplify the dynamics.

It is therefore not so surprising that, even with constraints from
supersymmetry, we know little about such theories, even if, for instance,
$(\mathcal{M},G,B)$ is just Euclidean space with flat metric and zero B-field:
in other words, we take a standard gauged linear sigma model and promote some
of its couplings to dynamical fields. These are the theories that we will
explore in this paper: they naturally arise from 6D SCFTs, and where an
effective description of the sort just sketched is valid, the theory is
described by a gauged linear sigma model fibered over some non-linear sigma
model. To emphasize the role of dynamic couplings, we refer to these as
\textit{dynamic GLSMs}.

Of course not all is lost! There are examples that are relatively
well-understood. The most venerable is surely Liouville theory~(see
\cite{Teschner:2001rv} for a review), an example of a non-compact conformal
field theory defined by a scalar Lagrangian with a potential. The field space
has a non-compact direction where the potential is exponentially small, and
this is essentially the reason why semi-classical reasoning based on the
Lagrangian is useful for describing aspects of the theory. In supersymmetric
theories we also have examples of 2d theories with exactly flat
directions, and these can receive important quantum corrections. For instance,
in the context of gauge theories with $(4,4)$ supersymmetry, the Higgs and
Coulomb branches which meet classically are found to be at infinite distance
in the quantum--corrected non-linear sigma model metric~\cite{Witten:1997yu}.
The result is that such a gauge theory may a priori correspond to
\textit{several distinct} SCFTs, labelled by the choice of
branch.\footnote{Some of these \textquotedblleft branches\textquotedblright%
\ may in fact be isolated points generated by quantum
dynamics~\cite{Harvey:2014zra}.} In addition, there are also a number of
interesting examples of $(2,2)$ and $(0,2)$ abelian gauge theories with
dynamical Fayet-Iliopoulos and $\theta$--angle terms, for instance references  \cite{Hori:2001ax,Adams:2006kb,Adams:2012sh,Quigley:2012gq,Melnikov:2012nm,Israel:2015aea}.

\subsection{Dynamical Gauge Coupling}

The previous sketches indicate, in broad terms, that there is a very large
class of 2d theories to be explored beyond simple elaborations
on gauged linear sigma models, but it should also be clear that it is not so
easy to find examples free of various pathologies such as local and global
gauge and diffeomorphism anomalies, nor is it easy to identify sufficient
conditions for the weakly--coupled picture of a GLSM fibered over some
non--linear sigma model to be a useful description of the dynamics. In
particular, since the classical non--linear sigma model metric will often have
singularities, how can we be sure that we do not get a singular SCFT? The
top--down perspective is invaluable when facing such questions, and it helps
us to find some reasonably firm footing. In the following sections we will
illustrate this with precise examples, but for now we will sketch out how the
2d structures just discussed naturally and sensibly emerge from
six dimensions.

The key players in this story are the bosonic degrees of freedom in the tensor
multiplet: the scalar field $t_{6D}$ and the anti-chiral two-form $B^{(-)}$ with anti-self-dual field strength $H=-\ast_{6}H$. For simplicity,
in the following subsection we concentrate on a single tensor multiplet. The
expectation value of $t_{6D}$ is a flat direction of the theory: the origin is
a non-Lagrangian SCFT, while for $\langle t_{6D}\rangle\neq0$ conformal
invariance is spontaneously broken, and $t_{6D}$ becomes the corresponding
Goldstone boson --- the dilaton. The corresponding low energy theory is a
gauge theory with Yang-Mills coupling $g_{6D}^{2}=1/\langle t_{6D}\rangle$, as
well as higher derivative terms suppressed by powers of $\langle t_{6D}%
\rangle$; i.e. it is weakly coupled for energy scales $E^{2}\ll\langle
t_{6D}\rangle$.

Now suppose that we compactify this theory on a four-manifold $M_{4}$ with a
smooth background metric and corresponding volume $V_{4}$. Reducing the
kinetic terms for the tensor scalar and the gauge field, we then find a
2d action that describes the physics away from the origin of the
tensor branch:%
\begin{align}
\mathcal{L}_{2d}  &  \supset\frac{V_{4}}{2}\partial_{\mu}t_{6D}\partial^{\mu
}t_{6D}+\frac{1}{4}V_{4}t_{6D}\tr\{F_{\mu\nu}F^{\mu\nu}\}\\
&  \supset\frac{1}{2}\partial_{\mu}t_{2d}\partial^{\mu}t_{2d}+\frac{t_{2d}%
}{4M_{\text{KK}}^{2}}\tr\{F_{\mu\nu}F^{\mu\nu}\}~.
\end{align}
In the second line we introduced the dimensionless field $t_{2d}=t_{6D}%
\sqrt{V_{4}}$ and the Kaluza-Klein scale $M_{\text{KK}}=V_{4}^{-1/4}$. For
energies $E\ll M_{\text{KK}}$ the contributions from the Kaluza-Klein towers
associated to $M_{4}$ will be suppressed, and the physics is well-described by
a 2d theory. If $t_{2d}$ were not a dynamical field but a
parameter, then by making $t_{2d}$ suitably large we could always find
energies $E$ in the following hierarchy of scales:%
\begin{equation}
\frac{M_{\text{KK}}^{2}}{t_{2d}}=g_{\text{2d}}^{2}\ll E^{2}\ll M_{\text{KK}%
}^{2}~.
\end{equation}
This would allow us to ignore the Kaluza-Klein excitations and define an
effective weakly coupled 2d theory. The Lagrangian would then be
a good starting point for exploring the low energy dynamics by applying
standard methods from the study of 2d gauge theories. Of course this is
exactly what we cannot do when $t_{2d}$ is a dynamical field.

There are several conceivable possibilities for the low--energy dynamics.
First, it may be that quantum corrections generate a potential which separates
$t_{2d}=0$ and $t_{2d}\neq0$ into two different branches. A second possibility
is that even if this potential is zero, corrections to the target space metric
may push the $t_{2d}=0$ origin to infinite distance. Finally, it may be that
$t_{2d}=0$ remains at finite distance, and the theory is intrinsically
strongly coupled. One of the main aims of this paper will be to present
evidence that it is the third possibility which is actually realized.

Along these lines, we can return to the anomaly matching argument introduced
in subsection \ref{sec:NHCreview}. There, we saw that the anomaly polynomial for
the $t_{2d}=0$ and $t_{2d}\neq0$ theories are actually the same. Now, in a
theory with only $\mathcal{N}=(0,1)$ supersymmetry, this allows us to match
just the gravitational anomaly. With $\mathcal{N}=(0,2)$ or more
supersymmetry, however, we can also extract the central charges of the
putative SCFTs. The exact match of central charges gives a strong indication
that these two sectors actually remain at finite distance. Indeed, we shall
often focus on the case of models with at least $(0,2)$ supersymmetry since
this is case where we can still utilize holomorphy to constrain the structure
of our models.

Consider, for example, the possible structure of quantum corrections to the
potential for $t_{2d}$ in a model with $(0,2)$ supersymmetry.
We expect this field to pair with some part of the reduction of
the two-form potential in six dimensions. This already constrains the
structure of our theory, because we get a compact, chiral boson, and so any
putative superpotential correction will be subject to constraints from the
associated shift symmetry. In other words, we expect these corrections to be
small when $t_{2d}$ is large. Indeed, such contributions to the potential are
generated by instanton corrections, i.e.,
from Euclidean D3-branes wrapped over four-manifolds of the form $\Sigma
\times\mathbb{P}^{1}$, with $\Sigma$ a Riemann surface in $M_{4}$ and
$\mathbb{P}^{1}$ an exceptional curve of the base $\mathcal{B}$. Experience
from the higher-dimensional case suggests that such instanton corrections will
in general correct the structure of the F-term data, but typically the modulus
$t_{2d}$ will not appear in isolation. Rather, it will be accompanied by
couplings to other modes of the model. Whereas this is usually problematic in
higher dimensions (for example in the context of enacting supersymmetric
moduli stabilization), in the present context it would suggest that we should
still expect some flat directions in field space. Said differently, we should
expect there is likely to still be a direction in field space which takes us
to the weak coupling limit described by the DGLSM. Of course, to really settle
this issue, one should actually perform the corresponding instanton
calculation, perhaps generalizing the analysis presented in reference
\cite{Ganor:1996xg}. We expect that the general methods developed for
calculating instanton corrections in F-theory (see e.g. \cite{Witten:1996bn,
Ganor:1996pe, Heckman:2008es, Marsano:2008py, Cvetic:2009mt, Donagi:2010pd})
can be suitably adapted to this purpose. We therefore leave this interesting
question for further work.

\subsubsection{Twisted Sectors}

Another interesting feature of such DGLSMs is the automatic appearance of
twisted sectors in any candidate CFT. In fact, this point was already noted in
the special case of the $A_{1}$ $(2,0)$ theory \cite{Ganor:1996xg}, though we
expect it to hold far more generally.

The reason is already apparent in the case of a single tensor multiplet
because of the condition $t_{2d}\geq0$, rather than having $t_{2d}$ valued on
the real line. This $\mathbb{Z}_{2}$ quotient means we should also expect
twisted sectors to be present in the full Hilbert space of the system. So in
particular, if we were to consider compactification of the Euclidean signature
theory on a genus $g$ Riemann surface, such sectors must also be taken into
account. In the broader context of 6D\ SCFTs, we typically have more than one
tensor multiplet. In such cases, we again expect multiple twisted sectors
stemming from the condition that we have non-negative values for all of the
tensor branch scalars.

In some cases, we can say more about the structure of these twisted sectors.
For example, if the 6D\ SCFT consists of a collection of just collapsing $-2$
curves, then there is a corresponding ADE\ classification of the possible
intersection pairings. The associated Weyl group for the root system is then
the orbifold group (that is, we pick a Weyl chamber), so all the twisted
sectors will be the conjugacy classes of the Weyl group. More generally,
however, the analogue of this Weyl group action is unknown for general
6D\ SCFTs.\footnote{There is one other case which is likely to follow a
similar Lie algebraic characterization. In the case of the configuration of
$k$ curves $1,2,...,2$, we also know that compactification on a circle leads
to a gauge theory with $\mathfrak{sp}(k)$ gauge symmetry. This would suggest
that for this theory, the conjugacy classes for the Weyl group of
$\mathfrak{sp}(k)$ describe the twisted sector states of such theories.} Though we leave a more complete analysis to future work,
let us point out that the inclusion of these twisted sectors is likely
to be essential in unraveling the structure of the theory in the strongly coupled $t = 0$ region of the NLSM.

\subsection{Chiral / Anti-Chiral Sectors}

One of the most striking features of the 6D tensor multiplet is the presence
of an anti-chiral two-form. Upon reduction to two dimensions, we now get
chiral and anti-chiral bosons. In the context of DGLSMs, one can again view
these modes as parameters which have now been promoted to dynamical fields. As
we explicitly show later, from a purely 2d perspective (i.e. without the
assistance of the 6D theory), this must be done with significant care.

Consider then, the reduction to two dimensions of a collection of anti-chiral
three-form field strengths. We label these as $H_{I}$ where the index
$I=1,...,T$ runs over the total number of tensor multiplets on the tensor
branch. Associated with this is a corresponding lattice of string charges
$\Lambda_{\text{string}}=H_{2}^{\text{cpct}}(\mathcal{B},Z)$ and a Dirac
pairing (just minus the geometric intersection pairing):%
\begin{equation}
A:\Lambda_{\text{string}}\times\Lambda_{\text{string}}\rightarrow%
\mathbb{Z}
,
\end{equation}
which we denote by $A^{IJ}$.\footnote{Our conventions for using raised indices
for this pairing are chosen to avoid possible confusions later on in the index
structure of our vertex operators. In the other literature
on F-theory realizations of 6D SCFTs, we typically would write $A_{IJ}$. So, $A^{IJ}_{here} = A_{IJ}^{there}$.
This is important because when the lattice $\Lambda_{\text{string}}$ is not self-dual,
the inverse of $A$ may not be integral.}

To keep track of this charge quantization condition in the 2d theory, it is
convenient to actually work in terms of the type IIB\ four-form potential with
self-dual field strength. Reduction on the collapsing $\mathbb{P}^{1}$'s of
the F-theory model then yields the basis of anti-self-dual three-form fluxes
in six dimensions.

Now, along these lines, we introduce a basis of self-dual and anti-self-dual
four-forms on the eight-manifold $M_{4}\times\mathcal{B}$ which are
Poincar\'{e} dual to compact four-cycles. Since all of the collapsing
$\mathbb{P}^{1}$'s are Poincar\'{e} dual (in the F-theory base $\mathcal{B}$) to
anti-self-dual two-forms, introduce the basis:
\begin{equation}
\Pi_{\alpha,K}={\pi}_{\alpha}\wedge\beta_{K}\text{ \ \ and \ \ }%
\widetilde{\Pi}_{\dot{\alpha},K}=\widetilde{\pi}_{\dot{\alpha}}\wedge\beta
_{K},
\end{equation}
where the $\pi$'s, $\widetilde{\pi}$'s and $\beta$'s satisfy the relations:%
\begin{equation}
\ast_{M_{4}}\pi_{\alpha}=-\pi_{\alpha}\text{, \ \ }\ast_{M_{4}}\widetilde{\pi
}_{\dot{\alpha}}=+\widetilde{\pi}_{\dot{\alpha}}\text{, \ \ }\ast
_{\mathcal{B}}\beta_{K}=-\beta_{K},
\end{equation}
so that $\alpha=1,...,b_{2}^{-}(M_{4})$ runs over a basis of anti-self-dual
two-forms and $\dot{\alpha}=1,...,b_{2}^{+}(M_{4})$ runs over a basis of
self-dual two-forms on $M_{4}$.

Reducing the five-form field strength yields the corresponding 2d currents:%
\begin{equation}
F^{(5)}=J^{\alpha,K}\wedge\Pi_{\alpha,K}+\widetilde{J}^{\dot{\alpha},K}%
\wedge\widetilde{\Pi}_{\dot{\alpha},K},
\end{equation}
where in our conventions, a self-dual one-form is associated with a
left-moving, (i.e. holomorphic) current $J^{\alpha,K}(z)$, and an anti-self-dual one-form is associated
with a right-moving (i.e. anti-holomorphic) current $\widetilde{J}^{\dot{\alpha},K}(\zb)$.
We may bosonize these in the usual fashion by writing these currents in terms
of derivatives of chiral bosons.  In Minkowski signature these take the form%
\begin{align}
J^{\alpha I}=\partial_{-}\varphi^{\alpha I}(x^-)\text{ \ \ and \ \ }\widetilde{J}^{\dot{\alpha}%
I}=\partial_{+}\widetilde{\varphi}^{\dot{\alpha}I}(x^+).
\end{align}
We do not expect a conventional Lagrangian description, but we can describe
the Kac-Moody-Virasoro primaries:%
\begin{equation}
\mathcal{O}_{Q_{L},Q_{R}}=\exp\left\{  iQ_{\alpha I}\varphi^{\alpha
I}+i\widetilde{Q}_{\dot{\alpha}I}\widetilde{\varphi}^{\dot{\alpha}I}\right\}
. \label{vertexops}%
\end{equation}
Associativity of the OPE requires that the charge vectors $(Q,\widetilde{Q})$
belong to a lattice $\Gamma_{n_{L},n_{R}}$ of signature $n_{L}-n_{R}$, and the
OPE is single--valued if and only if for all pairs $(Q,\widetilde{Q})$ and
$(Q^{\prime},\widetilde{Q}^{\prime})$ we have%
\begin{equation}
Q\cdot Q^{\prime}-\widetilde{Q}\cdot\widetilde{Q}^{\prime}\in{\mathbb{Z}}.
\end{equation}

Let us see how such operators come about from a higher-dimensional
perspective. It is actually instructive to consider both the 10D IIB
perspective as well as the construction directly in 6D terms. In 10D terms, we
can construct additional operators by evaluating periods of the IIB\ four-form
potential, or equivalently, periods of the anti-chiral two-forms. As a period
over the four-form potential, we write, for $S$ a four-cycle in $H_{4}%
^{\text{cpct}}(M_{4}\times\mathcal{B},\mathbb{Z})$:%
\begin{equation}
\mathcal{O}_{S}=\exp\left\{  i\int_{S}C\right\}  .
\end{equation}
This can also be stated in purely 6D terms by evaluating the integrals over
the F-theory base. For example, given a charge vector $v_{I}$ of the lattice
$\Lambda_{\text{string}}$, we have, for $\Sigma$ two-cycle in $H_{2}%
^{\text{cpct}}(M_{4},\mathbb{Z})$:%
\begin{equation}
\mathcal{O}_{\Sigma}=\exp\left\{  i\int_{\Sigma}v_{I}B^{I}\right\}  .
\end{equation}

To obtain a basis of vertex operators as in line (\ref{vertexops}), we first
expand in terms of the integral basis $e^{i}\in H^{2}(M_{4},\mathbb{Z})$
satisfying:%
\begin{equation}
\int_{M_{4}}e^{i}\wedge e^{j}=d^{ij}.
\end{equation}
Introduce vielbeins:%
\begin{equation}
\pi_{\alpha}=E_{\alpha i}e^{i}\text{ \ \ and \ \ }{\widetilde{\pi}}%
_{\dot{\alpha}}={\widetilde{E}}_{\dot{\alpha}i}e^{i},
\end{equation}
with:%
\begin{equation}
E_{\dot{\alpha}i}d^{ij}E_{\dot{\beta}j}=\delta_{\dot{\alpha}\dot{\beta}%
}\text{, \ \ }{\widetilde{E}}_{\alpha i}d^{ij}{\widetilde{E}}_{\beta
j}=-\delta_{\alpha\beta}\text{, \ \ }d_{ij}=E_{\dot{\alpha}i}E_{\dot{\alpha}%
j}-{\widetilde{E}}_{\alpha i}{\widetilde{E}}_{\alpha j}\text{,} \label{wyres}%
\end{equation}
we can now expand $B^{I}$ as:%
\begin{equation}
B^{I}=\varphi^{\alpha I}(z){\pi}_{\alpha}+\widetilde{\varphi}^{\dot{\alpha}%
I}({\overline{z}})\widetilde{\pi}_{\dot{\alpha}}.
\end{equation}
So, we can expand the dual of $\Sigma$ in
$H^{2}(M_{4},\mathbb{Z})$ as $q_{i}e^{i}$ to obtain:%
\begin{equation}
\mathcal{O}_{\Sigma}=\exp\left\{  iv_{I}q_{i}\int_{M_{4}}e^{i}\wedge
B^{I}\right\}  =\exp\left\{  iQ_{\alpha I}\varphi^{\alpha I}+i\widetilde{Q}%
_{\dot{\alpha}I}\widetilde{\varphi}^{\dot{\alpha}I}\right\}  ~,
\end{equation}
where%
\begin{equation}
Q_{\alpha I}=v_{I}\int_{\Sigma}{\pi}_{\alpha}=v_{I}q_{i}d^{ij}E_{\alpha
j}\text{, \ \ } \widetilde Q_{\dot{\alpha}I}=v_{I}\int_{\Sigma}\widetilde{\pi}%
_{\dot{\alpha}}=v_I q_{i}d^{ij}\widetilde{E}_{\dot{\alpha}j}~.
\end{equation}
The (Euclidean signature) OPE of two such operators takes the form%
\begin{equation}
\mathcal{O}_{\Sigma}(z,{\overline{z}})\mathcal{O}_{\Sigma^{\prime}}%
(0)=z^{\rho}{\overline{z}}^{\widetilde{\rho}}:\mathcal{O}_{\Sigma
+\Sigma^{\prime}}(0):~,
\end{equation}
where%
\begin{equation}
\rho=A^{IJ}Q_{\alpha I}Q_{\alpha J}^{\prime},\text{ \ \ }\widetilde{\rho
}=A^{IJ}\widetilde{Q}_{\dot{\alpha}I}\widetilde{Q}^{\prime}_{\dot{\alpha}J}~.
\end{equation}

Note that the OPE is single--valued because the potential phase under
$z\rightarrow e^{2\pi i}z$ is $e^{2\pi i(\rho-\widetilde{\rho})}$, and the
difference $\rho-\widetilde{\rho}$ is an integer. To see this, we expand (see
also \cite{Ganor:1996xg}):
\begin{equation}
\rho-\widetilde{\rho}=A^{IJ}\left(  Q_{\alpha I}Q_{\alpha J}^{\prime
}-\widetilde{Q}_{\dot{\alpha}I}\widetilde{Q}^{\prime}_{\dot{\alpha}J}\right)
=-A^{IJ}v_{I}v_{J}^{\prime}q_{i}d^{ij}q_{j}^{\prime} = -A^{IJ}v_{I}%
v_{J}^{\prime}\Sigma\cdot\Sigma^{\prime}\in\mathbb{Z}~. \label{leftright}%
\end{equation}
The second equality follows from (\ref{wyres}).

When this chiral sector is decoupled from the other degrees of freedom, then
the scaling dimensions of $\mathcal{O}_{\Sigma}$ are given by $(\frac{1}%
{2}\Vert Q\Vert^{2},\frac{1}{2}\Vert\widetilde{Q}\Vert^{2})$.
In that case we observe that we obtain additional chiral currents whenever one of these
vanishes and the other is $1$. This of course depends on the moduli of the
geometry, which determine the expansion of the $\pi_{\alpha}$ and
${\widetilde{\pi}}_{\dot{\alpha}}$ in terms of the integral basis $e^{i}$. So,
in particular, whenever the moduli are tuned so that:%
\begin{equation}
A^{IJ}v_{I}v_{J}\Sigma\cdot\Sigma=-2\text{, \ \ }\int_{\Sigma}\widetilde{\pi
}_{\dot{\alpha}}=0\text{, \ \ for all $\dot{\alpha},$}~
\end{equation}
we obtain additional left--moving non--abelian currents.

For instance, when $M_{4}$ is a K3 surface and $A^{IJ}v_{I}v_{J}=1$, we see that these
occur precisely at finite--distance orbifold singularities in the K3 moduli
space, where a $-2$ curve collapses to zero size. This would be the case where
we compactify the E-string theory on a K3 surface as it arises from collapsing
a single $-1$ curve in the F-theory base $\mathcal{B}$. Another case of
interest comes about from a collapsing $-1$ curve in $M_{4}$ for a $(2,0)$
theory. For other cases of interest such as those involving an NHC, we expect
to generically realize higher-spin currents rather than a Kac--Moody current
algebra. This is because the pairing $A^{IJ}v_{I}v_{J}$ reduces to an integer
$n\geq3$. It would be quite interesting to determine the structure of these
higher spin currents.


\section{$\mathcal{N}=(0,2)$ DGLSMs from Seven-Branes \label{sec:SEVEN}}

In the previous sections we sketched some general considerations on the
compactification of 6D\ SCFTs, and in particular the structure of DGLSMs
obtained from reduction on the tensor branch. To provide a more complete
characterization of the resulting 2d effective theories, in this section we
specialize to the case of DGLSMs obtained from the simple rank one
non-Higgsable clusters.

Along these lines, it is helpful to return to the F-theory construction of 6D
SCFTs. From this perspective, we are considering intersecting seven-branes
wrapped on K\"{a}hler threefolds of the form $M_{4}\times\mathbb{P}^{1}$,
where each $\mathbb{P}^{1}$ factor is associated with a collapsing curve in
the base $\mathcal{B}$ of a non-compact elliptically fibered Calabi-Yau threefold. The
field theory limit amounts to taking some collection of K\"ahler threefolds and
collapsing them to a lower-dimensional subspace, in this case a K\"ahler surface.

In the context of F-theory, decoupling gravity amounts to contracting the
K\"{a}hler manifold wrapped by a seven-brane to a lower-dimensional K\"{a}hler
manifold. For 6D\ SCFTs, this involves collapsing a $\mathbb{P}^{1}$ to a
point. For 4d theories a K\"ahler surface is collapsed to either a point or a
curve, and for 2d theories we have a K\"{a}hler threefold which either
collapses to a point, a curve or a K\"{a}hler surface. In the present context
where we first decouple gravity in six dimensions, we are contracting to a
K\"{a}hler surface. Note that in this case, it is acceptable for the K\"ahler surface to
have negative curvature. For further discussion on decoupling gravity in F-theory,
see for example references \cite{Bhardwaj:2015oru, Beasley:2008kw, Cordova:2009fg,
Heckman:2010pv}.

Thankfully, the effective field theory obtained from reduction of intersecting
seven-branes on K\"{a}hler threefolds has already been determined in the
context of compactifications of F-theory on a Calabi-Yau fivefold
\cite{Apruzzi:2016iac, Schafer-Nameki:2016cfr} (see also \cite{Vafa:1995fj, Dasgupta:1996yh, Gukov:1999ya, Gates:2000fj}).
As noted in reference \cite{Apruzzi:2016iac}, decoupling gravity can sometimes leave behind some
coupling to dynamical breathing modes. The symptom of this fact in 6D\ SCFTs
is that on the tensor branch, gauge couplings are dynamical.

Our plan in this section will be to assemble the various ingredients which
appear in the F-theory realization of DGLSMs. First of all, we review the
construction of reference \cite{Apruzzi:2016iac} (see also \cite{Schafer-Nameki:2016cfr})
on the dimensional reduction of intersecting seven-branes wrapped over K\"{a}hler threefolds.

After this, we turn to the mode content and couplings associated with minimal
couplings of the anti-chiral two-form to the other modes of the model. In
particular, for a general 6D\ SCFT, these come from the coupling (see for
example \cite{Sadov:1996zm}):%
\begin{equation}
L_{6D}\supset\mu^{I,V}\int_{6D}B_{I}\wedge X_{V}, \label{gsgs}%
\end{equation}
where $I$ is an index running over the tensor multiplets of the theory, and
$V$ is an index running over the irreducible vector multiplets of the theory.
As noted in reference \cite{Ohmori:2014kda}, in theories where the number of
irreducible gauge group factors is the same as the number of tensor
multiplets, factorization of the 6D anomaly polynomial uniquely fixes the form
of the couplings $\mu^{I,V}$ and the form of $X_{V}$. The structure of the
four-forms $X_{V}$ is fully determined by the condition that the Green-Schwarz
mechanism can actually cancel the anomalies of the theory. It depends on the
background field strengths for the gauge field, the tangent bundle, the
$\mathfrak{sp}(1)$ R-symmetry, and possible flavor symmetry field strengths.
Working through the reduction of line (\ref{gsgs}), we will find a number of
interaction terms between the GLSM\ sector and chiral extra sectors.

Finally, an added bonus of working in terms of the F-theory picture is that we
will also be able to provide a geometric parameterization of the space of
vacua associated with such DGLSMs.

\subsection{GLSM\ Sector}

To set our conventions, recall that an $\mathcal{N} = (0,2)$ GLSM is constructed with three sorts of
mulitiplets: a chiral scalar (CS) multiplet, a Fermi (F) multiplet and a
vector (V) multiplet.  In Wess-Zumino gauge these have superspace expansions:%
\begin{align}
\text{CS}  &  \text{: }\Phi=\phi+\sqrt{2}\theta^{+}\psi_{+}-i\theta
^{+}\overline{\theta}^{+}D_{+}\phi\\
\text{F}  &  \text{: }\Lambda=\lambda_{-}-\sqrt{2}\theta^{+}\mathcal{G-}%
\sqrt{2}\overline{\theta}^{+}E-i\theta^{+}\overline{\theta}^{+}D_{+}%
\lambda_{-}\\
\text{V}  &  \text{: }V=v_{-}-2i\theta^{+}\overline{\mu}-2i\overline{\theta
}^{+}\mu+2\theta^{+}\overline{\theta}^{+}\mathcal{D}\text{ \ \ and \ \ }%
\Xi=-i\theta^{+}\overline{\theta}^{+}v_{+}~,
\end{align}
where in the above, a subscript of $+/-$ on a greek letter indicates a spinor
index, while for a latin letter it indicates a vector index.
We sometimes denote the gaugino superfield strength
associated with $V$ by the variable $\Upsilon$.

Summarizing the discussion found in \cite{Apruzzi:2016iac}
(see also \cite{Schafer-Nameki:2016cfr}), the mode content
from seven-branes on $\mathbb{R}^{1,1}\times Y$ with $Y$ a K\"{a}hler
threefold include CS, Fermi and Vector multiplets which transform as
bundle-valued differential forms on the internal space:%
\begin{align}
\text{CS}\text{: }  &  \Phi_{(3,0)}\in\Omega^{(3,0)}(\text{ad}\mathcal{P)}\\
\text{CS}\text{: }  &  \mathbb{D}_{(0,1)}\in\Omega^{(0,1)}(\text{ad}%
\mathcal{P)}\\
\text{F}\text{: }  &  \Lambda_{(0,2)}\in\Omega^{(0,2)}(\text{ad}\mathcal{P)}\\
\text{V}\text{: }  &  V_{(0,0)}\in\Omega^{(0,0)}(\text{ad}\mathcal{P)}\text{,}%
\end{align}
where $\mathcal{P}$ is a principal $G$-bundle dictated by having a seven-brane
with gauge group $G$. Strictly speaking, this is really the mode content of an
eight-dimensional gauge theory packaged in terms of 2d supermultiplets.

Intersections of seven-branes translate to 6D\ hypermultiplets localized on a
K\"{a}hler surface, which also contribute CS\ and Fermi multiplets to the 2d
effective theory. In the present context where all seven-branes wrap the same
K\"ahler surface $M_{4}$, we have bundle-valued differential forms on the
internal space:%
\begin{align}
\text{CS}\text{: }  &  Q\in K_{M_{4}}^{1/2}\otimes\mathcal{R}_{1}%
\otimes\mathcal{R}_{2}^{\vee}\\
\text{CS}\text{:\ }  &  Q^{c}\in K_{M_{4}}^{1/2}\otimes\mathcal{R}_{1}^{\vee
}\otimes\mathcal{R}_{2}\\
\text{ F}\text{: }  &  \Psi\in\Omega^{(0,1)}(K_{M_{4}}^{1/2}\otimes
\mathcal{R}_{1}\otimes\mathcal{R}_{2}^{\vee})
\end{align}
where the $\mathcal{R}_{i}$ are bundles obtained from restriction of the
seven-branes to $M_{4}$.\footnote{We note again that when $M_4$ is not Spin, the twisting
will ensure that the fields transform in sections of well-defined bundles; thus, even if $K_{M_4}^{1/2}$
may not be a well--defined bundle, the tensor products, such as $K_{M_4}^{1/2} \otimes \mathcal{R}_1\otimes\mathcal{R}_2^\vee$,
will be well--defined.}  In addition to these localized modes, we also have
the reduction of the 6D tensor multiplet, that is, we have modes obtained from
the reduction of the volume of the $\mathbb{P}^{1}$ and its superpartner (in
six dimensions) the reduction of the four-form potential. As we have already
seen in section \ref{sec:TWOD}, this contributes chiral and anti-chiral currents which will
couple to the other modes of the GLSM. See Appendix \ref{app:FREE} for a summary of these
conditions from a purely 6D\ perspective.

Treating the modes of the tensor multiplet as a constant background, the
equations of motion for the bosonic field content is schematically of the form
\cite{Apruzzi:2016iac}:%
\begin{gather}
F_{(0,2)}=0\text{, \ \ }\overline{\partial}_{A}Q=0\text{, \ \ }\overline
{\partial}_{A}Q^{c}=0\text{, \ \ }\overline{\partial}_{A}\Phi_{(3,0)}%
=\delta_{M_{4}}\wedge\left(  Q^{c},Q\right) \\
\omega_{Y}\wedge\omega_{Y}\wedge F_{(1,1)}+\left[  \Phi_{(3,0)},\Phi
_{(3,0)}^{\dag}\right]  -\delta_{M_{4}}\wedge\left(  \left\langle Q^{\dag
},Q\right\rangle -\left\langle Q^{c},Q^{c\dag}\right\rangle \right)  =0,
\end{gather}
where $\omega_{Y}$ is the K\"{a}hler form on $Y$ a threefold. These equations are further
corrected in the presence of triple and quartic intersections of seven-branes.

Expanding around these background values, we find a zero mode spectrum, and a
corresponding GLSM. By itself, this gauge theory is typically anomalous and
must be accompanied by additional chiral sectors \cite{Apruzzi:2016iac}.

Now, implicit in our discussion is the assumption that the flux on the
$\mathbb{P}^{1}$ factor is trivial. In particular, this means that when we
reduce to six dimensions, any fluctuation associated with $\delta\Phi^{(3,0)}$
will automatically vanish. In this limit, then, we can simply switch off
$\Phi_{(3,0)}$.

With this in mind, we see that the dimensional reduction on a K\"{a}hler
surface will require us to specify a stable holomorphic vector bundle. When
6D\ hypermultiplets are present, we also need to specify global sections of
the associated bundles.

\subsubsection{Simple NHCs}

Let us now specialize further to the case of the simple rank one
non-Higgsable clusters. In all of these cases, we have no matter fields, and
the Lie algebra is simply laced. So, the full structure of the background
equations (for the GLSM\ sector) is given by:%
\begin{equation}
F_{(0,2)}=0\text{ \ \ and \ \ }\omega_{M_{4}}\wedge F_{(1,1)}=0,
\end{equation}
that is, we simply have the instanton equations of motion on a K\"{a}hler
surface. In particular, the zero modes are obtained by
expanding around the background defined by a stable holomorphic vector bundle.
We will shortly see how these moduli correspond to the target space of
dynamical fields localized on spacetime filling strings.

Now, to match to the case where we compactify our 6D\ SCFT, we could in
principle also consider switching on background values for various operators
in the CFT. Without additional data about the conformal fixed point, however,
this would simply amount to interpolating back from the tensor branch
description. Since one of our aims in this paper is to provide a complementary
perspective on these SCFTs, we shall adhere to the simplest case where all
background fields are switched off. Turning the discussion around, of course,
it is tempting to use the tensor branch description as a means to
\textit{specify} this additional data of the compactification for the 6D SCFT.
We leave a more complete treatment of this interesting possibility to future work.

Further restricting, then, to the special case where we have trivial
background fields activated, we can now count all the zero modes of the
system:%
\begin{align}
\text{CS}\text{: }  &  \delta\mathbb{D}_{(0,1)}\in H_{\overline{\partial}}%
^{1}(M_{4},\mathcal{O}_{M_{4}})\\
\text{F}\text{: }  &  \delta\Lambda_{(0,2)}\in H_{\overline{\partial}}%
^{2}(M_{4},\mathcal{O}_{M_{4}})\\
\text{V}\text{: }  &  \delta V_{(0,0)}\in H_{\overline{\partial}}^{0}%
(M_{4},\mathcal{O}_{M_{4}}).
\end{align}
Observe also that we can assemble the count of the zero modes into the
index
\begin{equation}
\#\delta V_{(0,0)}-\#\delta\mathbb{D}_{(0,1)}+\#\delta\Lambda_{(0,2)}=\dim
G\times\chi(M_{4},\mathcal{O}_{M_{4}}).
\end{equation}
Note that since the holomorphic Euler characteristic is typically non-zero,
we will need to supplement our GLSM\ by additional chiral
sectors \cite{Apruzzi:2016iac}.

We can also re-write this quantity in terms of the Pontryagin class $P$ and Euler character $\chi$
of the four-manifold:
\begin{equation}
\chi(M_{4},\mathcal{O}_{M_{4}})=\int_{M_{4}}\mathrm{Td}(M_{4})\mathrm{ch}%
(\mathcal{O}_{M_{4}})=\int_{M_{4}}\left(  \frac{c_{1}(M_{4})^{2}+c_{2}(M_{4}%
)}{12}\right)  =\frac{P(M_{4})+3\chi(M_{4})}{12}, \label{eq:indadj}%
\end{equation}
where $\mathrm{Td}(M_{4})$ is the Todd class of $M_{4}$, and the Chern characteristic of the trivial bundle
is $\mathrm{ch}(\mathcal{O}_{M_{4}})=1$.

A general analysis of interactions purely within the GLSM sector has been
given in references \cite{Apruzzi:2016iac, Schafer-Nameki:2016cfr} to which we
refer the interested reader. For purposes of exposition, we focus on the
interactions present in the simple NHC theories. In particular, we focus on
those interaction terms which are protected by $\mathcal{N}=(0,2)$
supersymmetry, i.e. roughly speaking the holomorphic F-terms.

We see that there are only a few holomorphic
interaction terms present for the GLSM\ sector. Indeed, using the results of
reference \cite{Apruzzi:2016iac}, we see that the Fermi multiplet\ which
transforms as a $(0,2)$ form on $M_{4}$ admits the expansion:%
\begin{equation}
\Lambda_{(0,2)}=\lambda_{-}^{(0,2)}-\sqrt{2}\theta^{+}\mathcal{G}%
_{(0,2)}\mathcal{-}\sqrt{2}\overline{\theta}^{+} F_{(0,2)}-i\theta
^{+}\overline{\theta}^{+}D_{+}\lambda_{-}^{(0,2)},
\end{equation}
where%
\begin{equation}
\mathbb{F}_{(0,2)}=[\mathbb{D}_{(0,1)},\mathbb{D}_{(0,1)}].
\end{equation}
There are no bulk zero modes from the $(3,0)$ form on $M_{4}\times
\mathbb{P}^{1}$, so these are all the holomorphic interaction terms from the GLSM\ sector.

\subsection{Chiral Currents}

As we have already remarked, one of the defining features of the DGLSM is that
some parameters are now promoted to fields. To further understand the
structure of the resulting field theories, we now turn to a discussion of the
possible supermultiplets in the case of $\mathcal{N}=(0,2)$ supersymmetry. For
simplicity, we focus on the case of a single tensor multiplet, as this is the
case most germane to our analysis of simple NHCs.

One of the important features of $(0,2)$ multiplets is that the number of
bosonic and fermionic degrees of freedom need not be the same. This will
persist in the context of supermultiplets constructed from reduction of the
anti-chiral two-form. Indeed, from our discussion of anti-chiral two-forms and
their reduction, we should expect the various chiral and anti-chiral bosons to
also assemble into supermultiplets.

The left-moving currents $J = \partial_- \varphi(x^-)$ are (0,2) SUSY singlets and
appear without superpartners.  On the other hand, the right-moving currents  $\widetilde{J} = \partial_+ \widetilde{\varphi}(x^+)$  appear as
top components of abelian $\mathcal{N} = (0,2)$ current multiplets.  These are easily described in
terms of CS multiplets of the form
\begin{align}
\text{CS}_{R}~ \text{:}~ \Phi_{R} = \widetilde{\varphi}_{R}+\sqrt{2}\theta^{+}\psi_{+}-i\theta
^{+}\overline{\theta}^{+}\partial_{+}\widetilde{\varphi}_{R}~.
\end{align}
The corresponding current supermultiplet is then:
\begin{align}
\Xi_{+} = \mathbf{D}_+ \Phi_{R}~.
\end{align}
For the explicit form of the resulting expressions, see Appendix \ref{app:FREE}.
%
Note that in the above, we implicitly assumed that $\widetilde{\varphi}_{R}$ was a complex
scalar.  As we will see shortly, this accounts for all but one of the right-moving bosons that
descend from the 6D anti-chiral two-form $B^{(-)}$:  the bosons and corresponding currents
can be packaged into complex multiplets.  The remaining scalar combines with
the descendant of the tensor scalar $t_{2d}$ into another CS multiplet.

%

Let us illustate how these multiplets arise in the context of 6D
$\mathcal{N}=(1,0)$ theories. First, reduction of the 6D vector
multiplet will clearly descend to some combination of 2d vector multiplets, CS
and Fermi multiplets. Similar considerations also hold for 6D hypermultiplets.

It is the reduction of the tensor multiplets which will lead us to the novel
multiplet structures expected in the context of a DGLSM. The explicit
reduction for the free tensor multiplet is carried out in Appendix \ref{app:FREE},
so here we summarize the salient features.

Our conventions for 6D superspace are adapted from reference
\cite{Buchbinder:2014sna}. Introduce coordinates $(x^{M},\,\theta^{s\alpha
})$; the Grassmann coordinates transform in a symplectic Majorana-Weyl spinor
and furnish the real representation $(\mathbf{4},\mathbf{2})$ of
$\mathfrak{so}(5,1)\times\mathfrak{sp}(1)_{6D,R}$. Here, $\alpha$ is a spinor
index and $s$ is a doublet index.

First, we introduce the superspace derivatives:
\begin{equation}
D_{\alpha}^{s}=\frac{\partial}{\partial\theta_{s}^{\alpha}}-i\theta
^{s\beta}\partial_{\alpha\beta}\quad\text{with}\quad\{D_{\alpha}^{s},D_{\beta
}^{t}\}=-2i\Omega^{st}\partial_{\alpha\beta}\,,
\end{equation}
where $\Omega^{st}$ is a two-index anti-symmetric tensor of $\mathfrak{sp}%
(1)_{6D,R}$. Now, since the twist treats the two components of the R-symmetry
doublet differently, we will find it convenient to explicitly indicate this
expansion into various pieces. We denote these two components by $s=1,2$.

We now turn to the construction of the on-shell free tensor multiplet. Recall
that the on-shell content consists of a real scalar $t_{6D}$, a left-handed
fermion $\xi_{\alpha}^{s}$ and the anti-self-dual 3-form flux $H_{MNL}$ in
bispinor representation $H_{\alpha\beta}=\gamma_{\alpha\beta}^{MNL}H_{MNL}$.
Our $\gamma$-matrix convention is given in Appendix \ref{app:FREE}.
As the multiplet is on-shell, these fields have to fulfill the equations of
motion
\begin{equation}
\partial_{M}\partial^{M}t_{6D}=0\,,\quad\partial^{\alpha\beta}\xi_{\beta}%
^{s}=0\quad\text{and}\quad\partial^{\alpha\gamma}H_{\gamma\alpha}=0\,.
\end{equation}

We implement the tensor multiplet as a real on-shell superfield $\mathcal{T}$
which is constrained by:%
\begin{equation}
D_{\alpha}^{(s}D_{\beta}^{t)}\mathcal{T}=0\,. \label{DDT}%
\end{equation}
In order to obtain the $(0,2)$ multiplets in two dimensions, we start from the
superspace expansion in 6D flat space.
\begin{align}
  \mathcal{T} = & t_{6D} - \theta^{2\alpha} \xi^1_\alpha + \theta^{1 \alpha} \xi^2_\alpha + \theta^{2 \alpha} \theta^{1 \beta} ( H_{\alpha \beta} - i \partial_{\alpha \beta} \phi ) - i \theta^{2\alpha} \theta^{2\beta} \theta^{1\gamma} \partial_{\beta \gamma} \xi^1_\alpha  + \nonumber \\
  &  i \theta^{2\alpha} \theta^{1\beta} \theta^{1\gamma} \partial_{\alpha\beta} \xi^2_\gamma - \frac{1}{2} \theta^{2 \alpha} \theta^{2 \beta} \theta^{1 \gamma} \theta^{1 \delta} \partial_{\alpha \gamma} \partial_{\beta \delta} \phi - i \theta^{2\alpha} \theta^{2\beta} \theta^{1\gamma} \theta^{1\delta} \partial_{\beta \delta} H_{\alpha \gamma} + \dots
\end{align}
which solves the constraint~(\ref{DDT}) on-shell.

To implement the twist, it is helpful to assemble the Grassmann coordinates
$\theta^{s\beta}$ as differntial forms on our K\"{a}hler surface. Following
our conventions of subsection \ref{ssec:TWISTSCFT}, we have the decomposition
of the Grassmann coordinates as in line (\ref{KahlerTwist}):%
\begin{equation}
\theta^{s\beta}\rightarrow\theta\oplus\theta_{i}\oplus\theta_{\overline
{ij}}~,
\end{equation}
where all $\theta$'s are associated with $-1/2$ charge spinors of
$\mathfrak{so}(1,1)$. The standard Grassmann coordinate of our $\mathcal{N}%
=(0,2)$ system is given by $\theta$, so we shall sometimes use the notation
$\theta^{+}$ to indicate this. Raising of differential form indices is
accomplished via the K\"{a}hler form.

Let us summarize the physical
content of the on-shell supermultiplets. The key point is that as we have
already remarked, the anti-chiral two-form splits up into $b_{2}^{+}$ internal
self-dual and $b_{2}^{-}$ anti-self-dual two forms, which are respectively
associated with right-moving and left-moving chiral bosons. Additionally, the
free field $t_{6D}$ can also be decomposed in terms of left--moving and right--moving parts:%
\begin{equation}
t_{2d}=t_{L}+t_{R}~.
\end{equation}
So all told, for each tensor multiplet we get $b_{2}^{+}+1$ right--movers, and
$b_{2}^{-}+1$ left--movers. Note that on a K\"{a}hler surface, we have:%
\begin{equation}
b_{2}^{+}=2h^{2,0}+1\text{ \ \ and \ \ }b_{2}^{-}=h^{1,1}-1~.
\end{equation}
The shift by one is due to the fact that the $(1,1)$ form proportional to the
K\"{a}hler form is actually self-dual rather than anti-self-dual. Summarizing,
from the reduction of a free tensor multiplet we obtain
\begin{align}
 2h^{2,0}+2  &&  \text{Right--Moving Currents} \\
 h^{1,1} &&\text{Left--Moving Currents}~.
\end{align}
Now, in keeping with the structure of $\mathcal{N}=(0,2)$ supersymmetry, we
should expect the right-movers to have fermionic partners, but with no partners
for the left-movers. Additionally, we expect (when there are non-trivial
one-cycles) 2d vector multiplets, which transform as one-forms on the internal
space. Retaining all of the differential form content, we get the following
multiplets:%
\begin{align}
\text{CS}_{R}  &  \text{:}\text{ }\mathbb{B}_{R}^{(2,0)}=B_{R}^{(2,0)}%
+\sqrt{2}\theta^{+}\psi_{+}^{(2,0)}-i\theta^{+}\overline{\theta}^{+}\partial_{+}%
B_{R}^{(2,0)}~,\\
\text{RS}_{R}  &  \text{:}\text{ }\operatorname{Re}\mathbb{\tau}_{R}%
=t_{R}+\frac{1}{\sqrt{2}}\theta^{+}\psi_{+}^{(0,0)}-\frac{1}{\sqrt{2}%
}\overline{\theta}^{+}\overline{\psi}_{+}^{(0,0)}-i\theta^{+}\overline{\theta
}^{+}\partial_{+}B_{\overline{i}}^{\overline{i}}~,\\
\text{V}\text{: }  &  \operatorname{Re}V^{(0,1)}=\operatorname{Re}\left(
v_{-}^{(0,1)}-2i\overline{\theta}^{+}\mu^{(0,1)}+2\theta^{+}\overline{\theta
}^{+}\mathcal{D}^{(0,1)}\right)  \text{ \ \ and \ \ } \nonumber\\
&  \operatorname{Re}%
(\Xi^{(0,1)})=\operatorname{Re}(-i\theta^{+}\overline{\theta}^{+}v_{+}^{(0,1)})\\
J_{L}  &  \text{:}\text{ }\partial_{-}t_{L}\text{ \ \ and \ \ }\partial_{-}B_{i\overline{j}%
}~,
\end{align}
where in the last line, we must take all $(1,1)$-forms orthogonal (using the
pairing defined by the wedge product) to the K\"{a}hler form.  The only possibly
mysterious ingredient here is the real scalar multiplet $\text{RS}_{R}$.  This is
just the real part of the CS multiplet $\tau_R$, but we choose to write the real part,
since that is what is naturally found in the 6D to 2d reduction, details of which
are in Appendix~\ref{app:FREE}.

Of course the bare compact boson fields such as $B^{(2,0)}_R$ are
more appropriately packaged in terms of current
supermultiplets. For the left-movers, there is not much to do, but for the right-movers, we see that there are $2h^{2,0}+2$ chiral currents.\footnote{We stress that this is the case for the reduction of the free tensor multiplet.  In the interacting case we will not be able to decompose $t_{\text{2d}}$ into left-- and right--moving parts; the corresponding degrees of freedom will combine with $B_{\bar{i}}^{\bar{i}}$ and fermionic superpartners into a chiral multiplet.  The axionic nature of $B_{\bar{i}}^{\bar{i}}$ naturally leads to selection rules and constraints on $t_{\text{2d}}$--dependent corrections to various F-terms.}

We also observe that the chiral bosons again assemble into an index:%
\begin{equation}
\#(\text{Right--Movers})-\#(\text{Left--Movers})=b_{2}^{+}-b_{2}^{-}=\tau(M_{4}),
\end{equation}
i.e., the signature of the four-manifold. This is in turn related to the
Pontryagin class as:%
\begin{equation}
\tau(M_{4})=\frac{P(M_{4})}{3}.
\end{equation}

Consider next the coupling of the GLSM\ sector to terms involving the
anti-chiral two-form. The minimal coupling between these two sectors is
controlled by the bosonic interaction:%
\begin{equation}
L_{2d}\supset\mu_{\text{GS}}\underset{6D}{\int}B\wedge X,
\end{equation}
where $X$ is a four-form constructed from the background gauge field strength,
the R-symmetry field strength, and the metric curvature.

Recall that the chiral bosons obtained from reduction transform as both
$(2,0)$ forms and as $(1,1)$ forms on the internal space. Now, since $X$ is a
function of characteristic classes (of holomorphic vector bundles),
we see that around a fixed background,
there will be no coupling between the $(2,0)$-form and the 2d vector
multiplet. There will, however, sometimes be a non-trivial coupling to the
$(1,1)$ components of the field strength. In the special case where we have a
trivial internal flux, or more generally when there are no abelian gauge
fields in the GLSM sector, these minimal couplings will also vanish.

But we do expect a rich class of non-minimal couplings between the vector
field and these chiral currents. To illustrate, consider the contribution to
$X$ proportional to Tr$(F_{6D}\wedge F_{6D})$. Now, we can split this field
strength up as:%
\begin{equation}
F_{6D}\rightarrow F_{2d}+ d_A \mathbb{D}+ d_A \mathbb{D}^{\dag
}+[\mathbb{D},\mathbb{D}]+[\mathbb{D}^{\dag},\mathbb{D}^{\dag}]+[\mathbb{D}%
,\mathbb{D}^{\dag}],
\end{equation}
where $d_A = d+A$ is the 2d gauge connection, and $\mathbb{D}$ is a $(0,1)$ form on the K\"{a}hler surface. So, we
certainly do have non-minimal couplings between the $(2,0)$ form and the
GLSM\ sector modes such as:%
\begin{align}
L_{2d}\supset &  \mu_{\mathrm{GS}} \underset{6D}{\int}B_{(2,0)}\wedge\text{Tr}%
(F_{2d}\wedge\lbrack\mathbb{D},\mathbb{D}])+B_{(2,0)}\wedge
\text{Tr}(d_A \mathbb{D\wedge} d_A \mathbb{D})+c.c\label{2dhol}\\
+ & \mu_{\mathrm{GS}} \underset{6D}{\int}B_{(1,1)}\wedge\text{Tr}(F_{2d}\wedge
\lbrack\mathbb{D},\mathbb{D}^{\dag}])+B_{(1,1)}\wedge\text{Tr}(d_A
 \mathbb{D\wedge}d_A \mathbb{D}^{\dag})+c.c.
\end{align}
We obtain the associated couplings to the zero modes by making the further
substitution:%
\begin{equation}
\mathbb{D\rightarrow D}_{\text{bkgnd}}+\delta\mathbb{D}\text{.}%
\end{equation}
As far as we are aware, the structure of such couplings is largely unexplored
in the context of 2d $\mathcal{N}=(0,2)$ models. Perhaps the biggest surprise
from a 2d perspective is that we would seem to require an integrality constraint from shifts of the form
$B \rightarrow B + 2 \pi$. Note that this is automatic from the topological structure of the 6D coupling.
It would be quite exciting to better understand this from a purely 2d perspective.
\subsection{Abelian Vector Multiplets}

Reduction of the B-field can also generate abelian vector multiplets. These
clearly come about from taking one leg of the $B$-field in the internal space
$M_{4}$, and one in the external space. The total number of such vector
multiplets on a K\"{a}hler surface is $2h^{1,0}$, so we denote these by
$\mathbb{B}_{(0,1)}^{\mu}$, i.e. they are vector fields on the 2d spacetime
and a $(0,1)$ form on the K\"{a}hler surface.

Consider next the coupling of these abelian vector multiplets to the other
modes of the system. Returning to the coupling:
\begin{equation}
L_{2d}\supset\mu_{\text{GS}}\underset{6D}{\int}B\wedge X,
\end{equation}
and performing the expansion:%
\begin{equation}
F_{6D}\rightarrow F_{2d}+ d_A \mathbb{D}+ d_A \mathbb{D}^{\dag
}+[\mathbb{D},\mathbb{D}]+[\mathbb{D}^{\dag},\mathbb{D}^{\dag}]+[\mathbb{D}%
,\mathbb{D}^{\dag}],
\end{equation}
we now get couplings to the GLSM\ modes such as:%
\begin{equation}
L_{2d}\supset\mu_{\text{GS}}\underset{6D}{\int}\mathbb{B}_{(0,1)}%
\wedge\text{Tr}(d_A \mathbb{D\wedge}[\mathbb{D},\mathbb{D}])+... \ .
\end{equation}

There are also background couplings to the background curvature and 2d
R-symmetry, as obtained by reduction of the appropriate field strengths of the
6D theory. We will return to these shortly when we turn to the calculation of
anomalies in the model.

\subsection{Spacetime Filling Strings}

One of the additional features of reducing the tensor multiplet is that we
also encounter spacetime filling strings.\ Again returning to the
6D\ Green-Schwarz term for a single tensor multiplet, we have:%
\begin{equation}
L_{2d}\supset\mu_{\text{GS}}\underset{6D}{\int}B\wedge X.
\end{equation}
To count the number of spacetime filling strings, we now need the explicit
form of $X$. Returning to the form of equation (\ref{IGS}) we see that for all
of the simple NHC\ theories, the Green-Schwarz term is fixed to be:%
\begin{equation}
L_{2d}\supset-n\int_{2d}B\int_{M_{4}}\left(  \frac{1}{4}\text{Tr}F^{2}%
+\frac{h_{G}^{\vee}}{n}\left(  c_{2}(R)+\frac{p_{1}(TM_{4})}{12}\right)
\right)  .
\end{equation}
Here, the term $\frac{1}{4}$Tr$F^{2}$ integrates to an integer (the
instanton number of the gauge bundle), and $h_{G}^{\vee}$ is the dual Coxeter
number for the group. The number of spacetime filling strings is given by
integrating $X$ over $M_{4}$:%
\begin{equation}
N_{\text{string}}=-\int_{M_{4}}\left(  \frac{1}{4}\text{Tr}F^{2}+\frac
{h_{G}^{\vee}}{n}\left(  c_{2}(R)+\frac{p_{1}(TM_{4})}{12}\right)  \right)  .
\end{equation}
The overall sign is fixed by our conventions for chirality of spinors and the
fact that we retain $(0,2)$ rather than $(2,0)$ supersymmetry in two
dimensions. We shall return to this point later when we analyze explicit
models, and count the number of effective strings. After applying the twist
appropriate for a K\"{a}hler surface, we have:%
\begin{equation}
N_{\text{string}}=-k_{\text{inst}}+\frac{2h_{G}^{\vee}}{n}\left(
\frac{P+3\chi}{12}\right)  ,
\end{equation}
where $k_{\text{inst}}$ is the instanton number for the gauge bundle. We can also
package the number of spacetime filling strings in terms of the holomorphic Euler characteristic:
\begin{equation}
N_{\text{string}}=-k_{\text{inst}}+\frac{2h_{G}^{\vee}}{n}\chi(M_{4}%
,\mathcal{O}_{M_{4}}). \label{NinstKinst}%
\end{equation}
Returning to our table of values for the ratio $h_{G}^{\vee}/n$, we see that
for all the simple NHCs but one, the quantity $N_{\text{string}}$ is an
integer. For $n=8$, however, $h_{G}^{\vee}/n=9/4$, so we would seem to
generate half integer quantization for the strings. This appears to be
compatible with the fact that of all the gauge groups for the SNHCs, $E_{7}$
is the only one for which matter in the fundamental representation can be in a
half hypermultiplet. We take this to mean that the $3-7$
strings generated in this way typically fill out half-hypermultiplets rather
than hypermultiplets.

Let us also comment on the relative signs in equation (\ref{NinstKinst}). This
is due to the anti-self-duality of the three-form field strength, which in
turn sets our convention for \textquotedblleft
anti-instantons\textquotedblright\ as being supersymmetric.
We shall return to this point later when we fix the
various signs and chiralities of all modes using a weakly coupled example.

Now, for all of the simple NHCs, the resulting effective field theory of
strings descends from a 4d $\mathcal{N}=2$ superconformal field theory with
flavor group $G$ as dictated by the presence of the seven-branes. For the
exceptional groups, these realize the Minahan-Nemeschansky theories
\cite{Minahan:1996fg, Minahan:1996cj}, and for the case $G=SU(3)$ we get the
$H_{2}$ Argyres-Douglas theory \cite{Argyres:1995jj}. The case $G=SO(8)$
realizes a weakly coupled model with $Sp$-type gauge theory
\cite{Seiberg:1994aj, Sen:1996vd, Banks:1996nj}.
In the limit where $M_{4}=\mathbb{C}^{2}$, there is also an $SU(2)_{L}\times SU(2)_{R}$
symmetry associated with moving the instanton around in these directions. When
we work on a curved background, we are simply activating a background field
strength in the 4d model.

At this point, it is perhaps worthwhile to note that for the other single
curve 6D\ SCFTs, i.e. the cases $n=7,9,10,11$, the interpretation of the
spacetime filling strings in terms of a compactified 4d theory is different.
In all these cases, the $\mathbb{P}^{1}$ wrapped by the D3-brane
has a marked point where matter fields (possibly strongly coupled)\ are
localized. So, to analyze these cases one would need to explicitly couple the
4d theory to 2d defects. In the case $n=5$ where we have a pure $F_{4}$ gauge
theory, we actually face the same issue even though no matter is present. The
reason is that this theory can be viewed as descending from an $E_{6}$ gauge
theory with a single hypermultiplet in the fundamental representation.
Activating a vev for this field triggers a Higgs branch flow to the $F_{4}$
gauge theory. There are, however, still cosmic string solutions from the
remnant of this breaking pattern. It would be interesting to study all of
these cases further.

\subsection{Parameter Space}
In the previous subsections we introduced the different sectors obtained from
compactifying our seven-brane on the manifold $M_{4}\times\mathbb{P}^{1}$. It
is also natural to ask whether our higher-dimensional perspective provides any
insight into the parameter space of the model. Our aim in this section will be
to study this question for the simple NHCs. Though it is tempting to interpret
all of these parameters as defining marginal couplings of the 2d theory,
strictly speaking this need not be the case, since a priori, we cannot exclude the possibility
that our ground state is not normalizable, as happens in Liouville theory, for example.
To keep this distinction clear, we
shall refer to the family of vacua we find as specifying a parameter space
rather than as a moduli space.

Consider first the parameters associated with a choice of background metric
for $M_{4}$. Since we are assuming we have a K\"{a}hler surface, the geometric
moduli will descend to parameters of the 2d theory.  Since the theories we consider
can be presented as 10D F-theory geometries, we expect to find a finite--dimensional
space of parameters that describe the complex structure and K\"ahler class of the surface.
These are easy to
characterize at the infinitesimal level since a small deformation of a compact
K\"{a}hler manifold remains K\"{a}hler~\cite{Kodaira:1960od}.\footnote{In
fact, for 2d surfaces it is true globally since a complex compact
surface is K\"{a}hler if and only if its first Betti number is even; see,
e.g.~\cite{Buchdahl:1999ck}.}

Hence, the geometric moduli will be a subset of
$H^{1,1}(M_{4},\mathbb{R})\oplus H^{1}(M_{4},T)$, where the first term counts
the K\"{a}hler deformations and the second the infinitesimal complex structure
deformations. Now, whereas the complex structure deformations automatically
come as complex parameters, the K\"ahler moduli are real. Since we are dealing
with an $\mathcal{N}=(0,2)$ supersymmetric theory, we expect that the K\"ahler
moduli will also be complexified. This is automatic when working on a 6D
supergravity background because the graviton multiplet contains a chiral
two-form $B^{+}$. There is then a natural complexification:%
\begin{equation}
\omega_{\mathbb{C}}= \omega +iB^{+},
\end{equation}
and the periods of $\omega_{\mathbb{C}}$ correspond to complexified parameters of
the 2d model.

In addition to these modes, we also expect a number of parameters from the
field theoretic degrees of freedom. These come about both from the GLSM, and
the moduli associated with reduction of the anti-chiral two-form. Indeed, the
moduli spaces of these two sectors are not really decoupled since, for
example, the instanton moduli space of the 4d field theory on $M_{4}$
translates into the directions transverse to the spacetime filling strings moving
inside of $M_{4}$. Physically, this means that some parameters that would have
been part of the moduli space in a higher-dimensional setting are really
associated with flat directions of a physical potential in two dimensions.

What then, remains of the moduli space of vacua from the field theory sector?
For one, we can see that the target space of the various chiral bosons
provides a natural class of such moduli. From the reduction of the self-dual
two-forms and the anti-self-dual two-forms, we get compact chiral /
anti-chiral bosons. Additionally, we have one non-compact real scalar, the
modulus $t$. This is quite analogous to the moduli space of the perturbative
heterotic string with target a torus, where in addition to the compact chiral
bosons, we have a non-compact mode from the dilaton. The parameter
spaces we have just sketched is really the one associated
with working around the free field limit. We expect that since we only have
$\mathcal{N}=(0,2)$ supersymmetry that there will be various quantum
corrections to the structure of this space. We leave the investigation of this
mathematically rich topic for future study.

\section{Anomalies \label{sec:ANOMO}}

In the previous section we discussed the general structure of 2d effective
theories obtained from compactifying an SNHC\ on a K\"{a}hler surface. In this
section we turn to some of the calculable quantities associated with these
theories. In particular, we compute the anomaly polynomial for these DGLSMs.

If the physical picture advocated in section \ref{sec:NHCreview} is to
hang together, we should be able to compute this anomaly polynomial in two
complementary ways. First, we can simply take the anomaly polynomial of the 6D
theory and dimensionally reduce.

Since there are no abelian flavor symmetries present in 6D, if we assume there
are no emergent symmetries in the IR, we can also directly compute the
infrared R-symmetry of the 2d model. In some sense, this is typically the best
one can hope for in compactifying a non-Lagrangian theory.

But since we also have a characterization of the tensor branch of our theory,
we can ask whether we obtain the same answer when we directly compute all
anomalies in the corresponding DGLSM. Our aim in this section will be to show
how reduction of the 6D Green-Schwarz terms breaks up into several ingredients
which all contribute to the calculation of the 2d anomaly polynomial. In all
of the SNHCs, we find a perfect match. This provides additional support for
the general physical picture we have developed, and also motivates further
study of the resulting weakly coupled branch of these DGLSMs.

In the rest of this section we show how to calculate the anomaly polynomial
for the 2d DGLSM\ directly in 2d terms. As a warmup, we first consider the
theory defined by a collapsing $-4$ curve. In this theory, all components of
the model admit a weakly coupled limit, and we can directly track the various
contributions. We then show how to perform an analogous computation when the
extra sector defined by D3-branes wrapped on the collapsing $\mathbb{P}^1$
is strongly coupled.

\subsection{The Weakly Coupled $SO(8)$ Theory}

The one simple NHC which defines a weakly coupled DGLSM (far on the tensor
branch) is given by a collapsing $-4$ curve. In this case, the bulk
seven-brane gauge group is $SO(8)$, and the D3-branes wrapped over
the collapsing $\mathbb{P}^{1}$ realize an $Sp(N)$ gauge theory. All together
then, we realize a quiver gauge theory with gauge group $SO(8)\times Sp(N)$,
and with matter fields from the $7-7$ strings, the $3-3$ strings and the $3-7$
strings. Additionally, the reduction of the tensor multiplet to a collection
of chiral / anti-chiral currents naturally couples to both gauge sectors. We
have already discussed this for the seven-brane sector, and it also follows
for the D3-brane sector since, for example, the volume of the collapsing
$\mathbb{P}^{1}$ directly sets the inverse gauge coupling squared for the
$Sp(N)$ gauge theory. Additionally, the various $3-3$ strings will necessarily
couple to these chiral bosons as well. One way to see this is to observe that
the $3-3$ mode content is essentially T-dual to that for the seven-brane
sector. So, the appropriate pullback maps onto the worldvolume of the D3-brane
lead to the analogous couplings for the D3-brane.

Let us discuss in more detail the resulting quiver gauge theory. Consider
first the $7-7$ strings. From the reduction of the $SO(8)$ seven-brane, we get
the zero modes given by a vector multiplet, $V_{(0,0)}$ some CS\ multiplets
$\mathbb{D}_{(0,1)}$ and Fermi multiplets $\Lambda_{(0,2)}$, all transforming
in the adjoint representation.

\begin{figure}[t!]
\begin{center}
\scalebox{1}[1]{
\includegraphics[scale=0.5]{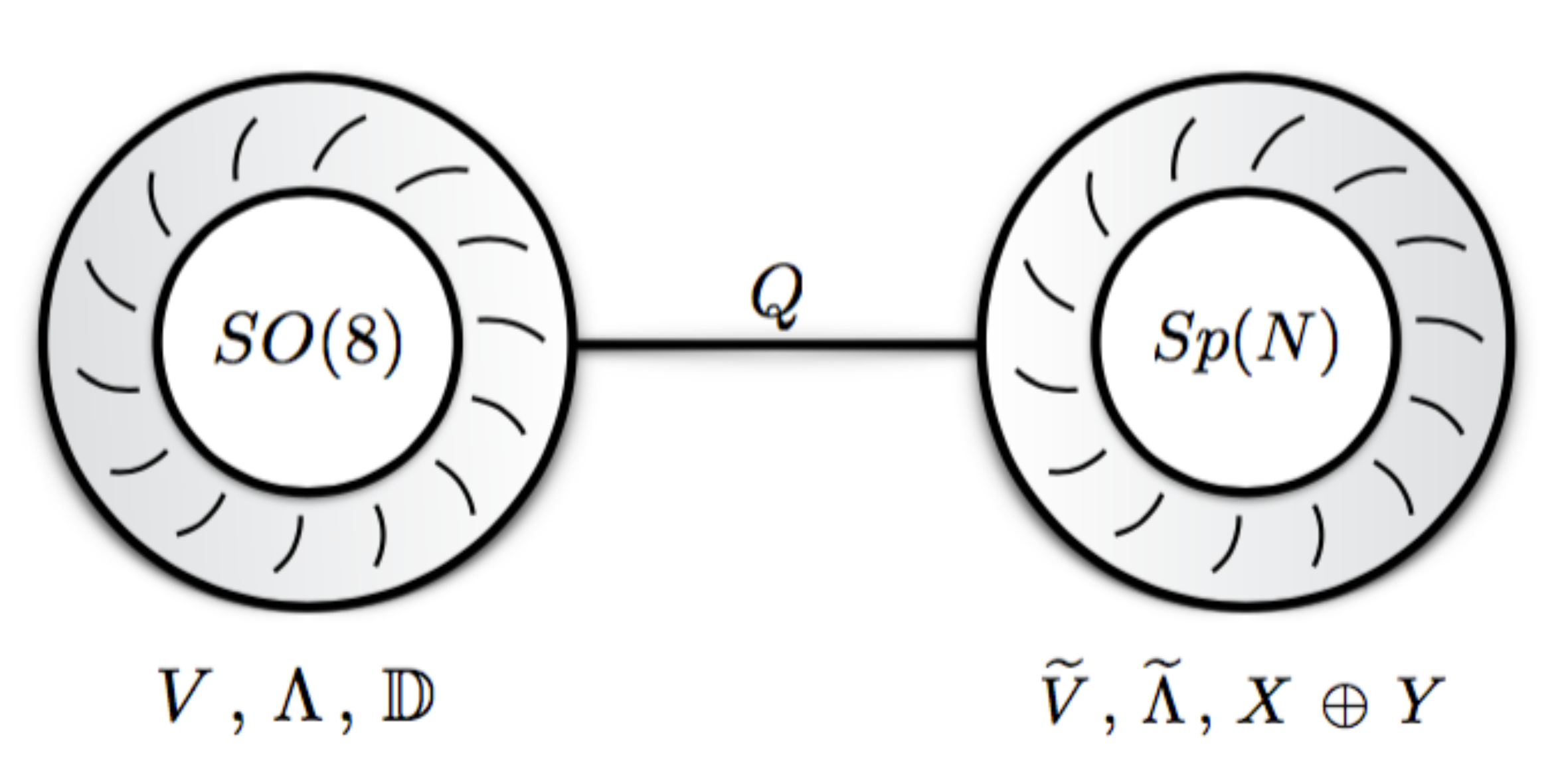}}
\end{center}
\caption{Depiction of the DGLSM defined by compactifying the 6D SCFT defined a $-4$ curve with $SO(8)$ gauge group
on a K\"ahler surface. In two dimensions, we obtain a GLSM with gauge group $SO(8) \times Sp(N)$, and matter
transforming in representations of the two gauge groups. Additionally, the GLSM sector is coupled to a collection of chiral and
anti-chiral currents which are depicted by a thickened shell around each quiver node.}
\label{fig:quiver2}
\end{figure}

Consider next the contribution from the D3-branes. The theory of spacetime
filling strings for this case has been determined in reference
\cite{Gadde:2015tra}, and also follows from the general discussion of extra
sectors given in reference \cite{Apruzzi:2016iac}. For the $3-3$ strings, we
have a vector multiplet in the adjoint representation of $Sp(N)$, so we denote
this contribution as $\widetilde{V}\oplus\widetilde{\Lambda}$.
Additionally, we have a $(0,4)$ hypermultiplet in the two-index anti-symmetric
representation which we can also write as two CS\ multiplets $X\oplus Y$.
Finally, we have $3-7$ strings transforming in the bifundamental
representation and as a $(0,4)$ half hypermultiplet $Q$. For a summary of the
associated quiver gauge theory, see figure \ref{fig:quiver2}.

So far, we have kept the number of D3-branes as a free parameter. However,
this is not really so, since we need to have a theory free of gauge anomalies.
Now, we have seen that the chiral / anti-chiral bosons from reduction of the
tensor multiplet do not contribute to the seven-brane gauge anomaly, so the
only way for us to cancel the $SO(8)$ gauge anomaly will be through a
combination of modes present in the quiver gauge theory.

Adding up the contribution from the $7-7$ strings and the $3-7$ strings, we
have the $SO(8)$ anomaly:%
\begin{align}
k_{\text{tot}}(SO(8))  &  =k_{7-7}(SO(8))+k_{3-7}(SO(8))\\
&  =-\mathrm{Ind}(\mathrm{\mathbf{adj}}(SO(8)))\times(h^{0,0}-h^{0,1}+h^{0,2})+\mathrm{Ind}(\mathrm{\mathbf{fund}}(SO(8)))\times2N. \nonumber
\end{align}
Here, $\mathrm{Ind}(\mathrm{\mathbf{adj}}(G))$ is the index of the adjoint
representation. In our normalization this is $2k$ for $SU(k)$ and $4k-4$ for
$SO(2k)$. The index of the fundamental representation of $SO(2k)$ is then $2$.
Setting $k_{\text{tot}}(SO(8))=0$, we can now solve for the total number of
D3-branes:%
\begin{equation}
N=\frac{\mathrm{Ind}(\mathrm{\mathbf{adj}}(SO(8)))}{2\mathrm{Ind}(\mathrm{\mathbf{fund}}(SO(8)))}\times
\frac{P+3\chi}{12}=\frac{P+3\chi}{4}.
\end{equation}
Note that in spite of appearances, this is always an integer since the
holomorphic Euler characteristic is given by:%
\begin{equation}
h^{0,0}-h^{0,1}+h^{0,2}=\chi(M_{4},\mathcal{O}_{M_{4}})=\frac{P+3\chi}{12}.
\end{equation}
We can also see that our formula for the total number of D3-branes agrees with
our general formula for the number of spacetime filling strings given in
equation (\ref{NinstKinst}):%
\begin{equation}
N_{\text{string}}=-k_{\text{inst}}+\frac{2h_{G}^{\vee}}{n}\chi(M_{4}%
,\mathcal{O}_{M_{4}}),
\end{equation}
where we set $k_{\text{inst}}=0$, $h_{G}^{\vee}=6$ and $n=4$.

Let us next verify that there is no gauge anomaly for the $Sp(N)$ sector.
Recall that here, we have in the $3-3$ sector the contribution from a $(0,4)$
vector multiplet, i.e. we have two gauginos in the adjoint
representation. Additionally, we have a $(0,4)$ hypermultiplet in the
two-index symmetric representation, i.e. two CS\ multiplets. Finally, we have
the $3-7$ sector which contributes eight $(0,4)$ half hypermultiplets in the
fundamental representation, i.e., eight CS multiplets in the fundamental.
Totalling up the contributions to the $Sp(N)$ anomaly, we have:%
\begin{align}
k_{\text{tot}}(Sp(N))  &  =k_{3-3}+k_{3-7} \\
   & =-2\times\mathrm{Ind}(\mathrm{\mathbf{adj}}(Sp(N)))+2\times\mathrm{Ind}(\mathrm{\mathbf{anti}}(Sp(N)))+8\times\mathrm{Ind}(\mathrm{\mathbf{fund}}(Sp(N))) \nonumber\\
&  =-2(4N + 4)+2(4N - 4)+16=0. 	\nonumber
\end{align}
Based on this, we conclude that we have indeed realized an anomaly free
spectrum, as expected.

Let us now turn to the interaction terms of the theory. We primarily
concentrate on the holomorphic terms, i.e. those associated with the
superpotential and E-potential (for the Fermi multiplets). To this end, it is
helpful to make manifest just the $(0,2)$ supersymmetry, so we write the
$(0,4)$ half hypermultiplet in the bifundamental as a pair of CS\ multiplets
$Q\oplus Q^{c}$ in conjugate representations of $U(4)\subset SO(8)$. To
maintain a parallel description of the various interaction terms in the two
sectors, we write all holomorphic data in terms of the associated E-potential
for the Fermi multiplets where they appear. Due to the effective $(0,4)$
structure of the system, the resulting system is given by the E-potentials for
the adjoint valued Fermi multiplets $\Lambda$ and $\widetilde{\Lambda}%
$:\footnote{Strictly speaking, here we use the fact that we have just
$\mathcal{N}=(0,2)$ supersymmetry rather than $\mathcal{N}=(0,4)$
supersymmetry in order to repackage all interaction terms using
E-potentials.}
\begin{align}
E_{\Lambda}  &  =\mathbb{D}\cdot T_{\text{SO(8)}}^{\text{adj}}\cdot
\mathbb{D}+Q^{c}\cdot T_{\text{SO(8)}}^{\text{fund}}\cdot Q\label{Epot}~,\\
E_{\widetilde{\Lambda}}  &  =X\cdot T_{\text{Sp(N)}}^{\text{symm}}\cdot
Y+Q\cdot T_{\text{Sp(N)}}^{\text{fund}}\cdot Q^{c}, \label{Epottilde}%
\end{align}
where the notation $A\cdot T_{G}\cdot B$ means that we take generators $T_{G}$
of the gauge group $G$ in the representation for $A$ and its conjugate
representation $B$, and trace over all other symmetry indices.

Now, in addition to these couplings, we also have, from the reduction of the
6D\ Green-Schwarz term, an interaction term with bosonic content obtained from
the reduction of line (\ref{2dhol}):%
\begin{equation}
L_{2d}\supset \mu_{\mathrm{GS}} \underset{6D}{\int}B_{(2,0)}\wedge\text{Tr}(F_{2d}%
\wedge\lbrack\mathbb{D},\mathbb{D}])+ B_{(2,0)}\wedge\text{Tr}(d_A
\mathbb{D\wedge}d_A \mathbb{D})+c.c.
\end{equation}
This sort of term is topological in 6D, and so we do not expect it to receive
quantum corrections. Additionally, since it continues to play an important
role in the low energy dynamics of the 2d system, we conclude that the
R-charge assignments for the associated bosonic modes $B_{(2,0)}$ and
$\mathbb{D}$ satisfy:%
\begin{equation}
R\left( B_{(2,0)}\right)  +2R(\mathbb{D)=}0.
\end{equation}
But, unitarity ensures that all such R-charges must also be non-negative.
Tracing through the rest of the relations in lines (\ref{Epot}) and
(\ref{Epottilde}), we see that the R-charge assignments in this case all
remain at their free-field values:%
\begin{equation}
R(\mathbb{D)=}R(Q^{c})=R(Q)=R(X)=R(Y)=0.
\end{equation}
Moreover, the R-charge assignments for the left-moving chiral bosons are
protected since they are associated with symmetry currents. This also extends
to all fermionic superpartners; each right--moving fermion in a chiral multiplet  has R-charge $-1$.
Additionally, based on the structure of the interactions, we see that all of the
left--handed Weyl fermions in Fermi multiplets have R-charge $-1$.

So, as we expected from the 6D analysis, there appear to be no emergent $U(1)$
flavor symmetries which could mix with a candidate R-symmetry. Said
differently, once we compute the anomaly polynomial, we will also be able to
extract the IR\ R-symmetry without appealing to $c$-extremization.

\subsubsection{Anomaly Polynomial}

We now turn to the calculation of the 2d anomaly polynomial for this system:%
\begin{equation}
I_{2d}=\frac{k_{RR}}{2}c_{1}(R_{2d,R})^{2}-\frac{k_{\text{grav}}}{24}%
p_{1}(TZ).
\end{equation}
Where:%
\begin{equation}
k_{RR}=\text{Tr}(\gamma_{3}R_{2d,R}R_{2d,R})\text{ \ \ and \ \ }%
k_{\text{grav}}=\text{Tr}(\gamma_{3}).
\end{equation}
Since we are at weak coupling, and all R-charge assignments are fixed by the
couplings of the model, it suffices to carefully count the various degrees of
freedom in the system.

We begin with the calculation of $k_{RR}$. This includes the sectors
associated with our GLSMs, as well as the contribution from the reduction of
the anti-chiral two-form, i.e., we obtain additional chiral bosons, as well as
additional vector multiplets from reduction on one-cycles. Finally, we also
need to include the coupling between the chiral bosons and the background
R-symmetry field strength, i.e., the corresponding 2d Green-Schwarz term.

Consider, then, the various contributions to $k_{RR}$. We have, in the obvious
notation:%
\begin{equation}
k_{RR}=k_{RR}^{7-7}+k_{RR}^{3-3}+k_{RR}^{3-7}+k_{RR}^{\text{Tensor}}%
+k_{RR}^{\text{GS}},
\end{equation}
where the last term is due to the fact that there is a direct coupling between
the chiral bosons and the $U(1)$ R-symmetry. Let us discuss each of these
terms. First, from the $7-7$ strings, we have:%
\begin{equation}
k_{RR}^{7-7}=\dim SO(8)\times\left(  -h^{0,0}+h^{0,1}-h^{0,2}\right)
=-28\times\left(  \frac{P+3\chi}{12}\right)  .
\end{equation}
Next, from the $3-3$ strings, we have:%
\begin{equation}
k_{RR}^{3-3}=-2\times\dim Sp(N)\times(R_{\widetilde{\Lambda}})^{2}%
+2\times\mathrm{\dim}(\mathbf{anti}(Sp(N)))\times(R_{X}-1)^{2}=-4N.
\end{equation}
From the $3-7$ strings, we have:%
\begin{equation}
k_{RR}^{3-7}=8\times\mathrm{\dim}(\mathbf{fund}(Sp(N)))\times(R_{Q}-1)^{2}=16N.
\end{equation}

Consider next the contribution from the reduction of the tensor multiplet
fields. Here,\ we have both fermionic partners of the complex chiral bosons,
$h^{2,0}+1$, all with R-charge $+1$, and gauginos for the $h^{1,0}$ modes
obtained from taking one leg of the $B$-field on $M_{4}$ and one in the
spacetime. The net contribution from these fermions to the anomaly is:%
\begin{equation}
k_{RR}^{\text{Tensor}}=(h^{0,0}-h^{1,0}+h^{2,0})=\left(  \frac{P+3\chi}%
{12}\right)  .
\end{equation}

Finally, we turn to the contribution from the reduction of the Green-Schwarz
term, i.e., terms associated with the exchange of the anti-chiral bosons. Now,
in 6D, the relevant interaction term is:%
\begin{equation}
L_{6D}\supset - n\int_{6D}B\wedge X, \label{READY2dGRR}%
\end{equation}
where for the $SO(8)$ SNHC, we have:%
\begin{equation}
X=\frac{1}{4}\text{Tr}F^{2}+\frac{6}{4}\left(  c_{2}(R)+\frac{p_{1}(T)}%
{12}\right)  .
\end{equation}
This yields the Green-Schwarz contribution to the 6D anomaly polynomial:%
\begin{equation}
I_{6D}\supset\frac{4}{2}X^{2}\text{.}%
\end{equation}

A similar term will be present in the 2d anomaly polynomial, i.e., for the
reducible contribution $c_{1}(R_{2d,R})^{2}$. To extract the form of this
contribution to the 2d anomaly polynomial, we need to track the coupling of
the chiral bosons to the relevant field strengths. Now, since we are
integrating characteristic classes (of holomorphic vector bundles), we see that only the chiral bosons
proportional to $(1,1)$ forms will actually contribute. Note there is at least
one such contribution, since we have the K\"{a}hler class of the manifold
itself. Observe, however, that this coupling does not involve the Pontryagin
class at all. What this means is that it is not correct to simply integrate
$X^{2}$ over $M_{4}$ to extract the contribution to $c_{1}(R_{2d,R})^{2}$.
Indeed, part of the point of our analysis in this section has been to see how
the 6D Green-Schwarz terms break up into separate contributions in the 2d
effective theory.

Instead, we must only include those terms directly associated with the
coupling of the 2d chiral bosons. In practical terms, this simply means we
make the substitution $X\rightarrow X_{\text{eff}}$ compatible with the twist
on $M_{4}$:%
\begin{equation}
X_{\text{eff}}=-\frac{6}{4}c_{1}(R_{2d,R})c_{1}(K_{M_{4}}),
\end{equation}
so that the contribution to the 2d anomaly polynomial is:%
\begin{equation}
I_{2d}\supset\frac{4}{2}\int_{M_{4}}X_{\text{eff}}^{2}=\frac{9}{2}%
\times(P+2\chi)\times c_{1}(R_{2d,R})^{2}.
\end{equation}
So in other words, the contribution to $k_{RR}$ is then:%
\begin{equation}
k_{RR}^{\text{GS}}=9(P+2\chi).
\end{equation}
We can now add up all of the contributions to $k_{RR}$, and we find:%
\begin{equation}
k_{RR}=\frac{3}{4}\left(  13P+27\chi\right)  ,
\end{equation}
which agrees with the general expression we presented in equation (\ref{krr})
from direct dimensional reduction of the 6D anomaly polynomial!

Consider next the contribution to $k_{\text{grav}}$. In this case there is no
contribution from a Green-Schwarz term, as the $p_{1}(T)$ term is irreducible.
Again, we split up the various contributions as:%
\begin{equation}
k_{\text{grav}}=k_{\text{grav}}^{7-7}+k_{\text{grav}}^{3-3}+k_{\text{grav}%
}^{3-7}+k_{\text{grav}}^{\text{Tensor}}.
\end{equation}
where the contributions from the GLSM\ sector are the same as for $k_{RR}$:%
\begin{align}
k_{\text{grav}}^{7-7}  &  =\dim SO(8)\times(-h^{0,0}+h^{0,1}-h^{0,2}%
)=-28\left(  \frac{P+3\chi}{12}\right)  .\\
k_{\text{grav}}^{3-3}  &  =-2\times\dim Sp(N)+2\times\mathrm{\dim
}(\mathbf{anti}(Sp(N)))=-4N\\
k_{\text{grav}}^{3-7}  &  =8\times\mathrm{\dim}(\mathbf{fund}(Sp(N))) = 16N.
\end{align}

For the reduction of the anti-chiral two-form, we need to include the
contributions from the fermionic superpartners, as well as the chiral bosons.
The fermionic contribution to the gravitational anomaly is the same as what we
already calculated for $k_{RR}$:%
\begin{equation}
k_{\text{grav}}^{\text{Tensor,}F}=h^{0,0}-h^{1,0}+h^{2,0}=\left(
\frac{P+3\chi}{12}\right)  .
\end{equation}
Additionally, we have $2h^{2,0}+2$ right-moving chiral bosons, and $h^{1,1}$
left-moving chiral bosons. The difference between these two sectors
contributions as the signature of the manifold:%
\begin{equation}
k_{\text{grav}}^{\text{Tensor,}B}=b_{2}^{+}-b_{2}^{-}=\frac{P}{3}.
\end{equation}
So in other words, the net contribution from reduction of the anti-chiral
two-form to the gravitational anomaly is:%
\begin{equation}
k_{\text{grav}}^{\text{Tensor}}=\left(  \frac{5P+3\chi}{12}\right)  .
\end{equation}
The net contribution to the gravitation anomaly is thus:%
\begin{equation}
k_{\text{grav}}=\frac{13P+27\chi}{12},
\end{equation}
which agrees with equation (\ref{kgrav}), which we obtained by direct
dimensional reduction of the $-4$ curve SNHC!

An interesting feature of this
case is that the ratio of $k_{RR}$ and $k_{\text{grav}}$ does not depend on
the details of the four-manifold:%
\begin{equation}
\frac{k_{RR}}{k_{\text{grav}}}=9.
\end{equation}

\subsection{Simple NHCs}

The simplifying feature of the $-4$ curve theory is that we essentially had a
standard GLSM\ coupled to a collection of chiral / anti-chiral bosons.
In this sense, the whole theory is just a DGLSM. We now turn to the other simple NHCs where we do
not a priori expect the spacetime filling strings to admit a description in
terms of a GLSM. We can, however, still apply the general logic used in our
analysis of the $SO(8)$ theory to calculate the resulting anomaly polynomial.

For the simple NHC$\ $defined by a $-n$ curve, we have a simply laced gauge
group $G$ which comes from a seven-brane on $\mathbb{R}^{1,1}\times
M_{4}\times\mathbb{P}^{1}$. Additionally, we now have some number of spacetime
filling strings which coupled to the seven-brane. Finally, we also have the
contribution from the reduction of the tensor multiplet.

To see that spacetime filling strings are really necessary, consider just
the GLSM\ sector from the seven-branes. The gauge anomaly for the gauge group
$G$ is:%
\begin{equation}
k_{7-7}(G)=-\mathrm{Ind}(\mathrm{\mathbf{adj}}(G))\times(h^{0,0}%
-h^{0,1}+h^{0,2})=-\mathrm{Ind}(\mathrm{\mathbf{adj}}(G))\left(  \frac
{P+3\chi}{12}\right)  . \label{77G}%
\end{equation}
The chiral bosons cannot contribute to anomaly cancellation, because this is a
non-abelian gauge group factor. So, we must assume the presence of additional
spacetime filling strings. Note that in weakly coupled terms, we need to add
some number of CS multiplets to cancel the anomaly. This means in particular
that we expect to cancel the anomaly with supersymmetric matter multiplets when
$P + 3 \chi > 0$.

Now, we can actually fix the total number of such strings using the structure
of the anomaly polynomial for these 2d effective theories. This has recently
been calculated in references \cite{Kim:2016foj, Shimizu:2016lbw,
DelZotto:2016pvm}, and in Appendix \ref{app:MN} we verify this
by directly dimensionally reducing on a $\mathbb{P}^1$ the 4d $\mathcal{N} = 2$ SCFTs (using
the values of the anomalies obtained in references \cite{Cheung:1997id, Aharony:2007dj,
Argyres:2007tq, Shapere:2008zf}) that we obtain the same result. For our purposes,
the presentation for the anomaly polynomial of $N$ spacetime filling strings
in flat space given in reference \cite{DelZotto:2016pvm} is most convenient.
Focussing on just the symmetries which survive in compactifying on a
K\"{a}hler surface, we have:%
\begin{equation}
I_{\text{string}}=N\times\left(  \frac{n}{4}\text{Tr}_{G}F^{2}-h_{G}^{\vee
}\left(  c_{2}(R)+\frac{1}{12}p_{1}(TZ)\right)  \right)  \text{,}%
\end{equation}
where in the above, we have introduced an overall relative minus sign between
the gauge field strength term and the background R-symmetry and tangent bundle
contributions. As we remarked before, this is due to the chirality conventions
for our system, which as we shall shortly verify, led to a self-consistent
prescription for supersymmetric anomaly cancellation.

Now, in our case where we can only activate a contribution from the 2d $U(1)$
R-symmetry and the background metric in two dimensions, the relevant
expression is given by:%
\begin{equation}
I_{\text{string}}=N\times\left(  \frac{n}{4}\text{Tr}_{G}F^{2}-h_{G}^{\vee
}\left(  -c_{1}(R_{2d,R})^{2}+\frac{1}{12}p_{1}(TZ)\right)  \right)  .
\end{equation}
Returning to equation (\ref{77G}), we see that the total gauge anomaly
is:%
\begin{equation}
k_{\text{tot}}(G)=k_{7-7}(G)+k_{\text{string}}(G)=-\mathrm{Ind}%
(\mathrm{\mathbf{adj}}(G))\left(  \frac{P+3\chi}{12}\right)  +N\times n,
\end{equation}
so the number of strings is fixed as:%
\begin{equation}
N=\mathrm{Ind}(\mathrm{\mathbf{adj}}(G))\times\frac{1}{n}\times\left(
\frac{P+3\chi}{12}\right)  .
\end{equation}
Reassuringly, this agrees with our answer from the previous subsection in the
special case $G=SO(8)$, where $\mathrm{Ind}(\mathrm{\mathbf{adj}}(G))=12$ and
$n=4$. We can also repackage this formula using the
fact that for all of the simply laced simple algebras (i.e. the gauge groups
of the simple NHCs), $\mathrm{Ind}(\mathrm{\mathbf{adj}}(G))=2h_{G}^{\vee}$. So, we
indeed recover the formula of equation (\ref{NinstKinst}):%
\begin{equation}
N=\frac{2h_{G}^{\vee}}{n}\times\left(  \frac{P+3\chi}{12}\right)  .
\end{equation}

Now, in spite of the fact that we do not have a Lagrangian description for
these spacetime filling strings, we can use the geometric picture to argue
that there are (at least in the UV) no additional flavor symmetries. Assuming
the absence of emergent abelian symmetries in the infrared, this means in
particular that we can fix the R-charge assignments of fields in terms of UV
data. Much as in our discussion of the weakly coupled model, we deduce that
the R-charges of the GLSM fields are frozen to their free field values. Similar
considerations hold for the chiral bosons of the system.

\subsubsection{Anomaly Polynomial}

Let us now turn to the calculation of the 2d anomaly polynomial. In this case,
we again see no evidence for additional abelian flavor symmetries. Thus, our
task reduces to calculating the coefficients in:%
\begin{equation}
I_{2d}=\frac{k_{RR}}{2}c_{1}(R_{2d,R})^{2}-\frac{k_{\text{grav}}}{24}%
p_{1}(TZ),
\end{equation}
where:%
\begin{equation}
k_{RR}=\text{Tr}(\gamma_{3}R_{2d,R}R_{2d,R})\text{ \ \ and \ \ }%
k_{\text{grav}}=\text{Tr}(\gamma_{3}).
\end{equation}

Now that we are at strong coupling, it will be more challenging to make
detailed checks of the proposal. However, we can still rely on anomalies as a
way to track this data. From this perspective, the main subtlety is to ensure
that we fix our chirality convention so that all anomalies are cancelled with
supersymmetric multiplets.

We begin with the calculation of $k_{RR}$. This includes the sectors
associated with our GLSM, the spacetime filling strings, as well as the
contribution from the reduction of the anti-chiral two-form, i.e., we obtain
additional chiral bosons, as well as additional vector multiplets from
reduction on one-cycles. Finally, we also need to include the coupling between
the chiral bosons and the background R-symmetry field strength, i.e., the
corresponding 2d Green-Schwarz term.

Consider, then, the various contributions to $k_{RR}$:%
\begin{equation}
k_{RR}=k_{RR}^{7-7}+k_{RR}^{\text{strings}}+k_{RR}^{\text{Tensor}}%
+k_{RR}^{\text{GS}},
\end{equation}
where the last term is due to the fact that there is a direct coupling between
the chiral bosons and the $U(1)$ R-symmetry. Let us discuss each of these
terms. First, from the $7-7$ strings, we have:%
\begin{equation}
k_{RR}^{7-7}=d_{G}\times\left(  -h^{0,0}+h^{0,1}-h^{0,2}\right)  =-d_{G}%
\times\left(  \frac{P+3\chi}{12}\right)  ,
\end{equation}
where $d_{G}$ is the dimension of the gauge group $G$.

Next, from the spacetime filling strings, we have the contribution:%
\begin{equation}
k_{RR}^{\text{strings}}=2N\times h_{G}^{\vee}=\frac{4\left(  h_{G}^{\vee
}\right)  ^{2}}{n}\times\left(  \frac{P+3\chi}{12}\right)  .
\end{equation}
Observe in particular that for $P + 3 \chi > 0$ this is a positive quantity, i.e., it is what we
would get from summing up the contribution from CS, rather than Fermi
multiplets, or (if supersymmetry had been broken) anti-CS multiplets.

The contribution from $k_{RR}^{\text{Tensor}}$ is the same as what we already
found for the $SO(8)$ model:%
\begin{equation}
k_{RR}^{\text{Tensor}}=(h^{0,0}-h^{1,0}+h^{2,0})=\left(  \frac{P+3\chi}%
{12}\right)  .
\end{equation}

Finally, there is the contribution from the Green-Schwarz term in two
dimensions. Using the same analysis presented near line (\ref{READY2dGRR}), we
see that now, $X_{\text{eff}}$ is given by:%
\begin{equation}
X_{\text{eff}}=-\frac{h_{G}^{\vee}}{n}c_{1}(R_{2d,R})c_{1}(K_{M_{4}}),
\end{equation}
so that the contribution to the 2d anomaly polynomial is:%
\begin{equation}
I_{2d}\supset\frac{n}{2}\int_{M_{4}}X_{\text{eff}}^{2}=\frac{\left(
h_{G}^{\vee}\right)  ^{2}}{2n}\times(P+2\chi)\times c_{1}(R_{2d,R})^{2}.
\end{equation}
So in other words, the contribution to $k_{RR}$ is then:%
\begin{equation}
k_{RR}^{\text{GS}}=\frac{\left(  h_{G}^{\vee}\right)  ^{2}}{n}(P+2\chi).
\end{equation}
The sum total of all these contributions is therefore:%
\begin{align}
k_{RR}  &  =-d_{G}\times\left(  \frac{P+3\chi}{12}\right)  +\frac{4\left(
h_{G}^{\vee}\right)  ^{2}}{n}\times\left(  \frac{P+3\chi}{12}\right)  +\left(
\frac{P+3\chi}{12}\right)  +\frac{\left(  h_{G}^{\vee}\right)  ^{2}}%
{n}(P+2\chi)\\
&  =(1-d_{G})\times\left(  \frac{P+3\chi}{12}\right)  +\frac{\left(
h_{G}^{\vee}\right)  ^{2}}{3n}\times\left(  4P+9\chi\right)  ,
\end{align}
which agrees with equation (\ref{krr}), i.e., our formula obtained from
dimensional reduction of the 6D\ SCFT anomaly polynomial!

Consider next the contributions to the gravitational anomaly. Just as in the
$SO(8)$ example, there is no Green-Schwarz term as the $p_1(T)$ term is irreducible.
We must, however, properly take into account the contributions from the chiral bosons.
Again, we split up the various contributions as:%
\begin{equation}
k_{\text{grav}}=k_{\text{grav}}^{7-7}+k_{\text{grav}}^{\text{strings}%
}+k_{\text{grav}}^{\text{Tensor}},
\end{equation}
where:%
\begin{align}
k_{\text{grav}}^{7-7}  &  =d_{G}\times(-h^{0,0}+h^{0,1}-h^{0,2})=-d_{G}\left(
\frac{P+3\chi}{12}\right)  \\
k_{\text{grav}}^{\text{strings}}  &  =2h_{G}^{\vee}N=\frac{4\left(
h_{G}^{\vee}\right)  ^{2}}{n}\times\left(  \frac{P+3\chi}{12}\right) \\
k_{\text{grav}}^{\text{Tensor}}  &  =\left(  \frac{5P+3\chi}{12}\right)  ,
\end{align}
which by inspection of equation (\ref{kgrav}) yields precisely the same answer
as obtained from dimensional reduction of the 6D\ SCFT answer!

We can also compute the ratio $k_{RR}/k_{\text{grav}}$, which in the general
case, does depend on the four-manifold:
\begin{equation}
\frac{k_{RR}}{k_{\text{grav}}}=\frac{n(d_{G}-1)(P+3\chi)-4\left(  h_{G}^{\vee
}\right)  ^{2}(4P+9\chi)}{n(d_{G}-1)(P+3\chi)-4\left(  h_{G}^{\vee}\right)
^{2}(P+3\chi)-4nP}.
\end{equation}
Indeed, it appears to be a special property of just the $-4$ curve theory that
all dependence on $P$ and $\chi$ vanishes. It would be quite interesting to
understand this better.

\subsection{Four-Manifold Constraints}

In our analysis of anomalies, we left the general choice of four-manifold
arbitrary. Of course, experience from higher-dimensional examples suggests
that the particular choice will impact whether we realize a supersymmetric 2d
theory at all.

As an example, the relative supersymmetries preserved by our DGLSM and our
spacetime filling strings may be misaligned, thus leading to a
non-supersymmetric matter spectrum. This depends on both the background
curvatures of the four-manifold in question, as well as the choice of
background flux from the seven-branes. Here, we focus on the case considered
in the rest of the paper, i.e., where the flux of the seven-brane is trivial.

Returning to the weakly coupled $SO(8)$ model, we see that to cancel the
$SO(8)$ gauge anomalies, it was necessary to add additional CS multiplets. The
total number of such CS multiplets is:%
\begin{equation}
N_{Q}\propto\frac{P+3\chi}{12},
\end{equation}
where the constant of proportionality is positive. So, to have a
supersymmetric matter spectrum we require:%
\begin{equation}
\frac{P+3\chi}{12}>0,
\end{equation}
a condition which is not always satisfied by a four-manifold.

Let us give some examples of both types. It is helpful to recall that the
quantity we are considering is also given by the holomorphic Euler
characteristic:%
\begin{equation}
\frac{P(M_{4})+3\chi(M_{4})}{12}=\chi(M_{4},\mathcal{O}_{M_{4}})=1-h^{0,1}%
+h^{0,2}.
\end{equation}
First of all, for all the del Pezzo surfaces $dP_{k}$ we have:%
\begin{equation}
\frac{P(dP_{k})+3\chi(dP_{k})}{12}=1,
\end{equation}
while for a K3 surface we have:%
\begin{equation}
\frac{P(K3)+3\chi(K3)}{12}=2.
\end{equation}

An additional class of examples comes from taking a Cartesian product of two
Riemann surfaces $M_{4}=\Sigma_{1}\times\Sigma_{2}$:%
\begin{equation}
\frac{P(\Sigma_{1}\times\Sigma_{2})+3\chi(\Sigma_{1}\times\Sigma_{2})}%
{12}=1-(g_{1}+g_{2})+g_{1}g_{2},
\end{equation}
where $g_{i}$ is the genus of $\Sigma_{i}$. So, we see that if both
$g_{1}=g_{2}=0$, this is positive, and it is non-negative if $g_{i}\geq1$ for
both $i$. In the case where one factor is a $\mathbb{P}^{1}$ and the other factor is a genus $g > 1$ Riemann surface,
we see that $\chi(M_{4},\mathcal{O}_{M_{4}})<0$, so in this case we do not realize a
supersymmetric matter spectrum. More broadly, one can consider activating
non-trivial background fluxes. This can lead to non-trivial bound states
between the seven-brane and spacetime filling strings. A related way to
preserve supersymmetry is to introduce defects, i.e. poles for some operators
in the internal directions. We leave the analysis of these more general
possibilities for future work.

Based on our discussion above, we conclude that for suitable four-manifolds, we realize
a supersymmetric matter spectrum. A
more subtle question has to do with whether the resulting infrared dynamics
will support a supersymmetric vacuum. One way to partially address this is to
calculate the Witten index and various examples of elliptic genera. As far as
we are aware, this has not been carried out for any DGLSMs, and only for some
examples of compactifications of $(2,0)$ theories (see e.g.
\cite{Gadde:2013sca})\ so we leave this task for future work.

Though we have mainly focussed on theories with a single tensor multiplet, an alternative way to provide evidence
for the existence of such fixed points for 6D SCFTs with a holographic dual
(see e.g. \cite{Apruzzi:2013yva, Gaiotto:2014lca, DelZotto:2014hpa, Apruzzi:2015wna}) is to consider the
resulting compactification on a four-manifold. Some examples along these lines have been worked out in references
\cite{Gauntlett:2000ng, Gauntlett:2006ux, Figueras:2007cn, Passias:2015gya, Karndumri:2015sia}),
where one finds that for a four-manifold of negative curvature,
there is a reduction from an $AdS_7$ solution to an $AdS_3$ solution. This is in line with the constraint $P + 3 \chi > 0$ we have observed. Note
also that this condition is somewhat weaker than what is required to get a weakly coupled gravity dual. Indeed, we expect that
compactification on a del Pezzo surface or a K3 surface should also yield a sensible 2d conformal fixed point. The holographic dual of
this case is likely to be a theory with higher spins in the bulk, which would be quite interesting to study further.

\section{Conclusions \label{sec:CONC}}

Compactifications of higher-dimensional SCFTs provide a general template for
understanding many non-trivial aspects of lower-dimensional quantum systems.
In this paper we have considered compactifications of 6D\ SCFTs on
four-manifolds, using the higher-dimensional string theory description as a
tool in the study of the resulting 2d effective field theories. In particular,
we have argued that the resulting theory can also be understood by studying
compactification on the tensor branch, where we obtain a dynamic GLSM, that
is, a GLSM\ in which some parameters and couplings have been promoted to
chiral / anti-chiral modes, and which is in turn typically coupled to a theory
of spacetime filling effective strings. We have provided general evidence for
this physical picture, by calculating, for example, the anomaly polynomial of
this DGLSM, finding perfect agreement with the answer expected from
compactification of the 6D\ SCFT. Turning the discussion around, we have also
used the 6D (and ultimately stringy) perspective to shed light on the dynamics
of DGLSMs. In the remainder of this section we discuss some avenues of future investigation.

One of the key simplifications we made in our analysis is to focus on the
simple NHCs. It would be quite interesting to extend this to all of the NHCs,
and to combine this with an analysis of compactification of the E-string
theory on a four-manifold. This would provide a class of tools to
systematically study more general compactifications of 6D SCFTs using the
classification results of \cite{Heckman:2013pva, Heckman:2015bfa}.

There has recently been much study of other compactifications of 6D\ SCFTs, and
in particular the case of the resulting 4d effective theories.
It is tempting to combine this thread with the present one to develop a generalized 2d / 4d
correspondence perhaps along the lines of reference \cite{Alday:2009aq}.
Indeed, we have seen that some aspects of these 2d theories are remarkably tractable, and moreover, that the
internal dynamics of these systems are often governed by a quasi-topological
field theory coupled to various defects. In that context, another
item of interest would be the calculation of the elliptic genera for our 2d effective
theories.

Another natural avenue of investigation involves the study of various defects
/ punctures in these theories. Now, in contrast to the case of
compactifications to four dimensions, the defects here will now be associated
with complex curves, leading to a rather rich generalization of the purely
algebraic data present for 4d theories.

The theories we have focussed on in this work are particularly simple because
they have a single tensor multiplet. It is natural to also consider the
opposite limit, i.e. where we take particular 6D\ SCFTs with a weakly coupled
holographic dual. Compactifications of these systems on a negative curvature
surface lead to $AdS_{3}$ vacua. It would be quite interesting to use the
present perspective to extract further information about the resulting 2d CFTs
thus obtained.

Finally, one of the original motivations for this work was to study a
particularly tractable class of examples in which the moduli present in
F-theory compactified on an elliptically fibered Calabi-Yau fivefold can be
systematically decoupled (see in particular \cite{Apruzzi:2016iac}). Having
developed further tools to study such systems, it would be exciting to
return to this more ambitious class of questions, and their application to
questions in the cosmology of super-critical strings.

\section*{Acknowledgements}

We thank F. Benini, C. Closset, M. Del Zotto, R. Donagi,
T.T. Dumitrescu, B. Haghighat, and D.R. Morrison for helpful
discussions. FA, FH and JJH thank the theory groups at Columbia University and
the ITS at the CUNY\ graduate center for hospitality during part of this work.
JJH also thanks the CCPP at NYU for hospitality during part of this work. The
work of FA, FH and JJH is supported by NSF CAREER grant PHY-1452037. FA, FH
and JJH also acknowledge support from the Bahnson Fund at UNC Chapel Hill as
well as the R.~J. Reynolds Industries, Inc. Junior Faculty Development Award
from the Office of the Executive Vice Chancellor and Provost at UNC Chapel
Hill. FA\ and FH\ also acknowledge support from NSF\ grant PHY-1620311.
The work of IVM was supported in part through the Faculty Assistance Grant
from the College of Science and Mathematics at James Madison University.
IVM also thanks the organizers of the ``Mathematics of String Theory''
program and the Institut Henri Poincar\'{e} for hospitality while some
of this work was carried out.


\newpage

\appendix

\section{Reduction of 6D Free Fields \label{app:FREE}}

This Appendix presents the detailed reduction of $(1,0)$ six dimensional superfields to two dimensions. While the main focus lies on the tensor multiplet, we also dimensionally reduce the vector- and hypermultiplet to provide a consistency check.

Before we discuss the actual dimensional reduction, let us review our superspace conventions. The superspace coordinates are $(x^M, \theta^{\alpha s})$ where $x^M$ are the coordinates of the 6D spacetime and the Grassmann coordinates $\theta^{\alpha s}$ represent a symplectic Majorana-Weyl spinor. Spinor indices $\alpha,\,\beta,\,\dots$ parameterize the fundamental/anti-fundamental of $SO(5,1)\sim SU^*(4)$. We use the convention that $\psi^\alpha_R$ is right-handed chiral and $(\psi_{L})_\alpha$ has the opposite chirality.
To avoid a redundancy in the notation, we either write $\psi_R$ or $\psi^\alpha$ and the same for left-handed spinors. Further, indices $s,\,t\, \dots$ represent the $Sp(1)\sim SU(2)$ $R$-symmetry. They are raised and lowered by totally antisymmetric $\Omega_{st}$, e.g. $\psi_s = \Omega_{st} \psi^t$ and $\psi^s = \Omega^{st} \psi_t$ with $\Omega^{su}\Omega_{ut} = \delta^s_t$. In most cases, we go to the explicit basis $s= (1,\,2)$ with
$\Omega_{12} = -1$ and $\Omega^{12} = 1$. Finally, there is complex conjugation which acts on pseudo-real symplectic Majorana-Weyl spinors as $\overline{\psi^\alpha_s} = \psi^{s \alpha} = \Omega^{st} \psi^\alpha_t$.

The 6D superspace derivatives
\begin{equation}
  D_\alpha^s = \frac\partial{\partial \theta^\alpha_s} - i \theta^{s \beta} \partial_{\alpha\beta}
\end{equation}
anti-commute with the supercharges
\begin{equation}
  Q^s_\alpha = \frac\partial{\partial \theta^\alpha_s} + i \theta^{s\beta}  \partial_{\alpha\beta}~,
\end{equation}
and they give rise to the algebra
\begin{equation}
  \{D_\alpha^s, D_\beta^t\} = - 2 i \Omega^{st} \partial_{\alpha\beta}\,,
\end{equation}
while for the supercharges
\begin{equation}
  \{Q_\alpha^s, Q_\beta^t\} = 2 i \Omega^{st} \partial_{\alpha\beta}
\end{equation}
holds. We use the abbreviation $\partial_{\alpha\beta} = \gamma_{\alpha\beta}^M \partial_M$ for partial derivatives contracted with an antisymmetric representation of the $\Gamma$-matrix sub-block $\gamma^M_{\alpha\beta}$. The full $8 \times 8$ matrices have the form
\begin{equation}
  \Gamma^M \Psi =
    \begin{pmatrix}
      0 & \gamma^M_{\alpha\beta} \\
      (\tilde \gamma^M)^{\alpha\beta} & 0
    \end{pmatrix}
    \begin{pmatrix}
      \psi_\beta \\ \psi^\beta
    \end{pmatrix}\,,\quad
  \tilde\gamma_M^{\alpha\beta} = \frac{1}{2} \epsilon^{\alpha\beta\sigma\delta} (\gamma_M)_{\sigma\delta}\,.
\end{equation}
Using this definition, we further find
\begin{equation}
  \Gamma^{MN} \Psi = \begin{pmatrix}
      (\gamma^{MN})_\alpha{}^\beta & 0 \\
      0 & (\tilde \gamma^{MN})^\alpha{}_\beta
    \end{pmatrix} \begin{pmatrix} \psi_\beta \\ \psi^\beta \end{pmatrix}
\end{equation}
with
\begin{equation}
  (\gamma^{MN})_\alpha{}^\beta = ( \gamma^{[M} \tilde\gamma^{N]} )_\alpha{}^\beta
    \quad \text{and} \quad
  (\tilde\gamma^{MN})^\alpha{}_\beta = - (\gamma^{MN})_\beta{}^\alpha\,.
\end{equation}
In six dimensions the chirality operator is defined by
\begin{equation}
  \Gamma_7 = - \Gamma^0 \Gamma^1 \Gamma^2 \Gamma^3 \Gamma^4 \Gamma^5\,.
\end{equation}
Finally, there is the completeness relation
\begin{equation}\label{eqn:completeness1}
  (\gamma^{MN})_\alpha{}^{\beta} (\gamma_{MN})_\delta{}^\sigma = 2 \delta_\alpha^\beta \delta_\delta^\sigma - 8 \delta^\sigma_\alpha \delta^\beta_\delta
\end{equation}
that is quite useful for several calculations.

\subsection*{Reduction on K\"ahler Surface}
In order to reduce the six dimensional theory on a K\"ahler surface $M_4$, we first split the coordinates according to $x^M = (y^+,\, y^-, \, z^i,\, \overline{z}^{\bar i})$. The first two directions are identified with the external $\mathbb{R}^{1,1}$, while $M_4$ has the complex coordinated $z^1$, $z^2$. Further, we fix the metric
\begin{equation}
  (d s)^2 = - 4 d y^+ d y^- \quad \text{or} \quad
  g_{+-} = -2\,, \quad g^{+-} = - \frac12
\end{equation}
of the external Minkowski space. On $M_4$ the only non-vanishing metric components are $g_{i\bar j}$.

In 6D, the covariantly constant spinor representing the parameter of supersymmetry transformations is denotes as $\eta^{s\alpha}$. It has the opposite chirality of the supercharge $Q_\alpha^s$. We use it to fix the + direction of the external coordinate system by requiring
\begin{equation}
  \eta^{s\alpha} \gamma^M_{\alpha\beta} \eta^{t\beta} \partial_M = \Omega^{st} \partial_+ \,.
\end{equation}
For the following calculations, we introduce a left-handed counterpart of $\eta^{s\alpha}$ defined as:
\begin{equation}\label{eqn:righttoleft}
  \eta^s_\alpha = \gamma^+_{\alpha\beta} \eta^{s \beta}\,,
\end{equation}
which immediately gives rise to $\eta^{s \alpha} \eta^t_\alpha = \Omega^{st}$. In addition to \eqref{eqn:righttoleft}, we further obtain the relations
\begin{equation}
  \tilde\gamma^{-\beta\alpha} \eta^s_\alpha = - \eta^s_\beta\,,
    \quad
  \gamma^-_{\beta\alpha} \eta^{s\alpha} = 0
    \quad\text{and}\quad
  \tilde\gamma^{+\beta\alpha} \eta^s_\alpha = 0
\end{equation}
by requiring a consistent inner product for $\eta^{s\alpha}$.

In the spinor representation, the generator of the $\mathfrak{u}(1)_{M_4}$, which arises after decomposing the holonomy group $U(2)_{M_4}$ of $M_4$ into $SU(2)\times U(1)_{M_4}$, reads\footnote{We normalized $J_{M_4}$ such that $[J_{M_4}, \Gamma^i] = \Gamma^i$ and $[J_{M_4}, \Gamma^{\bar i}] = - \Gamma^{\bar i}$ assuming the standard convention $\omega_{i\bar k}g^{\bar k j} = i \delta_i^j$ and $\omega_{\bar i k}g^{k \bar j} = - i \delta_{\bar i}^{\bar j}$.}
\begin{equation}
  J_{M_4} = -\frac{i}4 \omega_{MN} \Gamma^{MN}
\end{equation}
where
\begin{equation}
  \omega_{MN} = \frac{i}2 \Big( \eta^{1\beta} (\gamma_{MN})_\beta{}^\alpha \eta^2_\alpha + \eta^{2\beta} (\gamma_{MN})_\beta{}^\alpha \eta^1_\alpha \Big)
\end{equation}
By construction this choice of $\omega_{MN}$, results in $\mathfrak{u}(1)_{M_4}$ charges $-1$/$+1$ for $\eta^{1\alpha}$/$\eta^{2\alpha}$. This can be checked by applying the completeness relation \eqref{eqn:completeness1}. Further, we denote the generator of the $\mathfrak{u}(1)_R$ cartan subalgebra of the $SU(2)$ $R$-symmetry as $J_R$. Taking into account the branching $\mathbf{2}\rightarrow +1 \oplus -1$ of the fundamental in which $\eta^{s \alpha}$ transforms, we obtain the charges
\begin{equation}
  J_R \eta^{1\alpha} = \eta^{1\alpha} \quad\text{and}\quad
  J_R \eta^{2\alpha} = - \eta^{2\alpha}\,.
\end{equation}
The generator of the $U(1)_\mathrm{diag}$ which can be embedded $U(1)_{M_4} \times U(1)_R$ is $J_\mathrm{diag} = J_{M_4} + J_R$. Combining now the action of $J_{M_4}$ and $J_R$, we discussed so far, it is obvious that
\begin{equation}
  J_\mathrm{diag} \eta^{s\alpha} = 0
\end{equation}
holds. Thus, we indeed can identify $\eta^{s\alpha}$ with a covariant constant spinor on $M_4$. As a convenient crosscheck, we note that this spinor has perspective negative chirality
\begin{equation}\label{eqn:gammaz2D}
  \Gamma_{3,2d} = \Gamma^0 \Gamma^1 = \Gamma^- \Gamma^+ - \Gamma^+ \Gamma^-\,,
\end{equation}
e.g. $\Gamma_{3,2d} \eta = - \eta$, from a 2d perspective. This is equivalent to spin +1/2. The corresponding conserved supercharges have the opposite chirality and spin.

We now use $\eta^{s\alpha}$ and its left-handed counterpart to construct the arbitrary 6D spinors:
\begin{align}
  \psi^1_\alpha &= \Big( \frac14 \psi_{ij} \gamma^{ij} + \frac1{\sqrt{2}} \psi_i \gamma^{-i} + \psi \delta \Big){}_\alpha{}^\beta \eta^1_\beta &
  \psi^2_\alpha &= \Big( \frac14 \overline{\psi}{}_{\bar i\bar j} \gamma^{\bar i\bar j} + \frac1{\sqrt{2}} \overline{\psi}{}_{\bar i} \gamma^{-\bar i} + \overline{\psi} \delta \Big){}_\alpha{}^\beta \eta^2_\beta
  \label{eqn:leftspinorexp} \\
  \psi^{1\alpha} &= \Big( \frac14 \psi_{ij} \gamma^{ij} + \frac1{\sqrt{2}} \psi_i \gamma^{+i} + \psi \delta \Big){}_\beta{}^\alpha \eta^{1\beta} &
  \psi^{2\alpha} &= \Big( \frac14 \overline{\psi}{}_{\bar i\bar j} \gamma^{\bar i\bar j} + \frac1{\sqrt{2}} \overline{\psi}{}_{\bar i} \gamma^{+\bar i} + \overline{\psi}{} \delta \Big){}_\beta{}^\alpha \eta^{2\beta}
\label{eqn:rightspinorexp}
\end{align}
out of differential forms on the K\"ahler surface $M_4$. These expression are explicit realizations of the branching rules
\begin{align}
  (\mathbf{4}, \mathbf{2}) \rightarrow& (-1/2,\mathbf{1},2)\oplus (+1/2,\mathbf{2},+1)\oplus (-1/2,\mathbf{2},0) \oplus \nonumber \\
  &(-1/2,1,-2)\oplus (+1/2,2,-1)\oplus (- 1/2,1,0) \\
  (\overline{\mathbf{4}}, \mathbf{2}) \rightarrow& (+1/2,\mathbf{1},2)\oplus (-1/2,\mathbf{2},+1)\oplus (+1/2,\mathbf{2},0) \oplus \nonumber \\
  &(+1/2,1,-2)\oplus (-1/2,2,-1)\oplus (+1/2,1,0)
\end{align}
from $\mathfrak{so}(5,1)\times \mathfrak{su}(2)_{R_6D}$ to $\mathfrak{so}(1,1)\times \mathfrak{su}(2)_L \times \mathfrak{u}(1)_\mathrm{diag}$.
These equations provide an isomorphism between spinors in six dimensions and holomorphic/anti-holomorphic differential forms on $M_4$. We choose the normalization such that
\begin{align}
  \psi^{1\alpha} \phi^2_\alpha &= \psi \overline{\phi}{} + \psi_i \overline{\phi}{}^i - \frac12 \psi_{ij} \overline{\phi}{}^{ij} \nonumber \\
  \psi^{2\alpha} \phi^1_\alpha &= - \phi \overline{\psi} - \phi_i \overline{\psi}{}^i + \frac12  \phi_{ij}\overline{\psi}{}^{ij} = - \overline{\psi^{1\alpha} \phi^2_\alpha} \label{eqn:pairing1}
\end{align}
reproduces the natural pairing of $(p,q)$-forms applied to polyforms. To see how this works, consider the polyforms $\Psi = \psi_{(0,0)} + \psi_{(1,0)} + \psi_{(2,0)}$ and $\Phi$. We now write the pairing \eqref{eqn:pairing1} as
\begin{equation}\label{eqn:pairing1b}
  \psi^{1\alpha} \phi^2_\alpha = ( \Psi, \Phi) = ( \psi_{(0,0)}, \phi_{(0,0)} ) + ( \psi_{(1,0)}, \phi_{(1,0)} ) - ( \psi_{(2,0)}, \phi_{(2,0)} )\,.
\end{equation}
By remembering that the pairing in this equation utilizes the Hodge-star on $M_4$, we finally obtain
\begin{equation}\label{eqn:pairing1c}
  ( \Psi, \Phi ) \frac{\omega^2}2 = \left. \sigma(\Psi) \wedge * \overline{\Phi} \right|_\mathrm{top}\,.
\end{equation}
Here $\sigma$ is required to obtain the right sign for the two-form part in \eqref{eqn:pairing1b}. It is defined as $\sigma(\psi) = (-1)^{(n-1)n/2}$ for a form $\psi$ of degree $n$.

As a first application of the presented spinor decomposition, we reduce the 6D superspace derivatives
\begin{equation}
  D^2_\alpha = \frac\partial{\theta^{1\alpha}} - i \theta^{2\beta} \partial_{\alpha\beta}
    \quad \text{and} \quad
  D^1_\alpha = - \frac\partial{\theta^{2\alpha}} - i \theta^{1\beta} \partial_{\alpha\beta}\,.
\end{equation}
to the $(0,2)$ superspace derivative
\begin{align}
  \mathbf{D}_+ &= \eta^{1\alpha} D^2_\alpha =
  \frac\partial{\partial\ttwoD} - i \tbtwoD \partial_+ - i \sqrt{2} \overline{\theta}{}^i \partial_i
    \quad \text{and} \nonumber \\ \label{eqn:DtwoD}
    \overline{\mathbf D}_+ &= \eta^{2\alpha} D^1_\alpha = -
    \frac\partial{\partial\tbtwoD} + i \ttwoD \partial_+ + i \sqrt{2} \theta^{\bar i} \partial_{\bar i}
\end{align}
in 2d by contraction with $\eta^{1\alpha}$ or $\eta^{2\alpha}$, respectively. In doing the calculation, we decompose the Grassmann coordinates $\theta^{s\alpha}$ in the same way as $\psi^{s\alpha}$ in \eqref{eqn:rightspinorexp}. In order to decompose the partial derivative $\partial_{\alpha\beta}$, we further need the branching
\begin{align}
  \psi^{2\alpha} \gamma^M_{\alpha\beta} \phi^{1\beta} \lambda_M =&
  - \overline{\psi} ( \phi \lambda_+ - \sqrt{2} \phi^{\bar i}  \lambda_{\bar i} ) - \overline{\psi}^i ( \sqrt{2} \phi \lambda_i + \sqrt{2} \phi_{ji} \lambda^j - \phi_i \lambda_- ) \nonumber \\
  &+ \frac12 \overline{\psi}{}^{ij} ( \phi_{ij} \lambda_+ + 2 \sqrt{2} \phi_i \lambda_j ) \label{eqn:psigammaphi}
\end{align}
of a vector, say $\lambda_M$. Again, this decomposition is compatible with the corresponding branching
\begin{equation}
  (\mathbf{6},\mathbf{1}) \rightarrow (-1,\mathbf{1},0) \oplus (+1,\mathbf{1},0) \oplus (0,\mathbf{2},+1) \oplus (0,\mathbf{2},-1) = \lambda_+ \oplus \lambda_- \oplus \lambda_{(1,0)} \oplus \lambda_{(0,1)}
\end{equation}
of $\mathfrak{so}(5,1)\times \mathfrak{su}(2)_{R_6D} \supset \mathfrak{so}(1,1)\times \mathfrak{su}(2)_L \times \mathfrak{u}(1)_\mathrm{diag}$. We here explicitly write the spinor index for the Grassmann variables $\ttwoD$ and $\tbtwoD$ in 2d to be compatible with the standard conventions. Except for the last term, \eqref{eqn:DtwoD} looks exactly like the $(0,2)$ derivatives given in \cite{Witten:1993yc}. In order to get rid of this term, we perform the coordinate transformation
\begin{equation}\label{eqn:tildexM}
  \tilde x^M = \begin{pmatrix} x^+ & x^- & x^i + i \sqrt{2} \, \ttwoD
    \overline{\theta}{}^i & x^{\bar i} + i \sqrt{2}\,\tbtwoD \theta^{\bar i} \end{pmatrix}\,.
\end{equation}
The reduction of the supercharges to two dimensions works in the same way as for $D^s_\alpha$.

\subsection*{Vector multiplet}
As a consistency check of the compactification procedure on a K\"ahler manifold $M_4$ presented in the last subsection, we apply it to the 6D vector multiplet. It is fully determined in terms of the gauge potential $\mathcal{A}_{\alpha\beta}$ of the covariant derivative
\begin{equation}
  \mathcal{D}_\alpha^s = D_\alpha^s + i \mathcal{A}_\alpha^s
\end{equation}
and contains a vector $A_M$, a right-handed symplectic Majorana-Weyl fermion $\lambda^{s\alpha}$ and an auxiliary field $Y^{st}$ in the adjoint of $SU(2)$. Our starting point is the real superfield
\begin{equation}
  \mathcal{A}_{\alpha\beta} = A_{\alpha\beta} - 2 (\theta^{2\gamma} \lambda^{1\delta} - \theta^{1\gamma} \lambda^{2\delta})\epsilon_{\gamma\delta\alpha\beta} + \dots\,.
\end{equation}
In order to identify the $(0,2)$ superfield resulting after the compactification, we take a closer look at the components
\begin{equation}
  \mathcal{A}_- = \tilde\gamma^{\alpha\beta}_- \mathcal{A}_{\alpha\beta} = V + \overline{\theta}{}^{ij} \overline{\Lambda}{}_{ij} + \theta^{\bar i\bar j} \Lambda{}_{\bar i\bar j} + \dots
    \quad \text{and} \quad
  \mathcal{A}_{\bar i} = \tilde\gamma^{\alpha\beta}_{\bar i} \mathcal{A}_{\alpha\beta} = \mathds{A}_{\bar i} + \dots
\end{equation}
with the corresponding superfields
\begin{align}
  V &= A_- - 2 \tbtwoD \lambda - 2 \ttwoD \overline{\lambda}  + \dots \\
  \mathds{A}_{\bar i} &= A_{\bar i} + 2 \sqrt{2} \ttwoD \lambda_{\bar i} + \dots \\
  \Lambda_{\bar i\bar j} &= \lambda_{\bar i\bar j} + \dots
\end{align}
in two dimensions. These are exactly the superfields we studied in \cite{Apruzzi:2016iac}.
The gaugino $\lambda$ has negative chirality and therewith spin 1/2 (in our conventions it is a left--mover).
All superfields transform in bundle $\mathcal{E}$ which combines the adjoint action of the gauge algebra
and diffeomorphisms of the K\"ahler surface $M_4$. The latter are not dynamic. However,
they have to be taken into account when counting fermionic zero modes weighted by the chirality.
This zero mode count is given by the index $\chi(M_4, \mathcal{E})$ which reduces to $\chi(M_4, T^\ast)$ for
vanishing gauge field flux on $M_4$.

\subsection*{Hypermultiplet}
A hypermultiplet is captured by the on-shell superfield $\mathcal{Q}^s$ transforming in the fundamental of the $SU(2)$ $R$-symmetry. Further, it has to satisfy the superspace constraint
\begin{equation}
  \mathcal{D}^{(s}_\alpha \mathcal{Q}^{t)} = 0\,,
\end{equation}
where $\mathcal{D}^s_\alpha$ is the gauge covariant superspace derivative which fulfills the relation
\begin{equation}
  \{\mathcal{D}_\alpha^s, \mathcal{D}_\beta^t\} = -2 i \Omega^{st} D_{\alpha\beta}\,.
\end{equation}
It contains a complex scalar $q^s$ in the fundamental of $SU(2)$ and a left-handed Weyl-fermion $\chi_\alpha$.

To obtain the 2d superfield after the reduction, we start from the expansion
\begin{equation}
  \mathcal{Q}^s = q^s + \theta^{s\alpha} \chi_\alpha + \dots
\end{equation}
of the 6D superfield and express the Weyl-fermion in terms of antiholomorphic differential forms on the K\"ahler surface $M_4$. This gives rise to the expansion
\begin{equation}
  \chi_\alpha = \Big( \frac14 \chi_{\bar i\bar j} \gamma^{\bar i\bar j} + \frac1{\sqrt{2}} \chi_{-\bar i} \gamma^{-\bar i} + \chi \delta \Big){}_\alpha{}^\beta \eta^2_\beta\,.
\end{equation}

Calculating the term $\theta^{2\alpha}\chi_\alpha$ requires to introduce the $(2,0)$ form
\begin{equation}
  \mathbb{O}_{ij} = \frac12 \eta^{2\alpha} (\gamma_{ij})_\alpha{}^\beta \eta^2_\beta
    \quad \text{and its complex conjugate} \quad
  \overline{\mathbb{O}}_{\bar i\bar j} = \frac12 \eta^{1\alpha} (\gamma_{\bar i\bar j})_\alpha{}^\beta \eta^1_\beta\,.
\end{equation}
The normalization chosen here gives rise to
\begin{equation}
  (\mathbb{O}, \mathbb{O}) = \frac{1}{2} \mathbb{O}_{ij} \overline{\mathbb{O}}^{ij} = -1
\end{equation}
Using $\mathbb{O}$, we find
\begin{equation}
  \theta^{2\alpha} \chi_\alpha = \Big( \frac12 \overline{\theta}{}_{\bar i\bar j} \overline{\chi}{}  + \overline{\theta}{}_{\bar i} \overline{\chi}{}_{\bar j} + \frac12 \overline{\theta} \, \overline{\chi}{}_{\bar i\bar j} \Big)  \mathbb{O}^{\bar i\bar j}
\end{equation}
and finally
\begin{equation}
  \mathcal{Q}^1 = Q^1 + \theta^{\bar i} \Psi_{\bar i} + \dots
    \quad\text{and}\quad
  \mathcal{Q}^2 = Q^2 + \overline{\theta}{}_{\bar i} \Psi_{\bar j} \mathbb{O}^{\bar i\bar j} + \dots
\end{equation}
with the $(0,2)$ superfields
\begin{align}
  Q^1 &= q^1 + \ttwoD \chi + \dots & \Psi_{\bar i} = \chi_{\bar i} + \dots \\
  Q^2 &= q^2 + \frac{1}{2} \tbtwoD \chi_{\bar i\bar j} \mathbb{O}^{\bar i\bar j} + \dots\,.
\end{align}
where $Q^1$ is chiral and $Q^2$ is antichiral. The bundle assignments for these superfields are
\begin{equation}
  Q^1 \in K_{M_4}^{1/2} \otimes \mathcal{R}_1 \otimes \mathcal{R}_2^\vee \,, \quad
  Q^2 \in K_{M_4}^{-1/2} \otimes \mathcal{R}_1 \otimes \mathcal{R}_2^\vee \quad \text{and} \quad
  \Psi_{- \bar i} \in \Omega^{(0,1)}(K^{1/2} \otimes \mathcal{R}_1 \otimes \mathcal{R}_2^\vee)\,,
\end{equation}
if we assume that $q^s$ transforms in the bifundamental $\mathcal{R}_1 \otimes \mathcal{R}_2^\vee$. The square root of the canonical bundle $K^{1/2}$ in the bundle assignments is a consequence of the topological twist. In general, a superfield with charge $q$ under $J_R$ transforms in $K^{q/2}$. To make contact with the bundle assignments introduced in reference \cite{Apruzzi:2016iac},
we apply Serre duality to $Q^2$ resulting in
\begin{equation}
  (Q^2)^c \in K_{M_4}^{1/2} \otimes \mathcal{R}_1^\vee \otimes \mathcal{R}_2
\end{equation}
and suppress the $SU(2)$ $R$-symmetry indices. Zero modes are counted by the corresponding cohomology classes. If we count the massless fermions weighted by their chirality, we obtain the index $\chi(M_4, K_{M_4}^{1/2}\otimes\mathcal{R}_1 \otimes \mathcal{R}_2)$.  This matches perfectly with the results in \cite{Apruzzi:2016iac}. The SUSY transformation rules are given there.

\subsection*{Tensor multiplet}
We implement the tensor multiplet as a real, on-shell superfield $\mathcal{T}$ which is constrained by
\begin{equation}\label{eqn:constraintTM}
  D^{(s}_\alpha D^{t)}_\beta \mathcal{T} = 0\,.
\end{equation}
It contains a real scalar $t_{6D}$, the left-handed symplectic Majorana-Weyl fermion $\xi^s_\alpha$ and the anti-self-dual 3-form flux $H^{(-)}_{MNL}$, here in bispinor representation $H_{\alpha\beta}= \gamma_{\alpha\beta}^{MNL} H^{(-)}_{MNL}$. As the multiplet is on-shell, these fields have to fulfill the equations of motion
\begin{equation}\label{eqn:eomTM}
  \partial_M \partial^M t_{6D} = 0 \,, \quad
  \partial^{\alpha\beta} \xi^+_\beta = 0 \quad \text{and} \quad
  \partial^{\alpha\gamma} H_{\gamma\alpha} = 0\,.
\end{equation}
Remember that in six dimensions each 3-form can be decomposed into a self-dual and a anti-self-dual part
\begin{equation}
  H_{MNL} = H^{(+)}_{MNL} + H^{(-)}_{MNL} \quad\text{with}\quad
  H^{(\pm)}_{MNL} = \frac12 ( H_{MNL} \pm * H_{MNL} )\,.
\end{equation}
These two parts get mapped to spinor indices
\begin{equation}
  H^{\alpha\beta} = H^{(+)}_{MNL} (\tilde \gamma^{MNL})^{\alpha\beta}
    \quad\text{and}\quad
  H_{\alpha\beta} = H^{(-)}_{MNL} \gamma^{MNL}_{\alpha\beta}\,.
\end{equation}
Thus, we see that $H_{(\alpha\beta)}$ in the tensor multiplet's equations of motion~\eqref{eqn:eomTM} represent the anti-self-dual $H^{(-)}$. Because we are interested in massless fields in two dimensions, we expand it using all $b_2$ elements of $H^2(S, \mathbb{R})$. They decompose into $\dot\alpha=1,\dots,b_2^+$ self-dual and $\alpha=1,\dots,b_2^-$ anti-self-dual two forms.\footnote{At this point we overload our index convention. However, it will be clear from the context whether 6D spinor indices are used or we label a basis of $H^2(M_4, T^\ast)$ instead.}  Here, we use the same notation as in section \ref{sec:TWOD}. The splitting in self-dual and anti-self-dual cohomology representatives $\lambda_{\dot\alpha}$/$\lambda_\alpha$ give rise to the expansion
\begin{equation}
  H^{(-)} = \partial_{+} \varphi_L^{\dot \alpha}(x^+) \wedge \lambda_{\dot \alpha} + \partial_{-} \varphi_R^{\alpha}(x^-) \wedge \lambda_{\alpha}\,.
\end{equation}
Here the 2d scalars $\varphi_L^{\alpha}$/$\varphi_R^{\dot \alpha}$ have to be chiral/anti-chiral otherwise $H^{(-)}$ would violate the anti-self-duality constraint. Repeating this discussion for $H$ contributions with one or three legs in the internal space, we find that they have to vanish on-shell due to the self-duality constraint.

In order to obtain the $(0,2)$ multiplets in two dimensions, we again start from the superspace expansion
\begin{align}
  \mathcal{T} = & t_{6D} - \theta^{2\alpha} \xi^1_\alpha + \theta^{1 \alpha} \xi^2_\alpha + \theta^{2 \alpha} \theta^{1 \beta} ( H_{\alpha \beta} - i \partial_{\alpha \beta} \phi ) - i \theta^{2\alpha} \theta^{2\beta} \theta^{1\gamma} \partial_{\beta \gamma} \xi^1_\alpha  + \nonumber \\
  &  i \theta^{2\alpha} \theta^{1\beta} \theta^{1\gamma} \partial_{\alpha\beta} \xi^2_\gamma - \frac{1}{2} \theta^{2 \alpha} \theta^{2 \beta} \theta^{1 \gamma} \theta^{1 \delta} \partial_{\alpha \gamma} \partial_{\beta \delta} \phi - i \theta^{2\alpha} \theta^{2\beta} \theta^{1\gamma} \theta^{1\delta} \partial_{\beta \delta} H_{\alpha \gamma} + \dots
\end{align}
which solves the constraint~\eqref{eqn:constraintTM} on-shell. Going from the bispinor representation $H_{\alpha\beta}$ to the corresponding anti-self-dual 3-form, we obtain
\begin{align}
  \theta^{2\alpha} \gamma^{MNL}_{\alpha\beta} \theta^{1\beta} H^{(-)}_{MNL} = & - \overline{\theta} ( 6 \theta H_+{}_{\bar i}{}^{\bar i} - 6 \theta^{\bar i\bar j} H_{+\bar i\bar j}) - 12 \overline{\theta}{}^i \theta^{\bar j} H_{-\bar j i} \nonumber \\
  &+ \frac12 \overline{\theta}{}^{ij} ( 12 \theta H_{+ij} + 12 \theta_{ik} H_{+ j}{}^k )
\end{align}
and implement the branching
\begin{align}
  (\mathbf{10},\mathbf{1}) &\rightarrow (+1,\mathbf{3},0) \oplus (-1,\mathbf{1},2)
\oplus (-1,\mathbf{1}, 0) \oplus \xcancel{(0,\mathbf{2},1)} \oplus \xcancel{(0,\mathbf{2},-1)} \nonumber \\
  &= H_{-i\bar j} \text{ with } H_{-i}{}^i = 0\, \oplus H_{+ij} \oplus H_{+\bar i}{}^{\bar i} \oplus H_{+\bar i\bar j}\,.
\end{align}
from $\mathfrak{so}(5,1)\times \mathfrak{su}(2)_{6D,R}$ to $\mathfrak{so}(1,1)\times \mathfrak{su}(2)_L \times \mathfrak{u}(1)_\mathrm{diag}$.
The last products of irreps do not contribute because they correspond to three-from fluxes with only one or all three legs in the internal directions. As we already explained, such configurations are ruled out on-shell by the equations of motion. Finally, one finds
\begin{equation}\label{eqn:(0,2)expphi}
  \Phi = t_L + \operatorname{Re}\mathbb{\tau}_{R} - \overline{\theta}{}^i \Xi_i - \frac12 \overline{\theta}{}^{ij} \Xi_{ij} + 12 \theta^{\bar i} \overline{\theta}{}^j \mathcal{H}_{- \bar i j} + \dots + \text{c.c.}
\end{equation}
which allows us to identify the $(0,2)$ multiplets
\begin{align}
  \operatorname{Re}\mathbb{\tau}_{R}(\tilde x^M) &= t_R + \ttwoD \overline{\xi} - \tbtwoD \xi - 6 \tbtwoD \ttwoD H_{+\bar i}{}^{\bar i} &
    \Xi_i(\tilde x^M) &= \xi_i + i \tbtwoD \ttwoD \partial_+ \xi_i \nonumber \\
  \Xi_{ij}(\tilde x^M) &= \xi_{ij} + 12 \ttwoD H_{+ ij} + i \tbtwoD \ttwoD \partial_+ \xi_{ij} &
    \mathcal{H}_{-\bar i j}(\tilde x^M) &= H_{-\bar i j}.
\end{align}
We split $t_{6D} = t_L + t_R$ into left- and right-moving parts, as explained in section \ref{sec:SEVEN}. It is important that we use the coordinates $\tilde x^M$ which we introduce in \eqref{eqn:tildexM} for these multiplets. For $x^M$ the results would not be conclusive because the 2d superspace derivatives do not have the standard form. In principle, we face the same problem for the vector and hypermultiplet in the last subsections. But there we only considered the leading components which are always free from obstructions.

Only the multiplet $\operatorname{Re}\mathbb{\tau}_{R}$ appears somewhat exotic. It is real and contains the right-moving part of the tensor multiplets scalar as leading contribution. Its highest component is a chiral boson $H_+$ which is proportional to the K\"ahler form $\omega$ on $M_4$.  As we explained in section~\ref{sec:SEVEN}, it is, in fact, not mysterious:  it is simply the real part of the corresponding CS multiplet.  The Hodge number $2h^{2,0}$ counts the number of massless multiplets $\Xi_{ij}$ (weighted by 2 because they are complex, and we counting real degrees of freedom) and the $1$ corresponds to the real multiplet $\operatorname{Re}\mathbb{\tau}_{L}$. All $b_2^{(-)}$ anti-chiral bosons form singlets under supersymmetry and are captured by $\mathcal{H}_{-\bar i\bar j}$. Finally $\Xi_i$ hosts the two gauginos of the $U(1)^2$ abelian gauge theory arising from compactifying the 6D $B$-field with one leg in the internal and one leg in the external directions. Off-shell this multiplet also contains the corresponding field strength and an auxiliary field as bosonic degrees of freedom. On-shell all bosonic contributions vanish and we are left with the gauginos. From $\operatorname{Re}\mathbb{\tau}_{R}$ we can derive the abelian current multiplet
\begin{equation}\label{eqn:PSI}
  \Xi = \mathbf{D}_+ \operatorname{Re}\mathbb{\tau}_{R} = \xi + i \ttwoD ( H_{+\bar i}{}^{\bar i} + \partial_+ t_L ) - i \tbtwoD \ttwoD \partial_+ \xi
\end{equation}
which will be useful in deriving the superconformal current multiplet in the next subsection.

Finally we calculate the $(0,2)$ SUSY variation
\begin{equation}
  \delta_\epsilon = \overline{\epsilon}{}^+ \overline{Q}_+ - \epsilon^+ Q_+
\end{equation}
of the real multiplet $\operatorname{Re}\mathbb{\tau}_{R}$. It gives rise to
\begin{align}
  \delta_\epsilon t_R &= \overline{\epsilon}{}^+ \xi_+ - \epsilon^+ \overline{\xi}{}_+ &
  \delta_\epsilon \xi_+ &= - \epsilon^+ ( 6 H_{+\bar i}{}^{\bar i} + i \partial_+ t_R ) \nonumber \\
  H_{+\bar i}{}^{\bar i} &= \frac{i}6 ( \epsilon^+ \partial_+ \overline{\xi}{}_+ + \overline{\epsilon}{}^+ \partial_+ \xi{}_+) &
  \delta_\epsilon \overline{\xi}{}_+ &= - \overline{\epsilon}{}^+ ( 6 H_{+\bar i}{}^{\bar i} - i \partial_+ t_R )
\end{align}
and is compatible with $t_R$ and $i H_{+ \bar i}{}^{\bar i}$ being real.

\subsection*{Supercurrents for Tensor Multiplet}\label{sec:freetensorscurrents}
The chiral bosons arising from the reduction of the tensor multiplet make it difficult to write down an action for the $(0,2)$ multiplets discussed in the last subsection. But we still can explicitly calculate the energy momentum tensor of the two dimensional theory. To this end, we first consider the 6D real supercurrent
\begin{equation}
  J = \frac12 \mathcal{T}^2\,.
\end{equation}
It gives rise to the superfield
\begin{equation}
  V^{st}_{\alpha\beta} = D^{(s}_\alpha D^{t)}_\beta J = v^{st}_{\alpha\beta} + \dots
\end{equation}
which hosts the conserved $R$-symmetry current as its leading part $v^{12}_{\alpha\beta}$ \cite{Howe:1983fr}. We are interested in the component $V^{21}_{\alpha\beta}$. It is associated to the Cartan generator $J_R$ of $\mathfrak{su}(2)$ and thus the $R$-symmetry in 2d. The two antisymmetric spinor indices this multiplet carries can be mapped to a vector index
\begin{equation}
  V^{12}_M = \tilde\gamma_M^{\alpha\beta} V^{12}_{\alpha\beta} = \tilde\gamma_M^{\alpha\beta} v^{12}_{\alpha\beta} + \dots\,.
\end{equation}
In the $\mathbb{R}^{1,1}$ directions the leading components of this vector are:
\begin{align}
    V^{12}_+ &= \xi \overline{\xi} - \frac12 \xi_{ij} \overline{\xi}{}^{ij} + \dots \\
    V^{12}_- &= \xi_i \overline{\xi}{}^i.
\end{align}
These leading contributions are sufficient to figure out the full expression for $V^{12}_M$ in terms of the superfields we introduced in the last subsection. We obtain
\begin{equation}
  V_+^{12} = \overline{D}{}^+ \mathcal{T} D^- \mathcal{T} - \frac12 \Xi_{ij} \overline{\Xi}{}^{ij} = \Xi \overline{\Xi} - \frac12  \Xi_{ij} \overline{\Xi}{}^{ij}
    \quad \text{and} \quad
  V_-^{12} = - \Xi_i \overline{\Xi}^i~,
\end{equation}
where we have used \eqref{eqn:PSI}. It is straightforward to check that these two superfields represent conserved currents in 2d, namely they fulfill $\partial_- V_+^{12} = \partial_+ V_-^{12} = 0$. The latter is a singlet, while the former is the superconformal current of a 2d SCFT. It has exactly the form one would expect from the $(0,2)$ Sugawara construction of chiral currents \cite{Hull:1989py}.

\section{Effective Strings from 4d $\mathcal{N}=2$ SCFTs \label{app:MN}}

In this Appendix we provide further details on the spacetime filling strings
which are a necessary ingredient in the construction of our 2d theories. In
string theory terms, these effective strings arise from $N$ D3-branes wrapped on
the collapsing $\mathbb{P}^{1}$ factor of $M_{4}\times\mathbb{P}^{1}$, the
six-manifold wrapped by the seven-brane. Here, the $\mathbb{P}^1$ factor has
self-intersection $-n$ in the base of an F-theory model.

Now, in the limit where the volume of
the $\mathbb{P}^{1}$ is very large, we can approximate this system by the 4d
theory defined by D3-branes probing a seven-brane with gauge group $G$. For
the simple NHCs, this realizes 4d $\mathcal{N}=2$ superconformal field
theories where the seven-branes contribute a flavor symmetry $G$. Various aspects
of these theories have been studied for example in references
\cite{Minahan:1996fg,Minahan:1996cj,Argyres:1995jj,Seiberg:1994aj, Sen:1996vd,Banks:1996nj, Aharony:2007dj,
Argyres:2007tq,Shapere:2008zf, Noguchi:1999xq,Tachikawa:2009tt,
Heckman:2010fh, Heckman:2010qv, DelZotto:2016fju}.
In what follows we will include the free hypermultiplet describing motion of the center of mass for the
D3-branes inside the seven-brane.

In addition to the flavor symmetry localized on the seven-brane, we also have
the geometric isometries associated with directions both inside the
seven-brane, i.e., an $SO(4)_{\Vert}$ symmetry, and transverse to the
seven-brane, i.e. $U(1)_{\bot}$. The R-symmetry of the 4d $\mathcal{N}=2$ SCFT
is realized by $SU(2)_{R}\times U(1)_{\bot}$, where we use the isomorphism
$SO(4)_{\Vert}\simeq SU(2)_{L}\times SU(2)_{R}/%
\mathbb{Z}
_{2}$. The generator for the 4d $\mathcal{N}=1$ R-symmetry is given by the
linear combination:%
\begin{equation}
R_{\mathcal{N}=1}=\frac{1}{3}R_{\mathcal{N}=2}+\frac{2}{3}I^{(3)},
\end{equation}
where $R_{\mathcal{N}=2}$ is the generator associated with $U(1)_{\bot}$ and
$I^{(3)}$ is the generator of the Cartan subalgebra of $SU(2)_{R}$, with
eigenvalues $\pm1$ in the fundamental representation. For the theories in
question, the usual central charges%
\begin{equation}
a=\frac{3}{32}\left(  3\mathrm{Tr}R_{\mathcal{N}=1}^{3}-\mathrm{Tr}%
R_{\mathcal{N}=1}\right)  \ ,\qquad c=\frac{1}{32}\left(  9\mathrm{Tr}%
R_{\mathcal{N}=1}^{3}-5\mathrm{Tr}R_{\mathcal{N}=1}\right)  \ ,
\end{equation}
reduce to:
\begin{equation}
a=\frac{1}{8}N^{2}n+\frac{1}{12}Nh_{G}^{\vee}\ ,\qquad c=\frac{1}{8}%
N^{2}n+\frac{1}{8}Nh_{G}^{\vee}\ .
\end{equation}
Here, we include the contribution from the free hypermultiplet parameterizing
the center of mass degree of freedom for the stack of D3-branes inside of the seven-branes.

The anomalies for these theories have been computed in references
\cite{Argyres:2007cn, Cheung:1997id, Aharony:2007dj,
Shapere:2008zf, Tachikawa:2009tt}. It is convenient to assemble these expressions into a single
anomaly polynomial which depends on the background field strengths $F_{G}$,
$F_{L}$, $F_{R}$, $F_{\bot}$, and $T$, the background tangent bundle:%
\begin{align}
I_{4d} &  =\frac{k_{\bot GG}}{4}\left(  c_{1}(F_{\bot})\mathrm{Tr}F_{G}%
^{2}\right)  +\frac{k_{\bot\bot\bot}}{6}\left(  c_{1}(F_{\bot})^{3}\right)
+\frac{k_{\bot TT}}{24}\left(  c_{1}(F_{\bot})p_{1}(T)\right)  \\
&  +\frac{k_{\bot LL}}{4}\left(  c_{1}(F_{\bot})\mathrm{Tr}F_{L}^{2}\right)
+\frac{k_{\bot RR}}{4}\left(  c_{1}(F_{\bot})\mathrm{Tr}F_{R}^{2}\right)  .
\end{align}
where the explicit values are:%
\begin{align}
k_{\bot GG} &  =-Nn\text{, \ \ }k_{\bot\bot\bot}=-2Nh_{G}^{\vee}\text{,
\ \ }k_{\bot TT}=2Nh_{G}^{\vee}\text{, \ \ }\label{kfirst}\\
k_{\bot LL} &  =-N\left(  Nn-\frac{h_{G}^{\vee}}{3}\right)  \text{,
\ \ }k_{\bot RR}=N\left(  Nn+\frac{h_{G}^{\vee}}{3}\right)  ,\label{ksecon}%
\end{align}
which are in turn obtained by evaluating the anomalies:%
\begin{align}
k_{\bot GG}\times\delta^{AB} &  =\text{Tr}(R_{\mathcal{N}=2}T^{A}T^{B})\text{,
\ \ }k_{\bot\bot\bot}=\text{Tr}(R_{\mathcal{N}=2}^{3})\text{, \ \ }k_{\bot
TT}=-\text{Tr}(R_{\mathcal{N}=2})\text{,}\\
k_{\bot LL}\times\delta^{ab} &  =\text{Tr}(R_{\mathcal{N}=2}I_{L}^{a}I_{L}%
^{b})\text{, \ \ }k_{\bot RR}\times\delta^{ab}=\text{Tr}(R_{\mathcal{N}%
=2}I_{R}^{a}I_{R}^{b})\text{,}%
\end{align}
where in the above, we have introduced the Lie algebra generators $T^{A}$ for
the flavor group $G$ and generators $I_{L}^{a}$ and $I_{R}^{b}$ for
$SU(2)_{L}$ and $SU(2)_{R}$.

Next, consider the dimensional reduction of the 4d theory on a $\mathbb{P}%
^{1}$. In this case, the $SU(2)_{R}$ factor of $SO(4)_{\Vert}\simeq
SU(2)_{L}\times SU(2)_{R}/%
\mathbb{Z}
_{2}$ remains as a spectator, and we only use the abelian factor coming from
the $U(1)_{\bot}$ factor of the 4d $\mathcal{N}=2$ SCFT. This is required in
order to match to the symmetries preserved by the $\mathcal{N}=2$ SCFT. So, we
introduce the specific decomposition:%
\begin{equation}
R_{4d}^{\mathcal{N}=2}=R_{2d,R}\otimes K_{\mathbb{P}^{1}}^{1/2},
\end{equation}
where $K_{\mathbb{P}^{1}}$ is the canonical bundle on $\mathbb{P}^{1}$. The
overall twist is dictated by the same analysis presented, for example in
reference \cite{Benini:2015bwz}. The first Chern class splits up as:
\begin{equation}
c_{1}\left(  F_{\bot}\right)  =c_{1}(R_{2d,R})+\frac{1}{2}c_{1}(K_{\mathbb{P}%
^{1}})\ .
\end{equation}
To obtain the anomaly polynomial of the 2d effective strings, we now integrate
over the $\mathbb{P}^{1}$. To this end, we make the subsitutions
$p_{1}(T)\rightarrow p_{1}(TZ)$, and use the fact that:%
\begin{equation}
\int_{\mathbb{P}^{1}}c_{1}(K_{\mathbb{P}^{1}})=-2.
\end{equation}
The 2d anomaly polynomial is then given by:%
\begin{align}
I_{2d} &  =\frac{Nn}{4}\mathrm{Tr}F_{G}^{2}+Nh_{G}^{\vee}\left(
c_{1}(R_{2d,R})^{2}-\frac{1}{12}p_{1}(TZ)\right)  \\
&  +\frac{N}{4}\left(  Nn-\frac{h_{G}^{\vee}}{3}\right)  \mathrm{Tr}F_{L}%
^{2}-\frac{N}{4}\left(  Nn+\frac{h_{G}^{\vee}}{3}\right)  \mathrm{Tr}F_{R}%
^{2}.
\end{align}
On a curved background where $F_{L}$ and $F_{R}$ are generically broken, this
reduces to:%
\begin{equation}
I_{2d}=\frac{Nn}{4}\mathrm{Tr}F_{G}^{2}+Nh_{G}^{\vee}\left(  c_{1}%
(R_{2d,R})^{2}-\frac{1}{12}p_{1}(TZ)\right)  ,
\end{equation}
which is the expression used in section \ref{sec:ANOMO}.

\subsection*{Weakly Coupled Example}

To check that we have correctly fixed all normalizations, as well as all
relative signs, it is helpful to return to a weakly coupled example where all
contributions to the anomaly polynomial can be explicitly evaluated. Along these lines,
we consider the 4d $\mathcal{N}=2$ SCFT defined by $SU(2)$ gauge theory with
four flavors in the fundamental representation, and a decoupled free
hypermultiplet. This is the special case $N=1$ and $n=4$. In terms of
$\mathcal{N}=1$ multiplets, the matter content consists of four chiral
multiplets $Q\oplus Q^{c}$, another pair of chiral multiplets $H\oplus H^c$, a
chiral multiplet $\Phi$ in the adjoint representation of $SU(2)$, and a vector
multiplet $V$. In this presentation, we have a manifest $U(4)$ flavor symmetry
which enhances to $SO(8)$. In our normalization conventions for generators,
the fermionic components of each multiplet (denoted by $\psi$) have charge
assignments:%
\begin{equation}%
\begin{tabular}
[c]{|c|c|c|c|c|c|}\hline
& $I_{L}^{(3)}$ & $I_{R}^{(3)}$ & $U(1)_{\bot}$ & $U(4)$ & $SU(2)_{D3}%
$\\\hline
$\psi_{Q}$ & $0$ & $0$ & $-1$ & $\mathbf{4}$ & $\mathbf{2}$\\\hline
$\psi_{Q^{c}}$ & $0$ & $0$ & $-1$ & $\overline{\mathbf{4}}$ & $\mathbf{2}%
$\\\hline
$\psi_{H}$ & $+1$ & $0$ & $-1$ & $\mathbf{1}$ & $\mathbf{1}$\\\hline
$\psi_{H^c}$ & $-1$ & $0$ & $-1$ & $\mathbf{1}$ & $\mathbf{1}$\\\hline
$\psi_{\Phi}$ & $0$ & $-1$ & $+1$ & $\mathbf{1}$ & $\mathbf{3}$\\\hline
$\psi_{V}$ & $0$ & $+1$ & $+1$ & $\mathbf{1}$ & $\mathbf{3}$\\\hline
\end{tabular}
.
\end{equation}
Let us now verify that our formulae agree with the normalization conventions
used in lines (\ref{kfirst}) and (\ref{ksecon}). In the obvious notation, we
have:%
\begin{align}
k_{\bot GG} &  =k_{\bot GG}(\psi_{Q})+k_{\bot GG}(\psi_{Q^{c}}%
)=-2-2=-4\text{,}\\
k_{\bot\bot\bot} &  =k_{\bot\bot\bot}(\psi_{Q})+k_{\bot\bot\bot}(\psi_{Q^{c}%
})+k_{\bot\bot\bot}(\psi_{H})+k_{\bot\bot\bot}(\psi_{H^c})+k_{\bot\bot\bot}%
(\psi_{\Phi})+k_{\bot\bot\bot}(\psi_{V}) \nonumber \\
&  =-8-8-1-1+3+3=-12,\\
k_{\bot TT} &  =k_{\bot TT}(\psi_{Q})+k_{\bot TT}(\psi_{Q^{c}})+k_{\bot
TT}(\psi_{H})+k_{\bot TT}(\psi_{H^c})+k_{\bot TT}(\psi_{\Phi})+k_{\bot TT}%
(\psi_{V}) \nonumber \\
&  =+8+8+1+1-3-3=+12,\\
k_{\bot LL} &  =k_{\bot LL}(\psi_{H})+k_{\bot LL}(\psi_{H^c})=-1-1=-2\text{,}\\
k_{\bot RR} &  =k_{\bot RR}(\psi_{\Phi})+k_{\bot RR}(\psi_{V})=3+3=6\text{.}%
\end{align}
Returning to lines (\ref{kfirst}) and (\ref{ksecon}) and plugging in $N=1$ and
$n=4$, we obtain the same values.

\newpage

\bibliographystyle{utphys}
\bibliography{DGLSM}

\end{document}